\documentclass[a4paper,fleqn]{cas-dc}

\usepackage[numbers]{natbib}

\usepackage{lineno,hyperref}
\modulolinenumbers[5]
\usepackage{mathtools}
\usepackage{url}
\usepackage{tikz}
\usepackage{caption}
\usepackage{enumitem}
\usepackage{booktabs}
\usepackage{multirow,tabularx}
\usepackage{ amssymb }
\usepackage{upquote}
\usepackage{amsthm}
\usepackage{amsmath}
\usepackage{amsfonts}
\usepackage{stmaryrd}
\usepackage{arydshln}
\usepackage{listings}
\usepackage{tikz}
\usetikzlibrary{shapes}
\tikzstyle{line} = [draw, -stealth, thick]
\tikzstyle{elli}=[draw, ellipse, fill=yellow!50,minimum height=8mm, text width=3.9em, text centered]
\tikzstyle{block} = [draw, rectangle, fill=blue!20, text width=5em, text centered, minimum height=7mm, node distance=7em, rounded corners]
\tikzstyle{gblock} = [draw, rectangle, fill=green!20, text width=5em, text centered, minimum height=7mm, node distance=7em, rounded corners]
\tikzstyle{oblock} = [draw, rectangle, fill=orange!20, text width=5em, text centered, minimum height=7mm, node distance=7em, rounded corners]

\usepackage{array}
\newcolumntype{P}[1]{>{\centering\arraybackslash}p{#1}}
\newcolumntype{R}[1]{>{\tt\raggedleft\arraybackslash}p{#1}}

\newcommand{\rdfs}{\textsc{rdf}}
\newcommand{\iri}{\textsc{iri}}
\newcommand{\pp}{{\beta_{n}}}
\newcommand{\ppi}{{B_{n}}}
\newcommand{\ppp}{{\beta_{d}}}
\newcommand{\pppi}{{B_{d}}}
\newcommand{\PP}{{\Pi}}
\newcommand{\vars}{\mathcal{V}}
\newcommand{\fun}{\mathcal{F}}
\newcommand{\funs}{\mathcal{F}_s}
\newcommand{\preds}{\mathcal{P}_s}
\newcommand{\pat}{gp}
\newcommand{\tpat}{tp}
\newcommand{\gr}{G}
\newcommand{\ag}{\kod{D}}

\newcommand{\agfn}{I_{n}}

\newcommand{\dg}{\kod{G}_{\kod{d}}}

\newcommand{\vibl}{\kod{VIBL}}
\newcommand{\ibl}{\kod{IBL}}
\newcommand{\ible}{\kod{IBLe}}
\newcommand{\err}{err}
\newcommand{\seleq}{\sim}
\newcommand{\comp}{\simeq}

\newcommand{\sigvibl}{\sigma_{\kod{t}}}

\newcommand{\siginv}{({\sigvibl})^{-1}}

\newcommand{\context}{\mathit{cx}}
\newcommand{\povezano}{\leftrightarrow}
\newcommand{\definisemini}{\rightarrow}
\newcommand{\definisenimi}{\leftarrow}
\newcommand{\zadovoljava}{\Vdash}

\newcommand{\intrprt}{\mathcal{I}}
\newcommand{\model}[1]{(\mathfrak{D}, #1)}
\newcommand{\kod}[1]{\text{\texttt{#1}}}

\newcommand{\bool}{bool}

\newcommand{\wwwc}{\textsc{w3c}}

\newcommand{\fol}{\textsc{fol}}

\newcommand{\Eqdef}{:=}

\newcommand{\sparql}{\textsc{sparql}}

\newcommand{\error}{\texttt{err}}

\def\ojoin{\setbox0=\hbox{$\bowtie$}%
	\rule[+.02ex]{.25em}{.8pt}\llap{\rule[\ht0]{.25em}{.8pt}}}
\def\leftouterjoin{\mathbin{\ojoin\mkern-5.8mu\bowtie}}

\newcommand{\semd}[2]{\llbracket{#1}\rrbracket^{#2}}
\newcommand{\semdg}[3]{\llbracket{#1}\rrbracket^{#2}_{#3}}
\newcommand{\valuation}[2]{[[{#1}]]_{#2}}

\usepackage{ dsfont }

\newcommand{\specs}{\textsc{SpeCS}}
\newcommand{\afmu}{\textsc{afmu}}
\newcommand{\ts}{\textsc{ts}}
\newcommand{\jsag}{\textsc{jsag}}
\newcommand{\sa}{\textsc{sa}}
\newcommand{\qcan}{\textsc{QCan}}

\newcommand{\dataq}{\mathds{D}}

\newcommand{\semdq}[1]{\semd{#1}{\dataq}}

\theoremstyle{plain}
\newtheorem{thm}{Theorem}[section] % reset theorem numbering for each chapter
\newtheorem{lemma}[thm]{Lemma}

\newtheorem{innercustomgeneric}{\customgenericname}
\providecommand{\customgenericname}{}
\newcommand{\newcustomtheorem}[2]{%
	\newenvironment{#1}[1]
	{%
		\renewcommand\customgenericname{#2}%
		\renewcommand\theinnercustomgeneric{##1}%
		\innercustomgeneric
	}
	{
		\theoremstyle{definition}
		\endinnercustomgeneric
	}
}
\newcustomtheorem{customlemma}{Lemma}
\newcustomtheorem{customthm}{Theorem}

\providecommand{\customgenericname}{}

\newcustomtheorem{customthmproof}{}

\usepackage{hyperref}

\newtheorem{procedure}{Procedure}[section]

\usepackage{xpatch}
\makeatletter
\xpatchcmd{\@thm}{\thm@headpunct{.}}{\thm@headpunct{}}{}{}
\makeatother

\usepackage{xpatch}
\makeatletter
\AtBeginDocument{\xpatchcmd{\@thm}{\thm@headpunct{.}}{\thm@headpunct{}}{}{}}
\makeatother

\theoremstyle{definition}
\newtheorem{defn}{Definition}[section] % definition numbers are dependent on theorem numbers

\usepackage[most]{tcolorbox}
\def\startfig{%
	\begin{tcolorbox}[colback=gray!10!white,colframe=gray!75!white,coltitle=red!20!black,boxsep=-10pt,left*=2mm,right*=2mm,top=5mm,bottom=4mm]%
	}
\def\captionfig#1{%
	\vspace*{-2mm}
	\begin{tcolorbox}[colback=gray!25!white,colframe=gray!10!white,bottom=2mm,boxsep=-5pt]\vspace*{-2mm}\caption{#1}%
	}
\def\labelfig#1{%
	\label{#1}\end{tcolorbox}\vspace*{-2mm}\end{tcolorbox}%
}
\def\startsubfig{%
	\begin{tcolorbox}[colback=gray!5!white,colframe=gray!75!white,coltitle=red!20!black,boxsep=-4pt,left*=2mm,right*=2mm,top=2mm]%
	}
	\def\endsubfig{%
	\end{tcolorbox}\vspace*{-2mm}%
}
\def\endfig{%
	\end{tcolorbox}%
}
\def\startsubfigformula{%
	\begin{tcolorbox}[colback=gray!5!white,colframe=gray!75!white,coltitle=red!20!black,boxsep=-4pt,left*=2mm,right*=2mm,top=2mm,bottom=2mm]%
}%

\lstdefinestyle{sparqlgrammar}{
	morekeywords={Query,Vars,GPatt,TPatt,Subject,Predicate,Object,Expr,Cond,UnOp,BinOp,Projs,Proj,SubQuery},
	basicstyle=\ttfamily,
	keywordstyle=\color{blue},
	moredelim=**[is][\color{teal}]{@}{@},
}
\definecolor{LightGray}{rgb}{0.97,0.97,0.97}
\lstdefinestyle{SPARQL}{
	basicstyle=\ttfamily,
	backgroundcolor=\color{LightGray},
	columns=fullflexible,
	breaklines=false,
	sensitive=false,
	frame=bt,
	aboveskip=1em,
	belowskip=1em,
	xleftmargin=.5em,
	xrightmargin=.5em,
	framexleftmargin=.5em,
	framextopmargin=.5em,
	framexbottommargin=.5em,
	framexrightmargin=.5em,
	tabsize = 2,
	showstringspaces=false,
	morecomment=[l][\color{gray}]{\#},       % comments
	morecomment=[n][\color{blue}]{<http}{>}, % uris
	morestring=[b][\color{OliveGreen}]{\"},  % strings
	keywordsprefix=?,
	classoffset=0,
	keywordstyle=\color{Sepia},
	morekeywords={},
	classoffset=1,
	keywordstyle=\color{Purple},
	morekeywords={rdf,rdfs,owl,xsd,purl},
	classoffset=2,
	keywordstyle=\color{MidnightBlue},
	morekeywords={
		SELECT,CONSTRUCT,DESCRIBE,ASK,WHERE,FROM,NAMED,PREFIX,BASE,OPTIONAL,
		FILTER,GRAPH,LIMIT,OFFSET,SERVICE,UNION,EXISTS,NOT,BINDINGS,MINUS,a
	}
}

\usepackage{soul}
\definecolor{lightgrey}{rgb}{0.925, 0.925, 0.925}
\sethlcolor{lightgrey}
\makeatletter
\def\SOUL@hlpreamble{%
	\setul{}{3.5ex}% increased by 1ex
	\let\SOUL@stcolor\SOUL@hlcolor
	\dimen@\SOUL@ulthickness
	\dimen@i=-.75ex % increased by -0.25ex
	\advance\dimen@i-.5\dimen@
	\edef\SOUL@uldepth{\the\dimen@i}%
	\let\SOUL@ulcolor\SOUL@stcolor
	\SOUL@ulpreamble
}
\makeatother
\newcommand*{\codebox}[1]{{\hyphenchar\font=45\relax\hl{#1}}}

\makeatletter
\newcases{lrdcases}
{\quad}
{$\m@th\displaystyle{##}$\hfil}
{$\m@th\displaystyle{##}$\hfil}
{\lbrace}
{\rbrace}
\newcases{lrdcases*}
{\quad}
{$\m@th\displaystyle{##}$\hfil}
{{##}\hfil}
{\lbrace}
{\rbrace}
\makeatother

\begin{document}
	
	\let\WriteBookmarks\relax
	\def\floatpagepagefraction{1}
	\def\textpagefraction{.001}
	\shorttitle{S\&C of SpeCS}
	\shortauthors{M. Spasi\'c and M. Vujo\v{s}evi\' c Jani\v{c}i\'c}
	
	\title [mode = title]{Soundness and Completeness of SPARQL Query Containment Solver SpeCS}
	\tnotemark[1]
	
	\tnotetext[1]{This work was partially supported by Eurostars Project 3DFed (GA no. E!114681) and by the 
European Research Network on Formal Proofs (COST Action CA20111).}
	
	%	\tnotetext[2]{The second title footnote which is a longer text matter
		%		to fill through the whole text width and overflow into
		%		another line in the footnotes area of the first page.}

	\author[1,2]{Mirko Spasi\'c}[auid=000,bioid=1,
	orcid=0000-0002-9304-4007]
	\cormark[1]
	%	\fnmark[1]
	\ead{mirko@matf.bg.ac.rs}
	\ead[url]{www.matf.bg.ac.rs/~mirko}
	
	\address[1]{Faculty of Mathematics University of Belgrade, Studentski Trg 16, Belgrade, Serbia}
	
	\author[1]{ Milena Vujo\v{s}evi\'c Jani\v{c}i\'c}

	\address[2]{OpenLink Software, London, United Kingdom}
	
	\cortext[cor1]{Corresponding author}

\begin{abstract}
Tool \specs{} implements an efficient automated approach for reasoning about the \sparql{} query containment problem. In this paper, we prove the correctness of this approach. We give precise semantics of the core subset of \sparql{} language. We briefly discuss the procedure used for reducing the query containment problem into a formal logical framework. We prove that such reduction is both sound and complete for conjunctive queries, and also for some important cases of non-conjunctive queries containing operator \kod{union}, operator \kod{optional}, and subqueries. Soundness and completeness proofs are considered in both containment and subsumption forms.  

\end{abstract}
  
  
	\maketitle
	
	%%%%%%%%%%% The article body starts:
	
%\tableofcontents
	
%--------------------------------------------------------------------
\section{Introduction}
\label{sec:introduction}
\sparql{} (Simple Protocol and \rdfs{} Query Language) is a query language and data access protocol \cite{wwwc,prudhommeaux_sparql_2008} used for querying data in the form of Resource Description Framework (\rdfs) \cite{rdf,Lassila1999,Auer2011} within Semantic Web \cite{Kashyap2008,Hitzler2009,Hogan}. 
There is a large number of \rdfs{} data sources available \cite{lodc}. 
For achieving practical usability of these large amounts of data, executions of \sparql{} queries have to be highly optimized \cite{solving_with_specs}. 

\begin{figure}
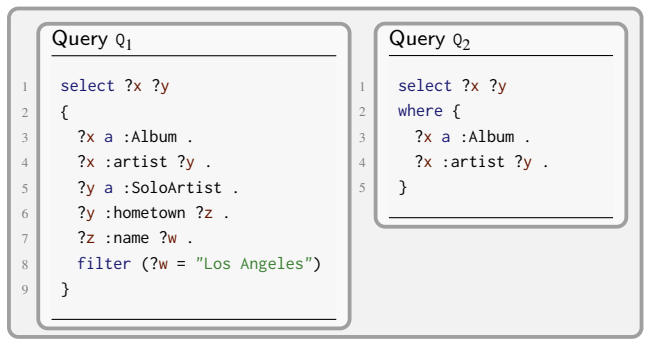

	\startfig
	\begin{footnotesize}
		\begin{minipage}{0.01\textwidth}
			~
		\end{minipage}
		\begin{minipage}{0.52\textwidth}
			\startsubfig
			Query $\kod{Q}_1$ \vspace*{-2mm}
			\begin{lstlisting}[style=sparql,numberstyle=\tiny\color{gray},numbers=left,numbersep=12pt,mathescape=true,escapeinside={(*}{*)}]
select ?x ?y
{
	?x a :Album .
	?x :artist ?y .
	?y a :SoloArtist .
	?y :hometown ?z .
	?z :name ?w .
	filter (?w = "Los Angeles")
}
			\end{lstlisting}
			\vspace*{-6px}
			\endsubfig
		\end{minipage}
		\begin{minipage}{0.01\textwidth}
			~
		\end{minipage}
		\begin{minipage}{0.42\textwidth}
	\startsubfig
	Query $\kod{Q}_2$ \vspace*{-2mm}
	\begin{lstlisting}[style=sparql,numberstyle=\tiny\color{gray},numbers=left,numbersep=12pt,escapeinside={(*}{*)}]
select ?x ?y
where {
	?x a :Album .
	?x :artist ?y .
}
	\end{lstlisting}
	\vspace*{-6px}
	\endsubfig
	\vspace*{38px}
	\end{minipage}
	\end{footnotesize}
	\vspace*{2px}
	\endfig
	\caption{$\kod{Q}_1$ (left) is a sub-query of $\kod{Q}_2$ (right). Additional filters that are present in $\kod{Q}_1$ make the results obtained by $\kod{Q}_1$ a subset of the results obtained by $\kod{Q}_2$. If a projection $?y$ is removed form $\kod{Q}_1$, then the query $\kod{Q}_1$ is subsumed by $\kod{Q}_2$.}
%	$\kod{Q}_1$ returns albums and their authors who are solo artists and whose hometown is Los Angeles, while $\kod{Q}_2$ has no such filters and, apart of the results of $\kod{Q}_1$, returns also all the other albums and their authors that are present in the given dataset.}
	\label{fig:containment1}
\end{figure}

One of the central problems for \sparql{} query optimizations is \textit{query containment}, namely deciding if each result
of one query is at the same time a result of another query despite of a dataset being queried, i.e.~this property should hold for any given dataset \cite{Chandra1977,chekol:hal-01767887,Pichler2014,Chekol:2012:SQC:2900728.2900730}. 
In the case of \sparql{} query language, many authors consider a weaker form of the containment relation, i.e.~\emph{subsumption} \cite{Perez:2009,Letelier:2013:SAO:2539032.2500130,Arenas2011,Pichler2014}.
Instead of the strict requirement that each result of the first query should be also a result of the second one, for the subsumption relation it is enough that for each result of the first query, there is a result of the second query in which it can be embedded (subsumed), i.e.~the result of the second query contains the same projections as in the result of the first query and eventually some additional ones.
An example of \sparql{} queries that are in the query containment relation and in the subsumption relation  is given in Figure \ref{fig:containment1}.
%An example of a query pair satisfying the subsumption relation can be obtained from Figure \ref{fig:containment1} by removing one projection from the query $\kod{Q}_1$\footnote{E.g.~if we remove projection \kod{?y} from the \kod{select} clause of $\kod{Q}_1$, it will return only albums, without their authors, but still, for each or them there is a result of query $\kod{Q}_2$ containing that particular album, together with additional information, i.e.~its author.}.

Many other important problems, like \textit{query equivalence} and \textit{query satisfiablity} \cite{chekol:hal-01767887,Pichler2014,Chekol:2012:SQC:2900728.2900730}, can be reduced to query containment \cite{solving_with_specs}. Additional applications of query containment solvers can be adopted from applications within relational databases \cite{levy1996,Friedman1999,fernandez1999,Gupta1994,LEVY1998165,Calvanese2000}.

For being practically usable, it is important that a query containment solver is efficient, that it covers a wide range of the \sparql{} language constructs and that it always gives correct answers. 
While efficiency and coverage of a solver can be experimentally assessed, e.g.~by using the relevant benchmarks \cite{Chekol2013,Saleem:2017:SSQ:3148011.3148017}, the correctness of an approach should be formally proved as experimental evaluation can only confirm the presence of correctness violations (and not the absence of correctness violations in general case \cite{dijkstra}).

Correctness of an approach includes \textit{soundness} and \textit{completeness}. 
In the context of query containment, soundness means that the approach cannot prove that two input queries are in the containment relation if they are not. Completeness guarantees that for each pair of queries that are in the containment relation the approach will confirm that. 

Correctness can be considered on two levels: \textit{procedure level} and \textit{implementation level}.
Correctness of the procedure level guarantees that the proposed approach is correct.  Unsound approaches are usually not useful and therefore soundness is considered as a basic requirement.
	For example, in the case of query containment, a trivial procedure that returns always true (stating for any two queries that these are in the containment relation) is not a sound one, because there exist two queries for which the procedure will claim the containment (as it always does), while they are not in the relation.
	On the other hand, completeness of this procedure is present as for any two queries satisfying the relation, the procedure will confirm that.
	Having completeness together with soundness is desirable, but it is usually achievable only on a subset of the considered problem.
Correctness of the implementation level guarantees that the procedure is correctly implemented, i.e.~that the software meets its specification. 

Frameworks like Isabelle \cite{isabelle} and Coq \cite{coq}, incorporate development of formally verified software. There are different solvers and tools developed within these frameworks \cite{compcert,MARIC20104333,Simplex2012}. However, such development is very expensive and time consuming. A step towards formally verified software is a formal correctness proof of the proposed procedure while correctness of the implementation is based on the standard software development techniques. There are several different approaches for deciding \sparql{} query containment, and for some of them, correctness of their procedures is proved.

The most important \sparql{} solvers that deal with query containment or query equivalence include \afmu{} \cite{chekol:hal-01767887}, \ts{} \cite{websolver}, \sa{} \cite{Letelier:2013:SAO:2539032.2500130}, \jsag{} \cite{Stadler2018}, \qcan{} \cite{qcan} and \specs{} \cite{solving_with_specs}.
\afmu{} solver uses an expressive modal logic i.e. $\mu$-calculus, and reduces \cite{10.1007/978-3-642-31365-3_13,Chekol:2012:SQC:2900728.2900730} the containment problem to a satisfiability problem in a fragment of the $\mu$-calculus without alternation \cite{Kozen:1982:RP:646236.682866}. 
It uses a tool for deciding satisfiablility  within $\mu$-calculus \cite{10.1007/11554554_21}.
\ts{} solver also uses a variant of $\mu$-calculus without alternation, specially designed for reasoning over finite trees \cite{ts2015}.
For \afmu{} and \ts{}, soundness and completeness of their procedures are proved \cite{chekol:hal-01767887,ts2015}. 
\sa{} solver introduces an algebra of query plans for a subset of \sparql{} graph patterns. The solver is a prototype used for theoretical examination of complexity of the query containment problem and its correctness is not considered. 
Similarly, \jsag{} solver is based on algebra expression trees that are constructed after \sparql{} queries and on a subgraph isomorphism solver. Correctness of its procedure is also not considered. 
On the other hand, \qcan{} is a tool (not a containment solver) for canonicalisation of monotone \sparql{} queries, i.e.~it reduces a query to an equivalent one in a canonical form.
It can be used for checking the equivalence of queries in a syntactic way by comparing the string representations of their canonical forms.
Soundness and completeness of this procedure is proved \cite{qcan, qcan_demo}.

\specs{} \cite{solving_with_specs} is an open source query containment solver which is based on transforming the query containment problem into a satisfiability problem \cite{BSST09HBSAT} in first order logic (\fol). \sparql{} and \fol{} have the same expressive power \cite{Angles:2008} so reducing query containment into \fol{} satisfiability problem is a promising approach. According to thorough evaluation on the two most important benchmarks, Query Containment Benchmark \cite{Chekol2013} and SQCFramework \cite{Saleem:2017:SSQ:3148011.3148017}, \specs{}  covers all widely used \sparql{} language constructs and it is highly efficient. Compared to other state-of-the-art solvers, \specs{} has a better performance and it supports a bigger number of language constructs \cite{solving_with_specs}. 

In this paper, we prove the correctness of the approach used by \specs. We give the precise semantics of the covered \sparql{} constructs and all definitions and proofs of lemmas that are necessary for proving correctness. 
All definitions are given recursively and proofs are done inductively, so when the approach is extended with additional language constructs, the definitions and proofs can also be incrementally extended. 

%\todo{Treba reci da postoji implementacija i citarti rad o eksperimentalnoj evaluaciji, ali sve to u jednom pasusu.}

\paragraph{Overview of the paper.}
%Section \ref{sec:related_work} presents the related work.
Section \ref{sec:sparql_and_qcp} defines a relevant subset of \sparql{} syntax and semantics, and formally defines problems of query containment, query subsumption, query equivalence and query satisfiability. 
Section \ref{sec:modeling} defines the signature of the used theory, the procedure for transforming conjunctive queries into \fol{} formulas and logical formulation of the containment and subsumption problem. It also gives proofs of lemmas that are connecting \sparql{} terms, variables, expressions, conditions, graph patterns and queries with corresponding terms, variables, expressions, conditions and formulas in \fol{}.
Section \ref{sec:correctness} gives necessary definitions and lemmas for proving correctness of the proposed modeling. It gives proofs of soundness and completeness of the proposed approach for both containment and subsumption problem. 
Section \ref{sec:modeling_ncq} deals with queries that are not conjunctive: queries that contain operators \kod{union} and \kod{optional} and that contain subqueries. 
Section \ref{sec:conclusions} gives final conclusions and presents possible directions for further work. 

%--------------------------------------------------------------------
%\section{Related Work}
%\label{sec:related_work}

%\input{relatedwork.tex}
%\todo{Ovde sada treba videti ko sta i kako dokazuje i to bi trebalo da je osnovni related work. razmotriti logicke frameworke za query containment u opstem slucaju, ne samo za sparql}

%--------------------------------------------------------------------
\section{SPARQL and Query Containment Problem}
\label{sec:sparql_and_qcp}
The Resource Description Framework (\rdfs{}), as a World Wide Web Consortium (W3C) recommendation, is an accepted data model for expressing information about World Wide Web resources in a flexible and extensible way.
It is represented as a directed graph consisting of triple statements, i.e.~\emph{\rdfs{} triples}, where each of them is composed of a node for the \emph{subject}, an edge for the \emph{predicate} connecting the subject to an object, and a node for the \emph{object}.
Each of these three parts can be identified by an internationalized resource identifier - \iri{} (generalisation of the uniform resource identifiers) \cite{RFC3987}, subject and object can be unidentified resources, i.e.~ \emph{blank nodes}, while the object can also be a literal value.
As we follow the standard definition of mutually disjoint, countable sets \kod{V} (a set which contains variables), \kod{I} (a set which contains \iri{}s), \kod{B} (a set which contains blank nodes) and \kod{L} (a set which contains literals) \cite{Gutierrez:2004:FSW:1055558.1055573, Perez:2009, Chekol:2012:SQC:2900728.2900730, Schmidt2010, Angles_negation},
an \rdfs{} triple is an element of the set $\kod{IB} \times \kod{I} \times \ibl{}$.\footnote{Like in \cite{chekol:hal-01767887}, we abbreviate any union of sets \kod{I}, \kod{B}, \kod{L} and \kod{V} as, for instance, $\ibl{} \Eqdef \kod{I} \cup \kod{B} \cup \kod{L}$.} 
Sets of \rdfs{} triples are called \emph{\rdfs{} graphs}, which can further form an \rdfs{} dataset, precisely introduced in the following definition.

Let $\kod{i}_k$ denote an \iri{} and let $\kod{\gr}_k$ denote an \rdfs{} graph. Let different \rdfs{} graphs have disjoint sets of blank nodes. According to \cite{solving_with_specs} we introduce the following definitions. 
\begin{defn}[\rdfs{} dataset]  
	\label{def:dataset}
	An \rdfs{} dataset \kod{D} is defined as a set containing a \emph{default graph} named $\dg$ and zero or more \emph{named graphs} $\langle \kod{i}_k, \kod{\gr}_k \rangle$: 
	$$\kod{D} \eqdef \{\dg, \langle \kod{i}_1, \kod{\gr}_1 \rangle, \ldots, \langle \kod{i}_n, \kod{\gr}_n \rangle \}$$ 
\end{defn}

\begin{defn}[Function $\mathit{df}$]  
	\label{def:fun_df}
	Function $\mathit{df}$ maps a dataset $\kod{D}$ to its default graph: 	$$\mathit{df}(\kod{D}) \eqdef \dg$$
\end{defn}

\begin{defn}[Function $names$]  
	\label{def:fun_names}
	Function $names$ maps a dataset $\kod{D}$ to the set of \iri{}s of its named graphs: $$names(\kod{D}) \eqdef \{\kod{i}_1, \ldots , \kod{i}_n\}$$
\end{defn}

\begin{defn}[Function $gr$]  
	\label{def:fun_gr}
	Function $gr$ maps a dataset and an $\iri$ into a graph corresponding to the $\iri$ within the given dataset:\footnote{For readability reasons, the first parameter is written in subscript.} 
	$$gr_\kod{\,D}(\kod{i}_k) \eqdef 
	\begin{cases}
		\kod{\gr}_k, & \text{if } \langle \kod{i}_k, \kod{\gr}_k \rangle \in \kod{D}\\
		\kod{\gr}_{\varnothing}, & \text{otherwise,}
	\end{cases}$$
	where $\kod{\gr}_{\varnothing}$ is an empty graph.
\end{defn}

\begin{defn}[Function $merge$]  
	\label{def:fun_merge}
	Function $merge$ maps a subset $\{\kod{\gr}_{k_1}, \ldots, \kod{\gr}_{k_m}\}$\footnote{Note that sets of blank nodes of different graphs $\kod{\gr}_{k_i}$ are disjoint.} of \rdfs{} graphs into a graph containing a union of their nodes: 
	$$merge\,(\{\kod{\gr}_{k_1}, \ldots, \kod{\gr}_{k_m}\})  \eqdef \bigcup_{i=1}^{m} \kod{\gr}_{k_i}$$
\end{defn}

\subsection{\sparql{} Syntax}
\label{subsec:syntax}

\sparql{} is a language for querying data stored in the relevant datasets, i.e.~\rdfs{} graphs.
The main feature of \sparql{} logic used in the evaluation of queries is a pattern matching facility, i.e.~a finding of relevant triple statements within a queried dataset based on a pattern specified by a query in question.

\wwwc{} precisely defined the \sparql{} query grammar in the \textsc{ebnf} notation \cite{wwwc}.
In this paper we discuss the simplified subset of this grammar, given in Figure \ref{fig:grammar1}, containing the most relevant language constructs used in practice.

\begin{figure*}
	\startfig
	\begin{center}
		\begin{normalsize}
			\startsubfig
			\begin{minipage}{0.47\textwidth}
				\begin{lstlisting}[style=sparqlgrammar]
  Query   ::= @'select'@ Vars
              (@'from'@ @iri@)*
              (@'from'@ @'named'@ @iri@)*
              @'where'@ @'{'@ GPatt @'}'@
  Vars    ::= @'*'@ | @var@+
  GPatt   ::= TPatt
            | GPatt @'.'@ GPatt
            | GPatt @'union'@ GPatt
            | GPatt @'minus'@ GPatt
            | GPatt @'diff'@ GPatt
            | GPatt @'optional'@ GPatt
            | GPatt @'filter'@ Cond
            |  @'{'@ GPatt @'}'@
            | @'{'@ SubQuery @'}'@
            | @'graph'@ @var@ @'{'@ GPatt @'}'@
            | @'graph'@ @iri@ @'{'@ GPatt @'}'@
  TPatt   ::= Subject Predicate Object
  Subject ::= @var@ | @iri@ | @blankNode@
				\end{lstlisting}
			\end{minipage}
			\vline
			\begin{minipage}{0.005\textwidth}
				~
			\end{minipage}
			\begin{minipage}{0.52\textwidth}
				\begin{lstlisting}[style=sparqlgrammar,escapeinside={(*}{*)}]
  Predicate ::= @var@ | @iri@
  Object    ::= @var@ | @iri@ | @blankNode@
              | @rdfLiteral@
  Cond      ::=  Expr @'='@ Expr
              | UnOp Cond  
              | Cond BinOp Cond
              | @'('@ Cond @')'@
              | @'isliteral'@ @'('@ Expr @')'@
              | ... (*  ~~ $//~other ~ built$-$in ~ functions$ *)
                    (*  ~~ $//~ and ~ relational ~ operators$ *)
  Expr      ::= @var@ | @iri@ | @rdfLiteral@
              | @'datatype'@ @'('@ Expr @')'@
              | ... (*  ~~ $//~other ~ built$-$in ~ functions$ *)
                    (*  ~~ $//~ and ~ arithmetic ~ operators$ *)
  UnOp      ::= @'!'@
  BinOp     ::= @'&&'@ | @'||'@ 
  SubQuery  ::= @'select'@ Vars
                @'where'@ @'{'@ GPatt @'}'@   
				\end{lstlisting}
			\end{minipage}
			\endsubfig
		\end{normalsize}
	\end{center}
	\vspace*{-7px}
	\endfig
	\caption{A core subset of the \sparql{} grammar, taken from \cite{solving_with_specs}.}
	\label{fig:grammar1}
\end{figure*}

Similar to other query languages, a \sparql{} query contains:
\begin{itemize}
	\item a \kod{select} clause, specifying a set of variables of interest, i.e.~\emph{distinguished variables}, (described by nonterminal \kod{Vars} in the grammar).
	Instead of listing them, this clause can contain symbol \kod{*}, meaning that the set of distinguished variables contains all the variables present in the graph pattern within \kod{where} clause.
	\item (optionally) \kod{from} and \kod{from named} clauses for precise specification of the relevant dataset.
	Normally, a query is executed against an \rdfs{} dataset denoted by $\dataq$.
	Using these two clauses, a query defines a different dataset on which query evaluation is performed, i.e.~\emph{query dataset} (Definition \ref{defn:query_dataset}).
	\item a \kod{where} clause containing a \emph{graph pattern} (described by nonterminal \kod{GPatt}) for pattern matching within the query dataset.
\end{itemize} 

\begin{defn}[Query dataset $\ag$]
	\label{defn:query_dataset} A query dataset $\ag$ which is specified by the \kod{from} and \kod{from named} clauses of a query executed against an \rdfs{} dataset $\dataq$, is defined in the following way:
	\begin{itemize}[noitemsep,nolistsep,topsep=0pt,leftmargin=4mm]
		\item[$\circ$] If a query does not contain neither \kod{from} nor \kod{from named} clauses, $$\ag \eqdef \dataq$$
		\item[$\circ$] Otherwise, $$\ag \eqdef \{\dg, \langle \kod{i}_{l_1}, gr_{\dataq}(\kod{i}_{l_1}) \rangle, \ldots, \langle \kod{i}_{l_n}, gr_{\dataq}(\kod{i}_{l_n}) \rangle \}$$ where
$$\dg = 
		\begin{cases}
			\kod{\gr}_{\varnothing} , & \text{if a query contains \kod{from named} clauses} \\
			& \text{and no \kod{from} clauses } \\
			merge\,(gr_{\dataq}(\kod{i}_{k_1}),\ldots,gr_{\dataq}(\kod{i}_{k_m})), \hspace*{-35mm}  & \\
			& \text{for \iri{}s }\kod{i}_{k_j} , j \in \{1, \ldots, m\}, \text{ apearing in}\\
			& \text{\kod{from} clauses of the query,} \\
		\end{cases}
$$
		and $i_{l_1}, \ldots, i_{l_n}$ are \iri{}s appearing in \kod{from named} clauses (if a query does not contain \kod{from named} clauses, than the dataset does not contain named graphs).
	\end{itemize}
\end{defn}

\noindent Note that if some \iri{} $\kod{k}$ mentioned in a \kod{from} or \kod{from named} clause does not belong to $names(\dataq)$, by Definition \ref{def:fun_gr} it holds $gr_{\dataq}(\kod{k}) = \varnothing$.

The simplest form of graph pattern is a \emph{triple pattern}, defined as a nonterminal \kod{TPatt} by the grammar, i.e.~an element of the set $\kod{VIB} \times \kod{VI} \times \vibl{}$. 
Elements of sets \kod{V}, \kod{I}, \kod{B} and \kod{L} correspond to the \rdfs{} terms, i.e.~\kod{var}, \kod{iri}, \kod{blankNode} and \kod{rdfLiteral} of the grammar given in Figure \ref{fig:grammar1}.
Graph patterns may contain different operators, like \kod{.}, \kod{union}, \kod{minus}, \kod{diff} \cite{Angles_negation}, \kod{optional}, \kod{filter}, braces and \kod{graph}, but also it can be a subquery (described by nonterminal \kod{SubQuery}).
A \kod{filter} clause can contain different conditions (described by nonterminal \kod{Cond} in the grammar), i.e.~a relational comparison of expressions and built-in functions (\kod{isliteral}, \kod{isuri}, \kod{contains}, \kod{bound}, \kod{regex}) and their logical combinations.
Expressions (described by nonterminal \kod{Expr} in the grammar) can be variables, \iri{}s, literals, and also arithmetic operators and built-in functions (\kod{datatype}, \kod{round}, \kod{abs}, \kod{str}, \kod{lcase}) over them.

\subsection{\sparql{} Semantics}
\label{subsec:semantics}

The following definition specifies variables appearing in a language construct.

\begin{defn}[Function $\kod{var}$]
	\label{def:var}
	Let \kod{t} be an \rdfs{} term, \kod{\tpat} be a triple pattern, \kod{E}, $\kod{E}_1$, $\kod{E}_2$ expressions (described by nonterminal \kod{Expr} in the grammar), \kod{R}, $\kod{R}_1$, $\kod{R}_2$ built-in conditions (described by nonterminal \kod{Cond} in the grammar), and \kod{\pat}, $\kod{\pat}_1$, $\kod{\pat}_2$ graph patterns.
	Variables appearing in the \rdfs{} term \kod{t}, triple pattern \kod{\tpat}, expression \kod{E}, condition \kod{R} and graph pattern \kod{\pat}, in notation $\kod{var}(\kod{t})$, $\kod{var}(\kod{\tpat})$, $\kod{var}(\kod{E})$, $\kod{var}(\kod{R})$ and $\kod{var}(\kod{\pat})$ respectively, are
	\begin{align*}
		\kod{var}(\kod{t}) \  \eqdef & \
		\begin{cases}
			\{\kod{t}\},& \kod{t} \in \kod{VB}\\
			\varnothing,& \kod{t} \in \kod{IL}\\
		\end{cases}     \\
		\kod{var}(\kod{\tpat}) \  \eqdef & \
		\begin{cases}
			\kod{var}(\kod{s}) \cup \kod{var}(\kod{p}) \cup \kod{var}(\kod{o}),
			& \kod{\tpat} \text{ is \codebox{$\kod{s p o}$}, } \kod{s} \in \kod{VIB}, \\
			& \kod{p} \in \kod{VI}, \kod{o} \in \kod{VIBL}\\
		\end{cases}     \\
		\kod{var}(\kod{E}) \  \eqdef & \
		\begin{cases}
			\kod{var}(\kod{t}),& \kod{E} \text{ is } \kod{t} \text{ and } \kod{t} \in \kod{VIL}\\
			\kod{var}(\kod{E}_1), & \kod{E} \text{ is } \kod{datatype(E}_1\kod{)}
		\end{cases}     \\
		\kod{var}(\kod{R}) \  \eqdef & \
		\begin{cases}
			\kod{var}(\kod{E}_1) \cup \kod{var}(\kod{E}_2),
			& \kod{R} \text{ is } \kod{E}_1 = \kod{E}_2\\
			\kod{var}(\kod{E}_1),
			& \kod{R} \text{ is } \kod{isliteral(E}_1\kod{)} \\
			\kod{var}(\kod{R}_1),
			& \kod{R} \text{ is } \kod{!R}_1 \text{ or } \\
			& \kod{R} \text{ is } \kod{(R}_1\kod{)}\\
			\kod{var}(\kod{R}_1) \cup \kod{var}(\kod{R}_2),
			& \kod{R} \text{ is } \kod{R}_1 \kod{\&\&} \kod{R}_2 \text{ or } \\
			& \kod{R} \text{ is } \kod{R}_1 \kod{||} \kod{R}_2\\
		\end{cases}     \\
		\kod{var}(\kod{\pat}) \  \eqdef & \
		\begin{cases}
			\kod{var}(\kod{\tpat}),& \kod{\pat} \text{ is } \kod{\tpat}\\
			\kod{var}(\kod{\pat}_1),& \kod{\pat} \text{ is } \kod{\pat}_1 \  \kod{filter} \ \kod{R} \text{ or } \\
			& \kod{\pat} \text{ is } \kod{\{\pat}_1\kod{\}} \text{ or }\\
			& \kod{\pat} \text{ is } \kod{\pat}_1 \  \kod{minus} \ \kod{\pat}_2 \text{ or }\\
			& \kod{\pat} \text{ is } \kod{\pat}_1 \  \kod{diff} \ \kod{\pat}_2 \text{ or }\\
			& \kod{\pat} \text{ is } \kod{graph i \{\pat}_1\kod{\}} \\
			\kod{var}(\kod{\pat}_1) \cup \kod{var}(\kod{\pat}_2),
			& \kod{\pat} \text{ is } \kod{\pat}_1 \  . \ \kod{\pat}_2 \text{ or }\\
			& \kod{\pat} \text{ is } \kod{\pat}_1 \  \kod{optional} \ \kod{\pat}_2 \text{ or } \\
			& \kod{\pat} \text{ is } \kod{\pat}_1 \  \kod{union} \ \kod{\pat}_2 \\
			\kod{var}(\kod{\pat}_1) \cup \{\kod{x}\},
			& \kod{\pat} \text{ is } \kod{graph x \{\pat}_1\kod{\}} \\
			\overline{\kod{dv}_\kod{sq}},& \kod{\pat} \text{ is } \kod{\{Q$_\kod{sq}$\}} \text{ and } \overline{\kod{dv}_\kod{sq}} \text{ is a}\\
			& \text{set of distinguished} \\
			& \text{variables of } \kod{Q$_\kod{sq}$}
		\end{cases}            
	\end{align*}
\end{defn}
\noindent
Note that we assume $\kod{var}(\kod{R}) \subseteq \kod{var}(\kod{\pat}_1)$ for each graph pattern $\kod{\pat} = \kod{\pat}_1 \  \kod{filter} \ \kod{R}$, as the opposite is not
computationally desirable \cite{perez2006}.

We follow the standard notation and set semantics of a \sparql{} query and of a graph pattern evaluation \cite{chekol:hal-01767887,perez2006,Perez:2009,Angles:2008}.\footnote{Different semantics are also possible, like the standard bag semantics \cite{Perez:2009,Angles:2008,Angeles:2016}.
A comparison \cite{Perez:2009} of set and bag semantics presents the reasons why set semantics is of fundamental importance in the development and implementation of a query language.}

Intuitively, a result of a \sparql{} query execution connects variables (and blank nodes which cannot be retrieved by the query) from the graph pattern of query to the values within a graph of the query dataset, i.e. to the values of \ibl{}.
Therefore, the function $\mu$ denotes a partial mapping from the set of variables and blank nodes \kod{VB} to the set \ibl{}.
In the following text, $dom(m)$ denotes the domain of a mapping $m$, while the function $\mu_{\kod{x} \rightarrow \kod{c}}$ denotes a mapping such that $dom(\mu_{\kod{x} \rightarrow \kod{c}}) \eqdef \{\kod{x}\}$ and $\mu_{\kod{x} \rightarrow \kod{c}}(\kod{x}) \eqdef \kod{c}$.

\begin{defn}[Compatible mappings $\comp$ ]
	\label{def:compatible_mappings}
	Mappings $m_1$ and $m_2$ are \emph{compatible}, denoted by $m_1 \comp m_2$, if for each ${x}$ such that ${x} \in dom(m_1) \cap dom(m_2)$ it holds that $m_1({x}) = m_2({x})$.
\end{defn}

\noindent Note that two mappings with disjoint domains are always compatible.
Also, if $m_1 \comp m_2$, then $m_1 \cup m_2$ is also a mapping, such that it holds:
\begin{align*}
	dom(m_1 \cup m_2) = & ~ dom(m_1) \cup dom(m_2)\\
	(m_1 \cup m_2)(x) = & \begin{cases}
		m_1(x), & x \in dom(m_1)\\
		m_2(x), & \text{otherwise}
	\end{cases}
\end{align*}
Furthermore, $m_1 \cup m_2$ is compatible with both $m_1$ and $m_2$.

\begin{defn}[Operations over sets of mappings]
	\label{def:union_join}
	Let $\Omega_1$ and $\Omega_2$ be sets of mappings.
	Operations \emph{union}, \emph{join}, \emph{difference}, \emph{minus} and \emph{left outer-join} are defined as follows \cite{Perez:2009}:
	\begin{align*}
		\hspace*{5mm}
		\Omega_1 \cup \Omega_2  \eqdef & \ \{ m \ | \ m \in \Omega_1 \ \text{or} \ m \in \Omega_2 \}&\\
		\Omega_1 \bowtie \Omega_2  \eqdef & \ \{ m_1 \cup m_2 \ | \ m_1 \in \Omega_1, m_2 \in \Omega_2, m_1 \comp m_2 \}&\\
		\Omega_1 \setminus \Omega_2  \eqdef & \ \{ m_1 \ | \ m_1 \in \Omega_1 \ \text{and for all} \ m_2 \in \Omega_2\text{,} &\\
		& ~~~ m_1 \text{ and } m_2 \text{ are not compatible}\} &\\
		\Omega_1 - \Omega_2  \eqdef & \ \{ m_1 \ | \ m_1 \in \Omega_1 \ \text{and for all} \ m_2 \in \Omega_2\text{,} &\\
		& ~~~ m_1 \text{ and } m_2 \text{ are not compatible, or } &\\
		& ~~~ dom(m_1) \cap dom(m_2) = \varnothing\}&\\
		\Omega_1 \leftouterjoin \Omega_2  \eqdef & \ (\Omega_1 \bowtie \Omega_2) \cup (\Omega_1 \setminus \Omega_2)&
	\end{align*}
\end{defn}

Let $\overline{x}$ denote a set containing variables $x_i$, for each $i \in \{1, \ldots, n\}$, i.e.~$\overline{x} = \{x_1, ..., x_n\}$.

\begin{defn}[Extension, restriction, projection]
	\label{def:extension_restriction}
	A mapping $m_1$ is an \emph{extension} of a mapping $m_2$ (a mapping $m_2$ is a \emph{restriction} of a mapping $m_1$), denoted by $m_1 \succeq m_2$ ($m_2 \preceq m_1$), if $m_1 \comp m_2$ and $dom(m_1) \supseteq dom(m_2)$.
	\emph{Projection} of a mapping $m$ to the set $\overline{x}$, denoted by $m_{\overline{x}}$, is a mapping such that $dom(m_{\overline{x}}) = dom(m) \cap \overline{x}$ and $m_{\overline{x}} \preceq m$.
\end{defn}

\noindent {Note that if sets $dom(m)$ and $\overline{x}$ are disjoint, $m_{\overline{x}}$ is an \emph{empty mapping}, i.e~a mapping with an empty domain.}

\begin{defn}[Projection operator]
	\label{def:projection}
	Let $\Omega$ be a set of mappings, and $\overline{x}$ be a set of variables.
	Then, a \emph{projection operator} over the sets $\overline{x}$ and $\Omega$, denoted $\PP_{\overline{x}}(\Omega)$, is defined as:
	$$\PP_{\overline{x}}(\Omega) \eqdef \{m_{\overline{x}} \;|\; m \in \Omega\}$$
\end{defn}

According to the \sparql{} semantics \cite{wwwc}, the following function is needed in Definition \ref{defn:expr_in_mu}.

\begin{defn}[Semantics of function \kod{dt}]
	\label{def:datatype}
	Let \kod{l} be a literal, function $\kod{dt}: \kod{L} \rightarrow \kod{I}$ is defined as follows:
	\begin{align*}
		\kod{dt(l)} \eqdef
		\begin{cases}
			\kod{i}\,, & \kod{l} \text{  is typed literal, with }\\
			& \kod{\^}\hspace*{-2pt}\kod{\^}\hspace*{-2pt}\kod{i} \text{ in suffix, \kod{i}} \hspace*{-1mm}\in\hspace*{-1mm} \text{\kod{I},}\\
			\kod{xsd:string}\,, & \kod{l} \text{  is untyped literal.} 
		\end{cases}
	\end{align*}
\end{defn}

A mapping may be naturally extended on \iri{s}, literals and variables and blank nodes outside of its domain, by introducing an additional constant named $\error$, as specified in the following definition.

\begin{defn}[Notation $\valuation{\cdot}{\mu}$]
	\label{defn:expr_in_mu}
	A value of an \rdfs{} term \kod{t}, an expression $\kod{E}$, and a triple pattern $\kod{\tpat}$, according to the mapping $\mu$, in notation 
	$\valuation{\kod{t}}{\mu}$, $\valuation{\kod{E}}{\mu}$, and $\valuation{\kod{\tpat}}{\mu}$ respectively, 	
	is a value from \ible, defined in the following way: 
	\begin{align*}	
	\valuation{\kod{t}}{\mu} & \eqdef \ 
	\begin{cases}
	\kod{t} \; ,\hphantom{---} & \kod{t} \in \kod{IL} \\
	\mu(\kod{t}) \; , \hphantom{--} & \kod{t} \in \kod{VB} \text{ and } \kod{t} \in dom(\mu)\\
	\error \; , \hphantom{--} & \kod{t} \in \kod{VB} \text{ and } \kod{t} \notin dom(\mu)\\
	\end{cases} \\
	\valuation{\kod{E}}{\mu} & \eqdef \ 
	\begin{cases}
	\valuation{\kod{t}}{\mu} \; ,\hphantom{--}& \kod{E} \ \text{is} \ \kod{t} \text{, } \kod{t} \in \kod{VIL} \\
	\kod{dt}(\valuation{\kod{E}_1}{\mu}),&  \kod{E} \text{ is } \kod{datatype(E}_1\kod{)} \text{ and } \valuation{\kod{E}_1}{\mu} \not= \error \\
	\error, & \kod{E} \text{ is } \kod{datatype(E}_1\kod{)} \text{ and } \valuation{\kod{E}_1}{\mu} = \error
	\end{cases} \\
	\valuation{\kod{\tpat}}{\mu} & \eqdef \
	\begin{cases}
	\valuation{\kod{s}}{\mu} \kod{ } \valuation{\kod{p}}{\mu} \kod{ } \valuation{\kod{o}}{\mu} \; ,  \\
	\hspace*{18mm} \kod{\tpat} \text{ is \codebox{$\kod{s p o}$} and } \kod{s} \in \kod{VIB},\\
	\hspace*{18mm} \kod{p} \in \kod{VI}, \kod{o} \in \vibl{}\\
	\end{cases}
	\end{align*}
\end{defn}

\noindent
A set \ible{} is an extension of the set \ibl{} containing this constant, i.e.~$\ible \Eqdef \ibl \cup \{ \error \}$.

\begin{defn}[Relation $\zadovoljava$]
	\label{defn:definition_filter}
	Let \kod{E}, $\kod{E}_1$ and $\kod{E}_2$ be expressions, \kod{R}, $\kod{R}_1$ and $\kod{R}_2$ built-in conditions.
	A mapping $\mu$ satisfies built-in condition \kod{R}, denoted $\mu \zadovoljava \kod{R}$, if:
	\begin{align*}
	\hspace*{3mm}
	\mu \zadovoljava \kod{R} \  \eqdef \ 
	\begin{cases}
	\valuation{\kod{E}_1}{\mu} \not= \error, \\ \valuation{\kod{E}_2}{\mu} \not= \error \text{ and } \\
	\valuation{\kod{E}_1}{\mu} = \valuation{\kod{E}_2}{\mu} \; , & \kod{R} \ \text{is} \ \kod{E}_1 = \kod{E}_2\\
	\text{if not } \mu \zadovoljava \kod{R}_1 \; , \text{ } & \kod{R} \ \text{is} \ \kod{!R}_1\\
	\mu \zadovoljava \kod{R}_1 \text{ and } \mu \zadovoljava \kod{R}_2 \; , & \kod{R} \ \text{is} \ \kod{R}_1 \kod{\&\&} \kod{R}_2\\
	\mu \zadovoljava \kod{R}_1 \text{ or } \mu \zadovoljava \kod{R}_2 \; , & \kod{R} \ \text{is} \ \kod{R}_1 \kod{||} \kod{R}_2\\
	\mu \zadovoljava \kod{R}_1 \; , & \kod{R} \ \text{is} \ \kod{(R}_1\kod{)}\\
	\valuation{\kod{E}_1}{\mu} \not= \error  \text{ and } \\
	\valuation{\kod{E}_1}{\mu} \in \kod{L}, & \kod{R} \text{ is } \kod{isliteral(E}_1\kod{)}
	\end{cases}		
	\end{align*}
\end{defn}

\begin{defn}[Evaluation of a graph pattern $\semdg{\cdot}{\ag}{\kod{\gr}}$]
	\label{definition_sem}
	Let $\ag$ be an \rdfs{} dataset, \kod{\gr} a graph within $\ag$, \kod{\tpat} a triple pattern, \kod{\pat}, $\kod{\pat}_1$, $\kod{\pat}_2$ graph patterns, and \kod{R} a built-in condition.
	An evaluation of a graph pattern over the graph $\kod{\gr}$, which is in this context called the \emph{active graph}, within the dataset $\ag$, 
	is defined recursively as follows:
	\begin{align*}
		\semdg{\kod{\tpat}}{\ag}{\kod{\gr}} &\eqdef \{ \mu \ | \ dom(\mu) = \kod{var}(\kod{\tpat}) \; \wedge  \valuation{\kod{\tpat}}{\mu} \in \kod{\kod{\gr}} \} \\
		\semdg{\kod{\pat}_1 \  . \ \kod{\pat}_2}{\ag}{\kod{\gr}} &\eqdef \semdg{\kod{\pat}_1}{\ag}{\kod{\gr}} \bowtie \semdg{\kod{\pat}_2}{\ag}{\kod{\gr}}\\
		\semdg{\kod{\pat}_1 \  \kod{union} \ \kod{\pat}_2}{\ag}{\kod{\gr}} &\eqdef \semdg{\kod{\pat}_1}{\ag}{\kod{\gr}} \cup \semdg{\kod{\pat}_2}{\ag}{\kod{\gr}} \\
		\semdg{\kod{\pat} \  \kod{filter} \ \kod{R}}{\ag}{\kod{\gr}} &\eqdef \{ \mu \in \semdg{\kod{\pat}}{\ag}{\kod{\gr}} \ | \ \mu \zadovoljava \kod{R} \}\\
		\semdg{\kod{\{\pat\}}}{\ag}{\kod{\gr}} &\eqdef \semdg{\kod{\pat}}{\ag}{\kod{\gr}} \\
		\semdg{\kod{\pat}_1 \  \kod{minus} \ \kod{\pat}_2}{\ag}{\kod{\gr}} & \eqdef \semdg{\kod{\pat}_1}{\ag}{\kod{\gr}} - \semdg{\kod{\pat}_2}{\ag}{\kod{\gr}} \\
		\semdg{\kod{\pat}_1 \  \kod{diff} \ \kod{\pat}_2}{\ag}{\kod{\gr}} & \eqdef \semdg{\kod{\pat}_1}{\ag}{\kod{\gr}} \setminus \semdg{\kod{\pat}_2}{\ag}{\kod{\gr}} \\
		\semdg{\kod{\pat}_1 \  \kod{optional} \ \kod{\pat}_2}{\ag}{\kod{\gr}} & \eqdef \semdg{\kod{\pat}_1}{\ag}{\kod{\gr}} \leftouterjoin \semdg{\kod{\pat}_2}{\ag}{\kod{\gr}} \\
		\semdg{\kod{graph x \{\pat\}}}{\ag}{\kod{\gr}} & \eqdef \hspace*{-3mm} \underset{\kod{i} \in names(\ag)} \bigcup \hspace*{-3mm} \left(\semdg{\kod{\pat}}{\ag}{gr_{\,\ag}(\kod{i})} \bowtie  \{ \mu_{\kod{x} \rightarrow \kod{i}} \} \right)\\
		\semdg{\kod{graph i \{\pat\}}}{\ag}{\kod{\gr}} & \eqdef \semdg{\kod{\pat}}{\ag}{gr_{\,\ag}(\kod{i})} \\
		\semdg{\kod{\{Q$_\kod{sq}$\}}}{\ag}{\kod{G}} & \eqdef \semd{\kod{Q}_\kod{sq}}{\ag} ~ \footnotemark
	\end{align*}
\end{defn}
\footnotetext{$\semd{\kod{Q}_\kod{sq}}{\ag}$ represents the evaluation of query $\kod{Q}_\kod{sq}$ over a dataset $\ag$, introduced later in Definition \ref{defn:query_evaluation}.}

The following lemma connects a domain of an evaluation of a graph pattern and variables of a graph pattern. 
\begin{lemma}
	\label{thm:pattern_domain}
	Let $\ag$ be an \rdfs{} dataset, \kod{\gr} a graph within $\ag$, \kod{\pat} a graph pattern which is \kod{union}-free, \kod{optional}-free and without subqueries, and $\mu$ a mapping such that $\mu \in \semdg{\kod{\pat}}{\ag}{\kod{\gr}}$.
	Then:
	\begin{center}
		$dom(\mu) = \kod{var}(\kod{\pat})$.
	\end{center}
	
	\begin{proof} 
		%\ref{proof:pattern_domain} is given in Appendix \ref{sec:appendix}. It is done by induction over the graph pattern $\kod{\pat}$.
		
		The lemma is proved by induction over graph pattern $\kod{\pat}$.
		
		\begin{description}
			\item \kod{\pat{}} is \kod{\tpat}\\
			By Definition \ref{definition_sem}, from $\mu \in \semdg{\kod{\tpat}}{\ag}{\kod{\gr}}$, it holds $dom(\mu) = \kod{var}(\kod{\tpat})$.
			
			\item $\kod{\pat}$ is $\kod{\pat}_1 . \kod{\pat}_2$\\
			By Definition \ref{definition_sem}, from $\mu \in \semdg{\kod{\pat}_1 . \kod{\pat}_2}{\ag}{\kod{\gr}}$, it holds
			$$\mu \in \semdg{\kod{\pat}_1}{\ag}{\kod{\gr}} \bowtie \semdg{\kod{\pat}_2}{\ag}{\kod{\gr}}.$$
			By Definition \ref{def:union_join}, there exist compatible mappings $\mu_1$ and $\mu_2$, such that $\mu = \mu_1 \cup \mu_2$,
			$$\mu_1 \in \semdg{\kod{\pat}_1}{\ag}{\kod{\gr}} \text{ and }\mu_2 \in \semdg{\kod{\pat}_2}{\ag}{\kod{\gr}}.$$
			Therefore, it holds $dom(\mu) = dom(\mu_1) \cup dom(\mu_2)$.
			By induction hypothesis, it holds
			$$dom(\mu_1) = \kod{var}(\kod{\pat}_1) \text{ and } dom(\mu_2) = \kod{var}(\kod{\pat}_2).$$
			Then, $dom(\mu) = \kod{var}(\kod{\pat}_1) \cup \kod{var}(\kod{\pat}_2)$, i.e.~by Definition \ref{def:var}, it holds $dom(\mu) = \kod{var}(\kod{\pat}_1 . \kod{\pat}_2)$.
			
			\item $\kod{\pat}$ is $\kod{\pat}_1 \kod{ filter } \kod{R}$\\
			By Definition \ref{definition_sem}, from $\mu \in \semdg{\kod{\pat}_1 \kod{ filter } \kod{R}}{\ag}{\kod{\gr}}$, it holds
			$\mu \in \semdg{\kod{\pat}_1}{\ag}{\kod{\gr}}$.
			By induction hypothesis, it also holds
			$dom(\mu) = \kod{var}(\kod{\pat}_1)$.
			Therefore, by Definition \ref{def:var}, it holds $dom(\mu) = \kod{var}(\kod{\pat}_1 \kod{ filter } \kod{R})$.
			
			\item $\kod{\pat}$ is $\kod{\{\pat}_1\kod{\}}$\\
			By Definition \ref{definition_sem}, from $\mu \in \semdg{\kod{\{\pat}_1\kod{\}}}{\ag}{\kod{\gr}}$, it holds
			$\mu \in \semdg{\kod{\pat}_1}{\ag}{\kod{\gr}}$.
			By induction hypothesis, it holds
			$dom(\mu) = \kod{var}(\kod{\pat}_1)$.
			Therefore, by Definition \ref{def:var}, it also holds $dom(\mu) = \kod{var}(\kod{\{\pat}_1\kod{\}})$.
			
			\item $\kod{\pat}$ is $\kod{\pat}_1 \; \kod{diff} \; \kod{\pat}_2$\\
			By Definition \ref{definition_sem}, from $\mu \in \semdg{\kod{\pat}_1 \; \kod{diff} \; \kod{\pat}_2}{\ag}{\kod{\gr}}$, it holds
			$$\mu \in \semdg{\kod{\pat}_1}{\ag}{\kod{\gr}} \setminus \semdg{\kod{\pat}_2}{\ag}{\kod{\gr}}.$$
			Therefore, by Definition \ref{def:union_join}, it holds $\mu \in \semdg{\kod{\pat}_1}{\ag}{\kod{\gr}}$.
			By induction hypothesis, it holds
			$dom(\mu) \hspace*{-3pt}=\hspace*{-3pt} \kod{var}(\kod{\pat}_1)$.
			Then, by Definition \ref{def:var}, it holds $dom(\mu) = \kod{var}(\kod{\pat}_1 \; \kod{diff} \; \kod{\pat}_2)$.
			
			\item $\kod{\pat}$ is $\kod{\pat}_1 \; \kod{minus} \; \kod{\pat}_2$\\
			By Definition \ref{definition_sem}, from $\mu \in \semdg{\kod{\pat}_1 \; \kod{minus} \; \kod{\pat}_2}{\ag}{\kod{\gr}}$, it holds
			$$\mu \in \semdg{\kod{\pat}_1}{\ag}{\kod{\gr}} - \semdg{\kod{\pat}_2}{\ag}{\kod{\gr}}.$$
			Therefore, by Definition \ref{def:union_join}, it holds $\mu \in \semdg{\kod{\pat}_1}{\ag}{\kod{\gr}}$.
			By induction hypothesis, it holds
			$dom(\mu) \hspace*{-3pt}=\hspace*{-3pt} \kod{var}(\kod{\pat}_1)$.
			Then, by Definition \ref{def:var}, it holds $dom(\mu) \hspace*{-2pt}=\hspace*{-2pt} \kod{var}(\kod{\pat}_1 \; \kod{minus} \; \kod{\pat}_2)$.
			
			\item $\kod{\pat}$ is $\kod{graph x \{\pat}_1\kod{\}}$\\
			By Definition \ref{definition_sem}, from $$\mu \in \semdg{\kod{graph x \{\pat}_1\kod{\}}}{\ag}{\kod{\gr}},$$ it holds
			$$\mu \in \underset{\kod{i} \in names(\ag)} \bigcup \left(\semdg{\kod{\pat}}{\ag}{gr_{\,\ag}(\kod{i})} \bowtie  \{ \mu_{\kod{x} \rightarrow \kod{i}} \} \right).$$
			Therefore, by Definition \ref{def:union_join} and induction hypothesis, it holds
			$$dom(\mu) = \kod{var}(\kod{\pat}_1) \cup \{\kod{x}\}.$$
			Then, by Definition \ref{def:var}, it holds
			$$dom(\mu) = \kod{var}(\kod{graph x \{\pat}_1\kod{\}}).$$
			
			\item $\kod{\pat}$ is $\kod{graph i \{\pat}_1\kod{\}}$\\
			By Definition \ref{definition_sem}, from $$\mu \in \semdg{\kod{graph i \{\pat}_1\kod{\}}}{\ag}{\kod{\gr}},$$ it holds
			$$\mu \in \semdg{\kod{\pat}_1}{\ag}{gr_{\,\ag}(\kod{i})}.$$
			Therefore, by induction hypothesis, it holds
			$$dom(\mu) = \kod{var}(\kod{\pat}_1).$$
			Then, by Definition \ref{def:var}, it holds
			$$dom(\mu) = \kod{var}(\kod{graph i \{\pat}_1\kod{\}}).$$
		\end{description}
	\end{proof}
\end{lemma}

\begin{defn}[Evaluation of a query $\semdq{\kod{Q}}$]
	\label{defn:query_evaluation}
	An \emph{evaluation} of a query \kod{Q} over a dataset $\dataq$ 
	is defined as 
	$$\semdq{\kod{Q}} \eqdef \PP_{\overline{\kod{dv}}}(\semdg{\kod{qpat}}{\ag}{\mathit{df}(\ag)})$$
	where within the query \kod{Q}, \kod{qpat} is a query pattern, $\overline{\kod{dv}}$ is a set of distinguished variables, and $\ag$ is a query dataset.
\end{defn}

In a case when a mapping $\mu$ from $\semdg{\kod{qpat}}{\ag}{\mathit{df}(\ag)}$ has a disjoint domain with $\overline{\kod{dv}}$, $\semdq{\kod{Q}}$ contains the empty mapping, corresponding to the empty \sparql{} solution. 
In a case when the set $\semdq{\kod{Q}}$ is an empty set, there is no result of a query evaluation.

Note that each query specifies the active graphs that are used for evaluation of its graph patterns. 
For each considered graph pattern (Figure \ref{fig:grammar1}) the active graph is equal to the default graph of the query dataset, except in case for $\kod{graph x \{ \pat{} \}}$ and $\kod{graph i \{ \pat{} \}}$.
These two constructs give a possibility to match the graph pattern $\kod{\pat}$ against named graphs in the query dataset \cite{Angles:2008}.

\begin{defn}[Relevant variables $\overline{\kod{rv}}$]
	\label{def:rv}
	For an \rdfs{} dataset $\dataq$, a query \kod{Q} and a mapping $\mu$ such that $\mu \in \semdq{\kod{Q}}$, all variables from $dom(\mu)$ are called \emph{relevant variables}. 
\end{defn}

\noindent There is a slight but important difference between relevant and distinguished variables. Although every relevant variable is at the same time a distinguished variable, in the opposite direction this does not hold. A projection variable appearing only in the \kod{select} clause and nowhere else in the query pattern is an example of a distinguished variable that is not relevant.
The following lemma connects the relevant variables with the corresponding query pattern and distinguished variables.

\begin{lemma}
	\label{thm:query_domain}
	Let \kod{Q} be a query, $\overline{\kod{dv}}$ a set of its distinguished variables, $\kod{qpat}$ its graph pattern which is \kod{union}-free, \kod{optional}-free and without subqueries, and $\overline{\kod{rv}}$ a set of its relevant variables.
	Then:
	\begin{center}
		$\overline{\kod{rv}}= \kod{var}(\kod{qpat}) \cap \overline{\kod{dv}}$.
	\end{center}
	\begin{proof} 
		%\ref{proof:query_domain} is given in Appendix \ref{sec:appendix}. It uses Definitions \ref{def:extension_restriction}, \ref{def:projection} and \ref{defn:query_evaluation} and Lemma \ref{thm:pattern_domain}.
		
		By Definition \ref{def:rv}, there is a mapping $\mu$ and a dataset $\dataq$ such that $\mu \in \semd{\kod{Q}}{\dataq}$ and $\overline{\kod{rv}} = dom(\mu)$.
		Then, by Definition \ref{defn:query_evaluation}, it holds
		$$\mu \in \PP_{\overline{\kod{dv}}}(\semdg{\kod{qpat}}{\ag}{\mathit{df}(\ag)}).$$
		By Definition \ref{def:projection}, there exists $\mu'$ such that $\mu = \mu'_{\overline{\kod{dv}}}$ and
		$$\mu' \in \semdg{\kod{qpat}}{\ag}{\mathit{df}(\ag)}.$$
		By Lemma \ref{thm:pattern_domain}, it holds $dom(\mu') = \kod{var}(\kod{qpat})$.
		Therefore, by Definition \ref{def:extension_restriction}, it holds $$dom(\mu) = dom(\mu') \cap \overline{\kod{dv}} = \kod{var}(\kod{qpat}) \cap \overline{\kod{dv}},$$
		i.e.
		$$\overline{\kod{rv}} = \kod{var}(\kod{qpat}) \cap \overline{\kod{dv}}.$$
	\end{proof}
\end{lemma}

\subsection{Containment, subsumption, equivalence and satisfiability}

\begin{defn}[Containment $\kod{Q}_1 \sqsubseteq \kod{Q}_2$]
	\label{defn:query_containment}
	Given two queries $\kod{Q}_1$ and $\kod{Q}_2$, $\kod{Q}_1$ is \emph{contained} in $\kod{Q}_2$, 
	if for every \rdfs{} dataset $\dataq$, it holds $\semdq{\kod{Q}_1} \subseteq \semdq{\kod{Q}_2}$.
\end{defn}

\noindent If $\kod{Q}_1$ is contained in $\kod{Q}_2$, then $\kod{Q}_2$ is called a \textit{super-query}, and $\kod{Q}_1$ is called a \textit{sub-query}. The \emph{query containment problem} is a problem to determine whether $\kod{Q}_1$ is contained in $\kod{Q}_2$.
The problem is undecidable if query $\kod{Q}_2$ contains projections \cite{Pichler2014}. 
Therefore, we assume that all variables in the graph pattern of $\kod{Q}_2$ appear in the \kod{select} clause.

\begin{defn}[Subsumption $\kod{Q}_1 \dot{\sqsubseteq} \kod{Q}_2$]
	\label{defn:query_subsumption}
	Given two queries $\kod{Q}_1$ and $\kod{Q}_2$, 
	$\kod{Q}_1$ is \emph{subsumed} by $\kod{Q}_2$, 
	if for every \rdfs{} dataset $\dataq$, for each mapping $\mu$ from $\semdq{\kod{Q}_1}$ there exists an extension $\mu'$ ($\mu' \succeq \mu$), such that $\mu'$ belongs to $\semdq{\kod{Q}_2}$.
\end{defn}

\noindent Subsumption relation can be considered as a weaker form of containment \cite{Perez:2009,Letelier:2013:SAO:2539032.2500130,Arenas2011,Pichler2014}, as $\kod{Q}_1 \sqsubseteq \kod{Q}_2$ implies $\kod{Q}_1 \dot{\sqsubseteq} \kod{Q}_2$ (a proof uses Definition \ref{defn:query_containment} and an extension $\mu' = \mu$). The same terminology concerning $\kod{Q}_2$ as a super-query and $\kod{Q}_1$ as a sub-query is used when subsumption relation is considered instead of query containment relation.
The \textit{subsumption problem} is a problem to determine whether $\kod{Q}_1$ is subsumed in $\kod{Q}_2$. Unlike query containment problem, query subsumption problem is not undecidable when query $\kod{Q}_2$ contains projections \cite{Pichler2014}.

\begin{defn}[Equivalence $\kod{Q}_1 \equiv \kod{Q}_2$]
	\label{defn:query_equivalence}
	Given two queries $\kod{Q}_1$ and $\kod{Q}_2$, 
	$\kod{Q}_1$ is \emph{equivalent} to $\kod{Q}_2$, 
	if for every \rdfs{} dataset $\dataq$, it holds $\semdq{\kod{Q}_1} = \semdq{\kod{Q}_2}$.
\end{defn}

\noindent Query equivalence corresponds to satisfying containment relation in both directions.
But, if two queries satisfy the subsumption relation in both directions, they do not have to be equivalent (if either of them contains \texttt{union} or projection operator) \cite{Pichler2014}.

\begin{defn}[Query satisfiability, unsatisfiability]
	\label{def:satisfiable}
	Query $\kod{Q}$ is \emph{satisfiable} if there exist a dataset $\dataq$ and a mapping $\mu$ such that $\mu \in \semdq{\kod{Q}}$. 
	Otherwise, \kod{Q} is \emph{unsatisfiable}.
\end{defn}

\noindent For each query it holds that an unsatisfiable query is its sub-query.

%--------------------------------------------------------------------
\section{Modeling the QC Problem of Conjunctive Queries}
\label{sec:modeling}

In this section, we model a subset of \sparql{} queries presented by the grammar in Figure \ref{fig:grammar1}, i.e. conjunctive queries \cite{Chandra1977,Chaudhuri1993} extended by \sparql{} negation, the operator \kod{graph} and built-in functions, while the rest of language constructs are covered in Section \ref{sec:modeling_ncq}.
We translate such queries into \fol{} formulas (Section \ref{subsec:transforming_queries}) that are used for reasoning about query relations (Section \ref{subsec:querycontainment}).
We have already presented the modeling in \cite{solving_with_specs}, while here, we reshow the definitions that are necessary to prove the correctness of our reduction.

\subsection{Theory signature}

The theory signature used for reasoning about queries $\kod{Q}_1$ and $\kod{Q}_2$, denoted by $\mathcal{L}$, is given in Definition \ref{def:signature} (Figure \ref{fig:definitions1}) as a tuple $$(\mathcal{C} \cup \fun, ~~ \mathcal{P} \cup \{ \ppp, \pp \}, ~~ ar).$$
For example, for queries presented in Figure \ref{fig:containment1}, the set $\mathcal{C}$ contains only constants $LosAngeles_l$ (corresponding to the literal \kod{"Los Angeles"}) and $a_i$, $Album_i$, $artist_i$, $SoloArtist_i$, $hometown_i$ and $name_i$ (corresponding to the \iri{}s \kod{a}, \kod{:Album}, \kod{:artist}, \kod{:SoloArtist}, \kod{:hometown} and \kod{:name}).
For the same pair of queries, the set $\fun$ is an empty set, while the set $\mathcal{P}$ contains only the equality predicate symbol. {Considering the grammar presented in Figure \ref{fig:grammar1}, the set $\fun$ can contain the symbol $\textit{datatype}$, corresponding to the \sparql{} function \kod{datatype} while the set $\mathcal{P}$ can also have the symbol \textit{isliteral}, corresponding to the \sparql{} predicate \kod{isliteral}, in cases where these built-in functions are used within the relevant queries.}
Predicate symbols $\ppp$ and $\pp$ model belonging of a triple to the graphs of an \rdfs{} dataset $\dataq$, e.g.: 
\begin{itemize}
	\item $\ppp(x_v,a_i,Album_i)$ models that the triple \codebox{$\kod{?x a :Album}$} belongs to the default graph of an \rdfs{} dataset $\dataq$,
	\item $\pp(x_v,a_i,Album_i, g_i)$ models that the triple \codebox{$\kod{?x a :Album}$} belongs to the named graph specified by the \iri{} \kod{g} of an \rdfs{} dataset $\dataq$, where $g_i$ is a corresponding constant to the \iri{} \kod{g}.
\end{itemize} 
\noindent A set of variables $\vars$ corresponds to the variables and blank nodes from the \sparql{} queries $\kod{Q}_1$ and $\kod{Q}_2$. For the query pair presented in Figure \ref{fig:containment1}, $\vars$ is equal to a set containing variables $x_v$, $y_v$, $z_v$ and $w_v$, corresponding to the \sparql{} variables \kod{?x}, \kod{?y}, \kod{?z} and \kod{?w}, respectively.

\subsection{Transforming Conjunctive Queries Into Formulas}
\label{subsec:transforming_queries}

The following auxiliary definition extends a function to a subset of its domain.

\begin{defn}[Function over sets]
	\label{def:function_over_sets}
	Let ${A}$ be a set of elements, and $f$ a function such that ${A} \subseteq dom(f)$.
	Function $f$ over set ${A}$, in notation $f({A})$, denotes the set $\{ f(e) \;|\; {e} \in {A} \}$.
\end{defn}

A translation of the \sparql{} queries into the corresponding \fol{} formulas is performed by function $\sigma$ recursively, \sparql{} construct by construct, using the auxiliary functions $\sigvibl$ and $\context$.
Their definitions are given in Figure \ref{fig:definitions1}.

Function $\sigvibl$ assigns a \fol{} constant to a \sparql{} \iri{} or a literal, and a \fol{} variable to a \sparql{} variable or a blank node.

Function $\context$ maps an active graph to the set of corresponding graph \iri{} constants.\footnote{The function $\context$ is not defined on the default graph of the \rdfs{} dataset $\dataq$ as there is no \iri{} associated to it.}
Note that, according to the semantics, the active graph $\kod{\gr}$ can be the default graph or a named graph of a query dataset $\kod{\ag}$.
If $\kod{\gr}$ is the default graph, it can be equal to the default graph of $\dataq$, an \rdfs{} merge of one or more graphs, or an empty graph $\kod{\gr}_{\varnothing}$.
These possibilities correspond to the cases present in Definition \ref{def:context}.

Definition \ref{def:sigma} presents a translation of the \sparql{} terms, expressions, conditions and graph patterns into the corresponding \fol{} formulas from the signature $\mathcal{L}$.
The full notation of the function $\sigma$ includes an active graph $\kod{\gr}$ used for matching the pattern within the query. For readability reasons, the active graph is denoted in superscript, i.e.~$\sigma^{\kod{\gr}}$, and is omitted when it is obvious from the context and for terms, expressions and conditions as these to do not depend on the active graph. Note that the active graph can be changed only in case of the operator \kod{graph}.

Definition \ref{def:phi} specifies formula $\Phi(\overline{v})$ corresponding to a query \kod{Q}, with a graph pattern \kod{qpat} specifying a query dataset $\ag$.
As a parameter, this formula has a set $\overline{v}$ of some variables from $\vars$, while $\overline{ov}$ denote other free variables in the formula $\sigma(\kod{qpat})$. If distinguished variables are used as a parameter $\overline{v}$, then, intuitively, variables $\overline{ov}$ within formula $\Phi(\overline{dv})$ denote variables form $\mathcal{V}$ that correspond to variables from \kod{V} that are used in the query $\kod{Q}$ but that are not selected.

\begin{figure*}
	\startfig
	\begin{footnotesize}
		\startsubfigformula
		\begin{defn}[Theory signature $\mathcal{L}$ corresponding to the que\-ries $\kod{Q}_1$ and $\kod{Q}_2$]
			\label{def:signature}
			~\\
			\begin{minipage}{0.64\textwidth}
				$$(\funs, \preds, ar)$$
			\begin{itemize}
				\item $\funs \Eqdef \mathcal{C} \cup \fun$ is a set of function symbols, where: 
				\begin{itemize}
					\item $\mathcal{C}$ is a set of constants (function symbols with arity 0) corresponding to the literals and \iri{}s that appear in the queries, and a constant $\err$, corresponding to $\error$ in \ible{}.
					\item $\fun$ is a set of function symbols corresponding to the built-in \sparql{} functions used in the queries $\kod{Q}_1$ and $\kod{Q}_2$.
				\end{itemize}
				\item $\preds \Eqdef \mathcal{P} \; \cup \; \{ \ppp, \pp \}  $ is a set of predicate symbols, where: 
				\begin{itemize}
					\item $\mathcal{P}$ is a set which contains the equality predicate symbol ($=$) and other predicate symbols corresponding to the boolean \sparql{} functions used in $\kod{Q}_1$ and $\kod{Q}_2$.
				\end{itemize}
			\end{itemize}
			\end{minipage}
			\begin{minipage}{0.34\textwidth}
			\begin{itemize}
				\item[]
					\begin{itemize}
						\item $\ppp$ and $\pp$ are predicate symbols that intuitively model belonging of a triple
						to the default graph of an \rdfs{} dataset $\dataq$, or to a named graph specified by an \iri, respectively.
					\end{itemize}
				\item $ar$ is an arity function:
				$$ar(\alpha) \Eqdef 
				\begin{cases}
					0\,, ~~\text{if } \alpha \in \mathcal{C},\\
					1\,, ~~\text{if } \alpha \in \fun \text{ and } \alpha \text{ is } datatype,\\
					1\,, ~~\text{if } \alpha \in \mathcal{P} \text{ and } \alpha \text{ is } isliteral,\\
					2\,, ~~\text{if } \alpha \in \mathcal{P} \text{ and } \alpha \text{ is } =,\\
					3\,, ~~\text{if } \alpha \in \preds \text{ and } \alpha \text{ is } \ppp,\\
					4\,, ~~\text{if } \alpha \in \preds \text{ and } \alpha \text{ is } \pp.
				\end{cases}$$
			\end{itemize}
			\end{minipage}
		\end{defn}
		\endsubfig
		\vspace*{3px}
		\begin{minipage}{0.34\textwidth}
		\startsubfigformula
		\begin{defn}[Function $\sigvibl: \kod{\vibl{}e} \rightarrow \vars \cup \mathcal{C}$]
			\label{def:sigmat}
			$$\sigvibl(\kod{t}) \Eqdef
			\begin{cases}
				c \ (c \in \mathcal{C})\,, &\text{if } \kod{t} = \kod{c} \text{ and } \kod{c} \in \kod{IL},\\
				v \ (v \in \vars)\,, &\text{if } \kod{t} = \kod{v} \text{ and } \kod{v} \in \kod{VB},\\
				\err\,, &\text{if } \kod{t} = \error .
			\end{cases}$$
		\vspace*{13mm}
		\end{defn}
		\endsubfig
		\end{minipage}
		\begin{minipage}{0.655\textwidth}
		\startsubfigformula
			\begin{defn}[Function $\context$]
				\label{def:context}
				\begin{align*}
					\context(\kod{\gr}) \Eqdef  \begin{cases}
					\begin{lrdcases}
						\sigvibl(\{ \kod{i}_{k_1}, \cdots, \kod{i}_{k_m} \}), & \kod{G} = merge(\kod{\gr}_{k_1}, ..., \kod{\gr}_{k_m}) \text{ and} \\ & gr_{\dataq}(\kod{i}_{k_j})=\kod{\gr}_{k_j}, j \in \{1, \cdots, m\}\\
						\varnothing, & \kod{\gr} = \kod{\gr}_{\varnothing}\\
					\end{lrdcases}, & \begin{tabular}{l} \text{if }  \kod{\gr} \text{ is the default graph} \\ \text{of the query dataset } \ag \end{tabular}\\
					\\[-8pt]
					\begin{lrdcases}
						~ \\
						\sigvibl(\{ \kod{i} \}), \hspace*{12mm} & gr_{\dataq}(\kod{i})=\kod{\gr} \hspace*{22mm} \\
						~
					\end{lrdcases}, & \begin{tabular}{l} \text{if } \kod{\gr} \text{ is a named graph} \\ \text{of the query dataset } \ag \end{tabular}
					\end{cases}
				\end{align*}
			\end{defn}
			\endsubfig
		\end{minipage}
		\startsubfigformula 
		\begin{defn}[Function $\sigma$]
			\label{def:sigma}
			~\\
			\begin{minipage}{0.33\textwidth}		
			\begin{align*}
				& \text{\textbf{Term} \kod{t}} \\
				\sigma(\kod{t}) & \Eqdef \sigvibl(\kod{t}), & \kod{t} \in \vibl{}
			\end{align*}
			\begin{align*}
				& \text{\textbf{Expression} \kod{E}} \\
				\sigma(\kod{E}) & \Eqdef \begin{cases}
					\sigvibl(\kod{E}_1), & \text{if } \kod{E} \text{ is } \kod{E}_1 \text{ and } \kod{E}_1 \in \kod{VILe} \\
					datatype(\sigma(\kod{E}_1)), & \text{if } \kod{E} \text{ is } \kod{datatype(E}_1\kod{)}
				\end{cases}
			\end{align*}
			\begin{align*}
				& \text{\textbf{Condition} \kod{R}} \\
				\sigma(\kod{R}) & \Eqdef \begin{cases}
					\sigma(\kod{E}_1) = \sigma(\kod{E}_2), & \text{if } \kod{R} \text{ is } \kod{E}_1 = \kod{E}_2\\
					\neg\;\sigma(\kod{R}_1), & \text{if } \kod{R} \text{ is } \kod{!R}_1 \\
					\sigma(\kod{R}_1) \wedge \sigma(\kod{R}_2), & \text{if } \kod{R} \text{ is } \kod{R}_1 \;\kod{\&\&}\; \kod{R}_2 \\
					\sigma(\kod{R}_1) \vee \sigma(\kod{R}_2), & \text{if } \kod{R} \text{ is } \kod{R}_1 \;\kod{||}\; \kod{R}_2 \\
					\sigma(\kod{R}_1), & \text{if } \kod{R} \text{ is } \kod{(R}_1\kod{)} \\
					isliteral(\sigma(\kod{E}_1)), & \text{if } \kod{R} \text{ is } \kod{isliteral(E}_1\kod{)}
				\end{cases}\\
			\end{align*}
			\end{minipage}
			\begin{minipage}{0.66\textwidth}		
			\begin{align*}
				& \text{\textbf{Pattern} \kod{\pat}} \\
				\sigma^{\kod{\gr}}(\kod{\pat}) & \Eqdef
				\begin{cases}
					\begin{lrdcases}
					\ppp(\sigma(\kod{s}), \sigma(\kod{p}), \sigma(\kod{o})), &
					\kod{\gr} \notin dom(\context)\\
					\underset{i_j \in \context(\kod{\gr})} \bigvee \pp(\sigma(\kod{s}), \sigma(\kod{p}), \sigma(\kod{o}), i_j), & \kod{\gr} \in dom(\context), \context(\kod{\gr}) \not= \varnothing \\
					\bot , & \kod{\gr} \in dom(\context), \context(\kod{\gr}) = \varnothing \\	
					\end{lrdcases}, \hspace*{-2mm} & \text{if } \kod{\pat} \text{ is }\text{\codebox{$\kod{s p o}$}}\\
					\\[-8pt]
					\begin{lrdcases}
						\sigma^{\kod{\gr}}(\kod{\pat}_1) \wedge \sigma^{\kod{\gr}}(\kod{\pat}_2) & \hspace*{40mm}
					\end{lrdcases}, \hspace*{-2mm} & \text{if } \kod{\pat} \text{ is }\kod{\pat}_1 \; . \; \kod{\pat}_2 \\
					\\[-8pt]
					\begin{lrdcases}
						\sigma^{\kod{\gr}}(\kod{\pat}_1) \wedge \sigma(\kod{R}_1) & \hspace*{42.3mm}
					\end{lrdcases}, \hspace*{-2mm} & \text{if } \kod{\pat} \text{ is }\kod{\pat}_1 \kod{\;filter R}_1\\
					\\[-8pt]
					\begin{lrdcases}
						\sigma^{\kod{\gr}}(\kod{\pat}_1) & \hspace*{51.3mm}
					\end{lrdcases}, \hspace*{-2mm} & \text{if } \kod{\pat} \text{ is }\{\kod{\pat}_1\}\\
					\\[-8pt]
					\begin{lrdcases}
						\sigma^{\kod{\gr}}(\kod{\pat}_1), & \kod{var}(\kod{\pat}_1) \cap \kod{var}(\kod{\pat}_2) = \varnothing\\
						\sigma^{\kod{\gr}}(\kod{\pat}_1) \wedge \big( \forall \overline{x} \ \neg \sigma^{\kod{\gr}}(\kod{\pat}_2) \big) , & \kod{var}(\kod{\pat}_1) \cap \kod{var}(\kod{\pat}_2) \not= \varnothing \hspace*{3.2mm}
					\end{lrdcases}, \hspace*{-2mm} & \text{if } \kod{\pat} \text{ is }\kod{\pat}_1 \kod{ minus } \kod{\pat}_2\\
					\\[-8pt]
					\begin{lrdcases}
						\sigma^{\kod{\gr}}(\kod{\pat}_1) \wedge \big( \forall \overline{x} \ \neg \sigma^{\kod{\gr}}(\kod{\pat}_2) \big) & \hspace*{32mm}
					\end{lrdcases}, \hspace*{-2mm} & \text{if } \kod{\pat} \text{ is }\kod{\pat}_1 \kod{ diff } \kod{\pat}_2, \\ 
					\\[-8pt]
					\begin{lrdcases}
						\underset{i \in \agfn}\bigvee \Big( \sigma^{gr_{\,\ag}(\siginv(i))}(\kod{\pat}_1) \wedge \sigma(\kod{x}) = i \Big), & \agfn \not= \varnothing \\
						\bot, & \agfn = \varnothing \hspace*{11mm}\\
					\end{lrdcases}, \hspace*{-2mm} & \text{if } \kod{\pat} \text{ is }\kod{graph x \{\pat}_1 \kod{\}} \\
					\\[-8pt]
					\begin{lrdcases}
						\sigma^{gr_{\,\ag}(\kod{i})}(\kod{\pat}_1), &  \sigma(\kod{i}) \in \agfn\\
						\bot, & \sigma(\kod{i}) \notin \agfn \hspace*{33.8mm}
					\end{lrdcases}, \hspace*{-2mm} & \text{if } \kod{\pat} \text{ is }\kod{graph i \{\pat}_1 \kod{\}} \\
				\end{cases}
			\end{align*}
			\end{minipage}
		\end{defn}
		\endsubfig
		\startsubfigformula
		\begin{defn}[Formula $\Phi(\overline{v})$ corresponding to \kod{Q}]
			\label{def:phi}
				\kod{qpat} - a graph pattern within the query \kod{Q};
				~~\ag{} - a query dataset; 
				~~$\overline{ov}$ - $var(\sigma(\kod{qpat})) \setminus \overline{v}$
			$$\Phi(\overline{v}) \Eqdef \exists \overline{ov}  \ \sigma^{\mathit{df}({\ag})}(\kod{qpat})$$
		\end{defn}
		\endsubfig
	\end{footnotesize}
	\vspace*{4px}
	\endfig
	\caption{Definitions taken from \cite{solving_with_specs} are given here in a short form: theory signature $\mathcal{L}$ corresponding to the queries $\kod{Q}_1$ and $\kod{Q}_2$, functions $\sigvibl$, $\context$ and $\sigma$ for transforming queries into formulas and formula $\Phi(\overline{x})$ corresponding to the query \kod{Q}. 
$\kod{E}_1$ and $\kod{E}_2$ are used for expressions, $\kod{R}_1$ and $\kod{R}_2$ for conditions, \kod{s}, \kod{p} and \kod{o} for subjects, predicates and objects, respectively ($\kod{s} \in \kod{VIB}, \kod{p} \in \kod{VI}, \kod{o} \in \vibl{}$), $\kod{\pat}_1$ and $\kod{\pat}_2$ for graph patterns, \kod{x} for a \sparql{} variable, \kod{i} for an \iri{}, while $\agfn$ is a set containing all graph \iri{} constants corresponding to the named graphs of the query dataset $\ag$. $\overline{x}$ stands for $var(\sigma(\kod{\pat}_2)) \setminus var(\sigma(\kod{\pat}_1))$.}
	\label{fig:definitions1}
\end{figure*}

\begin{figure}[h]
	\startfig
	\begin{small}
		\startsubfig
		\vspace*{-3mm}
		\begin{align*}
			\hspace*{-1mm}
			\exists \, z_v \, \exists \,  w_v ~ \Big(
			& \ppp(x_v,a_i,Album_i) \; \wedge \; \ppp(x_v,artist_i,y_v) \; \wedge \\
			&\ppp(y_v,a_i,SoloArtist_i) \; \wedge \; \ppp(y_v,hometown_i,z_v) \; \wedge\\
			&\ppp(z_v,name_i,w_v) \; \wedge \; w_v = LosAngeles_l ~ \Big)
		\end{align*}
		\endsubfig
	\end{small}
	\vspace*{2px}
	\endfig
	\caption{
		Formula $\Phi(x_v, y_v)$ corresponding to the upper query given in Figure \ref{fig:containment1}.
		For readability, each \iri, literal and variable is denoted by a corresponding subscript $i$, $l$ and $v$, respectively.}
	\label{fig:sigma_patern1}
\end{figure}

\begin{defn}[Function $var$ over formulas]
	\label{def:var_sigma}
	Let $\Phi$ be a \fol{} formula.
	$var(\Phi)$ denotes a set of free variables appearing in the formula $\Phi$.
\end{defn}

The following four lemmas connect functions $\kod{var}$, $\sigma$ and $var$.

\begin{lemma}
	\label{thm:dom_mu_nuT}
	Let $\kod{t}$ be a term.
	Then:
	\begin{center}
		$var(\sigma(\kod{t})) = \sigma(\kod{var}(\kod{t}))$.
	\end{center}
	\begin{proof}
		%\ref{proof:dom_mu_nuT} is given in Appendix \ref{sec:appendix}. It is done by induction over the term $\kod{t}$.
		
		The specified equality is proved by induction over a term $\kod{t}$.
		\begin{description}
			\item \kod{t}  is  $\kod{c}, \kod{c} \in \kod{IL}$
			\begin{align*}
				&\hspace*{20mm} var(\sigma(\kod{c})) \hspace*{-14mm} &  = \ \\
				&\text{(by Def \ref{def:sigma})} & = \ & var(\sigma_{\kod{t}} (\kod{c})) \\
				&\text{(by Def \ref{def:sigmat})} & = \ & var(c) \\
				&\text{(by Def \ref{def:var_sigma})} & = \ & \varnothing\\
				&\text{(by Def \ref{def:function_over_sets})} & = \ & \sigma(\varnothing)\\
				&\text{(by Def \ref{def:var})} & = \ & \sigma(\kod{var}(\kod{c}))
			\end{align*}
			
			\item \kod{t}  is  $\kod{v}, \kod{v} \in \kod{VB}$
			\begin{align*}
				&\hspace*{20mm} var(\sigma(\kod{v})) \hspace*{-14mm} & = \ \\
				&\text{(by Def \ref{def:sigma})} & = \ & var(\sigma_{\kod{t}} (\kod{v})) \\
				&\text{(by Def \ref{def:sigmat})} & = \ & var(v) \\
				&\text{(by Def \ref{def:var_sigma})} & = \ & \{ v \}\\
				&\text{(by Def \ref{def:function_over_sets})} & = \ & \sigma(\{ \kod{v} \})\\
				&\text{(by Def \ref{def:var})} & = \ & \sigma(\kod{var}(\kod{v}))
			\end{align*}
		\end{description}
	\end{proof}    
\end{lemma}

\begin{lemma}
	\label{thm:dom_mu_nuE}
	Let $\kod{E}$ be an expression.
	Then:
	\begin{center}
		$var(\sigma(\kod{E})) = \sigma(\kod{var}(\kod{E}))$.
	\end{center}
	\begin{proof}
		%\ref{proof:dom_mu_nuE} is given in Appendix \ref{sec:appendix}. It is done by induction over the expression $\kod{E}$.
		
		The specified equality is proved by induction over an expression $\kod{E}$.
		\begin{description}
			\item \kod{E} is $\kod{t}$
			\begin{align*}
				& \hspace*{15mm} var(\sigma(\kod{t})) \hspace*{-9mm} & = \ \\
				&\text{(by Lemma \ref{thm:dom_mu_nuT})}  & = \  & \sigma(\kod{var}(\kod{t})) 
			\end{align*}
			
			\item \kod{E}  is  $\kod{datatype(E}_1\kod{)}$
			\begin{align*}
				&var(\sigma(\kod{datatype(E}_1\kod{)})) \hspace*{-20mm}  & = \ \\
				&\text{(by Def \ref{def:sigma}} ) & = \ & var(datatype(\sigma(\kod{E}_1))) \\
				&\text{(by Def \ref{def:var_sigma})} & = \ & var(\sigma(\kod{E}_1)) \\
				&\text{(by induc.hyp.)} & = \ & \sigma(\kod{var}(\kod{E}_1)) \\
				&\text{(by Def \ref{def:var})} & = \ & \sigma(\kod{var}(\kod{datatype(E}_1\kod{)}))
			\end{align*}
		\end{description}
	\end{proof}    
\end{lemma}

\begin{lemma}
	\label{thm:dom_mu_nuR}
	Let $\kod{R}$ be a condition.
	Then:
	\begin{center}
		$var(\sigma(\kod{R})) = \sigma(\kod{var}(\kod{R}))$.
	\end{center}
	\begin{proof}
		%\ref{proof:dom_mu_nuR} is given in Appendix \ref{sec:appendix}. It is done by induction over the condition $\kod{R}$.
		
		The specified equality is proved by induction over a condition $\kod{R}$.
		\begin{description}
			\item \kod{R}  is  $\kod{E}_1$ = $\kod{E}_2$
			\begin{align*}
				& \hspace*{5mm} var(\sigma(\kod{E}_1 = \kod{E}_2)) \hspace*{-3mm}& = \ \\
				&\text{(by Def \ref{def:sigma}} ) & = \ & var(\sigma(\kod{E}_1) = \sigma(\kod{E}_2)) \\
				&\text{(by Def \ref{def:var_sigma})} & = \ & var(\sigma(\kod{E}_1)) \cup var(\sigma(\kod{E}_2)) \\
				&\text{(by Lemma \ref{thm:dom_mu_nuE})} & = \ & \sigma(\kod{var}(\kod{E}_1)) \cup \sigma(\kod{var}(\kod{E}_2)) \\
				&\text{(by Def \ref{def:function_over_sets})} & = \ & \sigma(\kod{var}(\kod{E}_1) \cup \kod{var}(\kod{E}_2)) \\
				&\text{(by Def \ref{def:var})} & = \ & \sigma(\kod{var}(\kod{E}_1 = \kod{E}_2))
			\end{align*}
			
			\item $\kod{R} \text{ is } \kod{!R}_1$
			\begin{align*}
				& \hspace*{15mm} var(\sigma(\kod{!R}_1)) \hspace*{-15mm}& = \  \\
				&\text{(by Def \ref{def:sigma}} ) & = \ & var(\neg \sigma(\kod{R}_1)) \\
				&\text{(by Def \ref{def:var_sigma})} & = \ & var(\sigma(\kod{R}_1)) \\
				&\text{(by induc.hyp.)} & = \ & \sigma(\kod{var}(\kod{R}_1)) \\
				&\text{(by Def \ref{def:var})} & = \ & \sigma(\kod{var}(\kod{!R}_1))
			\end{align*}
			
			\item $\kod{R} \text{ is } \kod{R}_1 \kod{\&\&} \kod{R}_2$
			\begin{align*}
				& \hspace*{2mm} var(\sigma(\kod{R}_1 \kod{\&\&} \kod{R}_2)) \hspace*{-10mm} & = \  \\
				&\text{(by Def \ref{def:sigma}} ) & = \ & var(\sigma(\kod{R}_1) \wedge \sigma(\kod{R}_2)) \\
				&\text{(by Def \ref{def:var_sigma})} & = \ & var(\sigma(\kod{R}_1)) \cup var(\sigma(\kod{R}_2))\\
				&\text{(by induc.hyp.)} & = \ & \sigma(\kod{var}(\kod{R}_1)) \cup \sigma(\kod{var}(\kod{R}_2))\\
				&\text{(by Def \ref{def:function_over_sets})} & = \ & \sigma(\kod{var}(\kod{R}_1) \cup \kod{var}(\kod{R}_2))\\
				&\text{(by Def \ref{def:var})} & = \ & \sigma(\kod{var}(\kod{R}_1 \kod{\&\&} \kod{R}_2))
			\end{align*}
			
			\item $\kod{R} \text{ is } \kod{R}_1 \kod{||} \kod{R}_2$\\
			Proof is analogous to the previous case.
			
			\item $\kod{R} \text{ is } \kod{(R}_1\kod{)}$\\
			Proof is analogous to the case $\kod{R} \text{ is } \kod{!R}_1$.
			
			\item \kod{R}  is  $\kod{isliteral(E)}$
			\begin{align*}
				&~~~var(\sigma(\kod{isliteral(E)})) = \hspace*{-28mm}&  \\
				&\text{(by Def \ref{def:sigma}} ) & = \ & var(isliteral(\sigma(\kod{E}))) \\
				&\text{(by Def \ref{def:var_sigma})} & = \ & var(\sigma(\kod{E})) \\
				&\text{(by induc.hyp.)} & = \ & \sigma(\kod{var}(\kod{E})) \\
				&\text{(by Def \ref{def:var})} & = \ & \sigma(\kod{var}(\kod{isliteral(E)}))
			\end{align*}
			
		\end{description}
	\end{proof}    
\end{lemma}

In the following text, $\exists \overline{a}$ abbreviates $\exists  a_1 ... \exists a_n$, and $\forall \overline{a}$ abbreviates $\forall a_1 ... \forall a_n$, when $n$ is clear from the context.

\begin{lemma}
	\label{thm:dom_mu_nu1}
	Let $\kod{\pat}$ be a graph pattern. Then, if all active graphs that are used for matching $\kod{\pat}$ are nonempty, it holds:
	\begin{center}
		$var(\sigma(\kod{\pat})) = \sigma(\kod{var}(\kod{\pat}))$.
	\end{center}
	\begin{proof}
		%\ref{proof:dom_mu_nu1} is given in Appendix \ref{sec:appendix}.	It is done by induction over the graph pattern $\kod{\pat}$.
		
		The specified equality is proved by induction over a graph pattern $\kod{\pat}$.
		\begin{description}
			\item \kod{\pat{}} is \kod{\tpat{}}, where \kod{\tpat{}} is \kod{s p o}\\
			Let $\kod{\gr}$ be an active graph for matching $\kod{\tpat{}}$.  \\
			If $\kod{\gr} \notin dom(cx)$:
			\begin{align*}
				& var(\sigma^{\kod{\gr}}(\kod{s\,p\,o})) \hspace*{-20mm}&  = \ \\
				&\text{(by Def \ref{def:sigma}} ) & = \ & var( \ppp(\sigma(\kod{s}), \sigma(\kod{p}), \sigma(\kod{o}))) \\
				&\text{(by Def \ref{def:var_sigma})} & = \ & \{ \sigma(\kod{s}), \sigma(\kod{p}), \sigma(\kod{o})\} \cap \vars \\
				&\text{(by Def \ref{def:sigmat})} & = \ & \{ \sigma(\kod{s}), \sigma(\kod{p}), \sigma(\kod{o})\} \cap \sigvibl(\kod{VB}) \\
				&\text{(by Def \ref{def:sigma})} & = \ & \{ \sigma(\kod{s}), \sigma(\kod{p}), \sigma(\kod{o})\} \cap \sigma(\kod{VB}) \\
				&\text{(by Def \ref{def:function_over_sets})} & = \ & \sigma(\{ \kod{s}, \kod{p}, \kod{o}\} \cap \kod{VB}) \\
				&\text{(by Def of } \cup) & = \ & \sigma( (\{ \kod{s} \} \cup  \{ \kod{p} \} \cup \{ \kod{o} \}) \cap \kod{VB}) \\
				&\text{(by dist. of } \cap  & = \ & \sigma( (\{ \kod{s} \} \cap \kod{VB}) \; \cup \\
				&\text{ over } \cup\text{)}&& (\{ \kod{p} \} \cap \kod{VB}) \cup (\{ \kod{o} \} \cap \kod{VB})) \\
				&\text{(by Def \ref{def:var})}  & = \ & \sigma(\kod{var}(\kod{s}) \cup  \kod{var}(\kod{p}) \cup  \kod{var}(\kod{o})  ))\\
				&\text{(by Def \ref{def:var})} & = \ & \sigma(\kod{var}(\kod{s p o}))\\
			\end{align*}
			
			If $\kod{\gr} \in dom(cx)$, as graph \kod{\gr} is a nonempty graph, by Definition \ref{def:context}, it holds $cx(\kod{\gr}) \not= \varnothing$.
			
			\begin{align*}
				& var(\sigma^{\kod{\gr}}(\kod{s p o})) \hspace*{-28mm}&  \\
				&\text{(by Def \ref{def:sigma}} ) & = \ & var( \hspace*{-3mm} \underset{i_j \in \context(\kod{\gr})} \bigvee \hspace*{-3mm} \pp(\sigma(\kod{s}), \sigma(\kod{p}), \sigma(\kod{o}), i_j)) \\
				&\text{(by Def \ref{def:var_sigma})} & = \ & \hspace*{-3mm} \underset{i_j \in \context(\kod{\gr})} \bigcup \hspace*{-2mm} var( \pp(\sigma(\kod{s}), \sigma(\kod{p}), \sigma(\kod{o}), i_j)) \\
				&\text{(by Def \ref{def:context})} & = \ & \{ \sigma(\kod{s}), \sigma(\kod{p}), \sigma(\kod{o})\} \cap \vars \\
				&\text{(by Def \ref{def:sigmat})} & = \ & \{ \sigma(\kod{s}), \sigma(\kod{p}), \sigma(\kod{o})\} \cap \sigvibl(\kod{VB}) \\
				&\text{(by Def \ref{def:sigma})} & = \ & \{ \sigma(\kod{s}), \sigma(\kod{p}), \sigma(\kod{o})\} \cap \sigma(\kod{VB}) \\
				&\text{(by Def \ref{def:function_over_sets})} & = \ & \sigma(\{ \kod{s}, \kod{p}, \kod{o}\} \cap \kod{VB}) \\
				&\text{(by Def of } \cup) & = \ & \sigma( (\{ \kod{s} \} \cup  \{ \kod{p} \} \cup \{ \kod{o} \}) \cap \kod{VB}) \\
				&\text{(by dist. of } \cap  & = \ & \sigma( (\{ \kod{s} \} \cap \kod{VB}) \; \cup \\
				&\text{ over } \cup\text{)} && (\{ \kod{p} \} \cap \kod{VB}) \cup (\{ \kod{o} \} \cap \kod{VB})) \\
				&\text{(by Def \ref{def:var})}  & = \ & \sigma(\kod{var}(\kod{s}) \cup  \kod{var}(\kod{p}) \cup  \kod{var}(\kod{o})  ))\\
				&\text{(by Def \ref{def:var})} & = \ & \sigma(\kod{var}(\kod{s p o}))
			\end{align*}
			
			\item $\kod{\pat}$  is $\kod{\pat}_1 . \kod{\pat}_2$
			\begin{align*}
				& var(\sigma(\kod{\pat}_1 . \kod{\pat}_2)) \hspace*{-4mm} & = \  \\
				&\text{(by Def \ref{def:sigma}} ) &  = \ & var( \sigma(\kod{\pat}_1) \wedge \sigma(\kod{\pat}_1))\\
				&\text{(by Def \ref{def:var_sigma})} & = \ & var( \sigma(\kod{\pat}_1)) \cup var(\sigma(\kod{\pat}_1)) \\
				&\text{(by induc.hyp.)} \hspace*{-3mm}& = \ & \sigma(\kod{var}(\kod{\pat}_1)) \cup \sigma(\kod{var}(\kod{\pat}_1)) \\
				&\text{(by Def \ref{def:function_over_sets})} & = \ & \sigma(\kod{var}(\kod{\pat}_1) \cup \kod{var}(\kod{\pat}_2)) \\
				&\text{(by Def \ref{def:var})} & = \ & \sigma(\kod{var}(\kod{\pat}_1 . \kod{\pat}_2))
			\end{align*}
			
			\item $\kod{\pat}$ is $\kod{\pat}_1 \kod{ filter } \kod{R}$
			\begin{align*}
				& var(\sigma(\kod{\pat}_1 \kod{ filter } \kod{R})) \ =  \hspace*{-15mm} &  \\
				&\text{(by Def \ref{def:sigma}} ) &  = \ & var( \sigma(\kod{\pat}_1) \wedge \sigma(\kod{R}))\\
				&\text{(by Def \ref{def:var_sigma})} & = \ & var( \sigma(\kod{\pat}_1)) \cup var(\sigma(\kod{R})) \\
				&\text{(by induc.hyp.)} \hspace*{-3mm} & = \ & \sigma(\kod{var}(\kod{\pat}_1)) \cup \kod{var}(\sigma(\kod{R})) \\
				&\text{(by Lemma \ref{thm:dom_mu_nuR})} & = \ & \sigma(\kod{var}(\kod{\pat}_1)) \cup \sigma(\kod{var}(\kod{R})) \\
				&\text{(by Def \ref{def:function_over_sets})} & = \ & \sigma(\kod{var}(\kod{\pat}_1) \cup \kod{var}(\kod{R})) \\
				&\text{(as } \kod{var}(\kod{R}) \hspace*{-1mm}\subseteq\hspace*{-1mm} \kod{var}(\kod{\pat}_1) \text{)} \hspace*{-3mm} & = \ & \sigma(\kod{var}(\kod{\pat}_1)) \\
				&\text{(by Def \ref{def:var})} & = \ & \sigma(\kod{var}(\kod{\pat}_1 \kod{ filter } \kod{R}))
			\end{align*}
			
			\item $\kod{\pat}$ is $\kod{\{\pat}_1\kod{\}}$
			\begin{align*}
				& \hspace*{8mm} var(\sigma(\kod{\{\pat}_1\kod{\}}) \hspace*{-8mm} & = \  \\
				&\text{(by Def \ref{def:sigma}} ) &  = \ & var( \sigma(\kod{\pat}_1))\\
				&\text{(by induc.hyp.)} \hspace*{-3mm} & = \ & \sigma(\kod{var}(\kod{\pat}_1)) \\
				&\text{(by Def \ref{def:var})} & = \ & \sigma(\kod{var}(\kod{\{\pat}_1\kod{\}}))		
			\end{align*}
			
			\item $\kod{\pat}$ is $\kod{\pat}_1 \; \kod{diff} \; \kod{\pat}_2$
			\begin{align*}
				& var(\sigma(\kod{\pat}_1 \; \kod{diff} \; \kod{\pat}_2))  = \hspace*{-30mm} \\
				&\text{(by Def \ref{def:sigma}} ) &  = \ & var(\sigma(\kod{\pat}_1) \wedge \forall \overline{x} \ \neg \sigma(\kod{\pat}_2))\\
				&\text{(by Def \ref{def:var_sigma}} ) &  = \ & var(\sigma(\kod{\pat}_1)) \\
				&\text{(by induc.hyp.)} &  = \ & \sigma(\kod{var}(\kod{\pat}_1))\\
				&\text{(by Def \ref{def:var})} & = \ & \sigma(\kod{var}(\kod{\pat}_1 \; \kod{diff} \; \kod{\pat}_2))
			\end{align*}
			
			\item $\kod{\pat}$ is $\kod{\pat}_1 \; \kod{minus} \; \kod{\pat}_2$
			
			If $\kod{var}(\kod{\pat}_1) \cap \kod{var}(\kod{\pat}_2) = \varnothing$ then:
			\begin{align*}
				& var(\sigma(\kod{\pat}_1 \; \kod{minus} \; \kod{\pat}_2))  = \hspace*{-30mm} \\
				&\text{(by Def \ref{def:sigma}} ) &  = \ & var(\sigma(\kod{\pat}_1))\\
				&\text{(by induc.hyp.)} &  = \ & \sigma(\kod{var}(\kod{\pat}_1))\\
				&\text{(by Def \ref{def:var})} & = \ & \sigma(\kod{var}(\kod{\pat}_1 \; \kod{minus} \; \kod{\pat}_2))
			\end{align*}		
			Otherwise:
			\begin{align*}
				& var(\sigma(\kod{\pat}_1 \; \kod{minus} \; \kod{\pat}_2))  = \hspace*{-30mm} \\
				&\text{(by Def \ref{def:sigma}} ) &  = \ & var(\sigma(\kod{\pat}_1) \wedge \forall \overline{x} \ \neg \sigma(\kod{\pat}_2))\\
				&\text{(by Def \ref{def:var_sigma}} ) &  = \ & var(\sigma(\kod{\pat}_1)) \\
				&\text{(by induc.hyp.)} &  = \ & \sigma(\kod{var}(\kod{\pat}_1))\\
				&\text{(by Def \ref{def:var})} & = \ & \sigma(\kod{var}(\kod{\pat}_1 \; \kod{minus} \; \kod{\pat}_2))
			\end{align*}
			
			\item $\kod{\pat}$ is $\kod{graph x} \ \{ \kod{\pat}_1 \}$\\
			
			If $\agfn = \varnothing$, there is no active graph for matching $\kod{\pat}_1$. Then, by Definition \ref{definition_sem}, this case is equivalent to a case when the active graph is empty, which does not correspond to the assumption of this lemma. Therefore, we can assume that $\agfn \not= \varnothing$.
			
			From $\agfn \not= \varnothing$, according to Definition \ref{def:sigma}, it holds that for all $i \in \agfn$ graph $gr_{\,\ag}(\siginv(i))$ is a nonempty one.
			\begin{align*}
				&var(\sigma(\kod{graph x} \  \{ \kod{\pat}_1 \})) = \hspace*{-30mm}&  \\
				&\text{(by Def \ref{def:sigma}} ) \hspace*{-2mm}&  = & var(\underset{i \in \agfn}\bigvee ( \sigma^{gr_{\,\ag}(\siginv(i))}(\kod{\pat}_1) \\
				&&& \hspace*{7mm} \wedge \sigma(\kod{x}) = i ))\\
				&\text{(by Def \ref{def:var_sigma}} ) &  = & \underset{i \in \agfn}\bigcup ( var(\sigma^{gr_{\,\ag}(\siginv(i))}(\kod{\pat}_1)) \\
				&&& \hspace*{7mm} \cup var(\sigma(\kod{x}) = i ))\\
				&\text{(by induc.hyp.)} \hspace*{-5mm}&  = & \underset{i \in \agfn}\bigcup ( \sigma(\kod{var}(\kod{\pat}_1))\\
				&&& \hspace*{7mm} \cup var(\sigma(\kod{x}) = i ))\\
				&\text{(as $\agfn \cap \vars$ is $\varnothing$)} \hspace*{-2mm}& = & \sigma(\kod{var}(\kod{\pat}_1)) \cup \{ \sigma(  \kod{x})\} \\
				&\text{(by Def \ref{def:function_over_sets})} & = & \sigma(\kod{var}(\kod{\pat}_1) \cup \{ \kod{x} \}) \\
				&\text{(by Def \ref{def:var})} \hspace*{-2mm} & = & \sigma(\kod{var}(\kod{graph x} \  \{ \kod{\pat}_1 \}))
			\end{align*}
			
			\item $\kod{\pat}$ is $\kod{graph i} \; \{ \kod{\pat}_1 \}$\\
			If $\sigma(\kod{i}) \notin \agfn$, by Definition \ref{def:sigma}, it holds $\kod{i} \notin names(\ag)$, i.e.~by Definition \ref{def:dataset} $gr_\kod{\,D}(\kod{i}) = \varnothing$. Then, according to Definition \ref{def:sigma}, an active graph that is used for matching $\kod{\pat}_1$ is empty, which does not correspond to the assumption of this lemma. Therefore, we can assume that $\sigma(\kod{i}) \in \agfn$.

			From $\sigma(\kod{i}) \in \agfn$, according to Definition \ref{def:sigma}, the graph $gr_{\,\ag}(\kod{i})$ is a nonempty graph.
			\begin{align*}
				&var(\sigma(\kod{graph i} \; \{ \kod{\pat}_1 \})) =  \hspace*{-35mm}&  \\
				&\text{(by Def \ref{def:sigma}} )  &  = & var( \sigma^{gr_{\,\ag}(\kod{i})}(\kod{\pat}_1))\\
				&\text{(by induc.hyp.)}  &  =  & \sigma(\kod{var}(\kod{\pat}_1)) \\
				&\text{(by Def \ref{def:var})}  & = & \sigma(\kod{var}(\kod{graph i} \; \{ \kod{\pat}_1 \}))
			\end{align*}
		\end{description}
	\end{proof}    
\end{lemma}

\begin{lemma}
	\label{thm:var_phi}
	Let $\overline{rv}$ be a set of variables from $\vars$ that correspond to the relevant variables of a query \kod{Q} and let $\Phi(\overline{rv})$ be a formula that corresponds to \kod{Q}.
	Then:
	\begin{center}
		$var(\Phi(\overline{rv})) = \overline{rv}$.
	\end{center}
	\begin{proof}
		%\ref{proof:var_phi} is given in Appendix \ref{sec:appendix}. It uses Definitions \ref{def:function_over_sets}, \ref{def:var_sigma} and \ref{def:phi} and Lemmas \ref{thm:query_domain} and \ref{thm:dom_mu_nu1}.
		
		Let \kod{qpat} and $\ag$ be a query pattern and a query dataset of the query $\kod{Q}$, respectively.
		Let $\overline{ov}$ denote $var(\sigma(\kod{qpat})) \setminus \overline{rv}$.
		\begin{align*}
			& \hspace*{12mm} var(\Phi(\overline{rv})) \hspace*{-4mm} &=\; \\
			&\text{(by Def \ref{def:phi})} &=\;&  var(\exists \overline{ov} \  \sigma^{\mathit{df}(\ag)}(\kod{qpat})) \\
			& \text{(By Def \ref{def:var_sigma})} &=\;&  var(\sigma^{\mathit{df}(\ag)}(\kod{qpat})) \setminus \overline{ov}\\
			&\text{(by Def \ref{def:phi})} &=\;& var(\sigma^{\mathit{df}(\ag)}(\kod{qpat})) \setminus\\
			&&& (var(\sigma^{\mathit{df}(\ag)}(\kod{qpat})) \setminus \overline{rv}) \\
			&&=\;& var(\sigma^{\mathit{df}(\ag)}(\kod{qpat})) \cap \overline{rv} \\
			& \text{(as } \sigma(\overline{\kod{rv}}) = \overline{rv} \text{)} &=\;&  var(\sigma^{\mathit{df}(\ag)}(\kod{qpat})) \cap \sigma(\overline{\kod{rv}})\\
			&\text{(by Lemma \ref{thm:dom_mu_nu1})} &=\;&  \sigma(var(\kod{qpat})) \cap \sigma(\overline{\kod{rv}})\\
			&\text{(by Def \ref{def:function_over_sets})} &=\;& \sigma(var(\kod{qpat}) \cap \overline{\kod{rv}})\\
			&\text{(by Lemma \ref{thm:query_domain})} &=\;& \sigma(\overline{\kod{rv}})\\
			& \text{(as } \sigma(\overline{\kod{rv}}) = \overline{rv} \text{)} &=\;& \overline{rv}
		\end{align*}
	\end{proof}    
\end{lemma}

\subsection{Modeling the Containment Relation}
\label{subsec:querycontainment}

Modeling the query containment problem and subsumption problem include defining formulas $\Theta$ and $\Psi$, and relation $\seleq$, for containment, and relation $\dot{\seleq}$ for subsumption. These are defined in \cite{solving_with_specs} and given in a short form in Figure \ref{fig:definitions2}.

Definition \ref{def:theta} specifies formula $\Theta$, based on a query $\kod{Q}_1$ (a candidate for sub-query), where $\overline{rv_1}$ denote variables from $\vars$ that correspond to the relevant variables of $\kod{Q}_1$, while $\Phi_1(\overline{rv_1})$ is a formula that corresponds to $\kod{Q}_1$ according to Definition \ref{def:phi}.

Definition \ref{def:psi} specifies formula $\Psi$, based of queries $\kod{Q}_1$ and $\kod{Q}_2$, where $\overline{rv_1}$ denote variables from $\vars$ that correspond to the relevant variables of $\kod{Q}_1$, while $\Phi_1(\overline{rv_1})$ and $\Phi_2(\overline{rv_1})$ are formulas that correspond to $\kod{Q}_1$ and $\kod{Q}_2$, respectively according to Definition \ref{def:phi}.

\begin{figure}
	\startfig
	\begin{footnotesize}
		\begin{minipage}{0.49\textwidth}
		\startsubfigformula
		\begin{defn}[Formula $\Theta$]
			\label{def:theta}
			$$\neg \big(\exists \overline{rv_1} \ \Phi_1(\overline{rv_1})\big)$$
		\end{defn}
		\endsubfig
		\end{minipage}
		\begin{minipage}{0.49\textwidth}
		\startsubfigformula 
		\begin{defn}[Formula $\Psi$]
			\label{def:psi}
			$$\forall \overline{rv_1} \ \big( \Phi_1(\overline{rv_1}) \Rightarrow \Phi_2(\overline{rv_1}) )$$
		\end{defn}
		\endsubfig
		\end{minipage}
		\startsubfigformula
		\begin{defn}[Relations $\seleq$ and $\dot{\seleq}$]
			\label{def:set_equality}
			~\\
			\begin{minipage}{0.49\textwidth}
			$$\kod{Q}_1 \seleq \kod{Q}_2 \text{~~~iff~~~} \overline{\kod{rv}_1} = \overline{\kod{rv}_2}$$
			\end{minipage}
			\begin{minipage}{0.49\textwidth}
			$$\kod{Q}_1 \dot{\seleq} \kod{Q}_2  \text{~~~iff~~~} \overline{\kod{rv}_1} \subseteq \overline{\kod{rv}_2}$$
			\end{minipage}
		\end{defn}
		\endsubfig
	\end{footnotesize}
	\vspace*{4px}
	\endfig
	\caption{Formula $\Theta$ (based on $\kod{Q}_1$), formula $\Psi$ (based on $\kod{Q}_1$ and $\kod{Q}_2$) and relations $\seleq$ and $\dot{\seleq}$ (according to \cite{solving_with_specs}).}
	\label{fig:definitions2}
\end{figure}

Relations $\seleq$ and $\dot{\seleq}$ are given within Definition \ref{def:set_equality}. These relations connect queries $\kod{Q}_1$ and $\kod{Q}_2$ based on a relation between their relevant variables $\overline{\kod{rv}_1}$ and $\overline{\kod{rv}_2}$, respectively.

The following lemma connects the relevant variables of $\kod{Q}_1$ and variables of $\kod{Q}_2$ if relation $\seleq$ holds.

\begin{lemma}
	\label{thm:set_equality}
	Let $\kod{Q}_1$ and $\kod{Q}_2$ be queries, such that $\kod{Q}_2$ does not contain projections and $\kod{Q}_1 \seleq \kod{Q}_2$ holds.
	Let $\overline{\kod{rv}_1}$ be a set of relevant variables of $\kod{Q}_1$ and $\kod{qpat}_2$ a query pattern of $\kod{Q}_2$.
	Then:
	\begin{center}
		$\overline{\kod{rv}_1} = \kod{var}(\kod{qpat}_2)$.
	\end{center}
	\begin{proof}
		%\ref{proof:set_equality} is given in Appendix \ref{sec:appendix}. It uses Definitions \ref{defn:query_evaluation} and \ref{def:set_equality} and Lemma \ref{thm:query_domain}.
		
		By Definition \ref{def:set_equality}, from $\kod{Q}_1 \seleq \kod{Q}_2$, it holds
		$$\overline{\kod{rv}_1} = \overline{\kod{rv}_2}.$$
		Then, by Lemma \ref{thm:query_domain}, it holds
		$$\overline{\kod{rv}_1} = \kod{var}(\kod{qpat}_2) \cap \overline{\kod{dv}_2}.$$
		By Definition \ref{defn:query_evaluation}, as $\kod{Q}_2$ does not contain projections, it holds
		$$\overline{\kod{rv}_1} = \kod{var}(\kod{qpat}_2).$$
	\end{proof}
\end{lemma}

The following lemma is dual to Lemma \ref{thm:set_equality}, but considering relation $\dot{\seleq}$ instead of $\seleq$ and without a restriction regarding projections of $\kod{Q}_2$.

\begin{lemma}
	\label{thm:set_equality_subsumption}
	Let $\kod{Q}_1$ and $\kod{Q}_2$ be queries, such that $\kod{Q}_1 \dot{\seleq} \kod{Q}_2$.
	Let $\overline{\kod{rv}_1}$ be a set of relevant variables of $\kod{Q}_1$ and $\kod{qpat}_2$ and $\overline{\kod{dv}_2}$ a query pattern and a set of distinguished variables of $\kod{Q}_2$, respectively.
	Then:
	\begin{center}
		$\overline{\kod{rv}_1} \subseteq \kod{var}(\kod{qpat}_2) \cap \overline{\kod{dv}_2}$.
	\end{center}
	\begin{proof}
		%\ref{proof:set_equality_subsumption} is given in Appendix \ref{sec:appendix}. It uses Definition \ref{def:set_equality} and Lemma \ref{thm:query_domain}.
		
		By Definition \ref{def:set_equality}, from $\kod{Q}_1 \dot{\seleq} \kod{Q}_2$, it holds
		$$\overline{\kod{rv}_1} \subseteq \overline{\kod{rv}_2}.$$
		Then, by Lemma \ref{thm:query_domain}, it holds
		$$\overline{\kod{rv}_1} \subseteq \kod{var}(\kod{qpat}_2) \cap \overline{\kod{dv}_2}.$$
	\end{proof}
\end{lemma}

The following lemma simplifies the formula $\Psi$ in cases when the relation $\seleq$ holds and $\kod{Q}_2$ does not contain projections.

\begin{lemma}
	\label{thm:psi}
	Let $\kod{Q}_1$ and $\kod{Q}_2$ be queries, let $\overline{rv_1}$ denote variables from $\vars$ that correspond to the relevant variables of the query $\kod{Q}_1$, and let $\Phi_1(\overline{rv_1})$ correspond to the query $\kod{Q}_1$.
	Let $\kod{qpat}_2$ and $\ag_2$ be a query pattern and a query dataset of $\kod{Q}_2$, respectively.
	If $\kod{Q}_2$ does not contain projections and $\kod{Q}_1 \seleq \kod{Q}_2$ holds, then formula $\Psi$ is equal to:
	$$\forall \overline{rv_1} \ \big( \Phi_1(\overline{rv_1}) \Rightarrow \sigma^{\mathit{df}(\ag_2)}(\kod{qpat}_2)\big).$$
	\begin{proof}
		%\ref{proof:psi} is given in Appendix \ref{sec:appendix}. It uses Definitions \ref{def:function_over_sets}, \ref{def:phi}, \ref{def:set_equality} and \ref{def:psi} and Lemmas \ref{thm:query_domain} and \ref{thm:dom_mu_nu1}.
		
		By Definition \ref{def:psi}, $\Psi$ is defined as
		$$\forall \overline{rv_1} \ \big( \Phi_1(\overline{rv_1}) \Rightarrow \Phi_2(\overline{rv_1})),$$
		i.e.~by Definition \ref{def:phi}
		$$\forall \overline{rv_1} \ \big( \Phi_1(\overline{rv_1}) \Rightarrow \exists \overline{ov_2}  \ \sigma^{\mathit{df}({\ag_2})}(\kod{qpat}_2)\big),$$
		where $\overline{ov_2}$ is equal to $var(\sigma(\kod{qpat}_2)) \setminus \overline{rv_1}$.
		By Definition \ref{def:set_equality}, as $\kod{Q}_1 \seleq \kod{Q}_2$, it holds
		$\overline{\kod{rv}_1} = \overline{\kod{rv}_2}$, i.e.~by Lemma \ref{thm:query_domain},
		$$\overline{\kod{rv}_1} = \kod{var}(\kod{qpat}_2)  \cap \overline{\kod{dv}_2}.$$
		As $\kod{Q}_2$ does not contain projections, it holds 
		$$\overline{\kod{rv}_1} = \kod{var}(\kod{qpat}_2),$$
		i.e.~
		$$\kod{var}(\kod{qpat}_2) \setminus \overline{\kod{rv}_1} = \varnothing.$$
		Therefore, by Definition \ref{def:function_over_sets}, it holds
		$$\sigma(\kod{var}(\kod{qpat}_2) \setminus \overline{\kod{rv}_1}) = \varnothing,$$
		i.e.~
		$$\sigma(\kod{var}(\kod{qpat}_2)) \setminus \sigma(\overline{\kod{rv}_1}) = \varnothing,$$
		i.e.~by Lemma \ref{thm:dom_mu_nu1},
		$$var(\sigma(\kod{qpat}_2)) \setminus \overline{rv_1} = \varnothing.$$
		Therefore, $\overline{ov_2}$ is equal to an empty set.
		Then, the formula $\Psi$ is reduced to
		$$\forall \overline{rv_1} \ \big( \Phi_1(\overline{rv_1}) \Rightarrow \sigma^{\mathit{df}(\ag_2)}(\kod{qpat}_2)\big).$$
	\end{proof}
\end{lemma}

We propose the following reductions of the containment problem and the subsumption problem.
\begin{procedure}[Logical formulation of the \sparql{} query containment problem]
	\label{thm:main}
	Let $\kod{Q}_1$ and $\kod{Q}_2$ be queries, $\Theta$ be a formula generated from $\kod{Q}_1$ (Def \ref{def:theta}) and $\Psi$ be a formula generated from $\kod{Q}_1$ and $\kod{Q}_2$ (Def \ref{def:psi}).
	The query containment problem, i.e.~if $\kod{Q}_1 \sqsubseteq \kod{Q}_2$ holds, is reduced to a satisfiability of a  disjunction of the following two conditions:
	\begin{itemize}%[noitemsep,topsep=0pt]
		\item[(1)] $\Theta$ is valid
		\item[(2)] $\kod{Q}_1 \seleq \kod{Q}_2$ holds and $\Psi$ is valid.
	\end{itemize}
\end{procedure}
\noindent Correctness of the procedure \ref{thm:main} is a direct consequence of proofs for theorems \ref{thm:soundness} (Section \ref{subsec:soundness}) and \ref{thm:completeness} (Section \ref{subsec:completeness}).

\begin{procedure}[Logical formulation of the \sparql{} query subsumption problem]
	\label{thm:main_subsumption}
	Let $\kod{Q}_1$ and $\kod{Q}_2$ be queries, $\Theta$ be a formula generated from $\kod{Q}_1$ (Def \ref{def:theta}) and $\Psi$ be a formula generated from $\kod{Q}_1$ and $\kod{Q}_2$ (Def \ref{def:psi}).
	The query subsumption problem, i.e.~if $\kod{Q}_1 \dot{\sqsubseteq} \kod{Q}_2$ holds, is reduced to a satisfiability of a disjunction of the following two conditions:
	\begin{itemize}%[noitemsep,topsep=0pt]
		\item[(1)] $\Theta$ is valid
		\item[(2)] $\kod{Q}_1 \dot{\seleq} \kod{Q}_2$ holds and $\Psi$ is valid.
	\end{itemize}
\end{procedure}

\noindent Correctness of the procedure \ref{thm:main_subsumption} is a direct consequence of proofs for theorems \ref{thm:soundness_subsumption} (Section \ref{subsec:soundness}) and \ref{thm:completeness_subsumption} (Section \ref{subsec:completeness}).

%--------------------------------------------------------------------
\section{Correctness of the Proposed Modeling}
\label{sec:correctness}

In this section, we prove the correctness of our modeling. We start with introducing necessary definitions and proving auxiliary lemmas (Section \ref{subsec:aux}) and continue by proving soundness (Section \ref{subsec:soundness}) and completeness (Section \ref{subsec:completeness}) of the proposed reduction. 

\subsection{Interpretations and Models}
\label{subsec:aux}

There is a standard definition of $\mathcal{L}$-structure over a signature $\mathcal{L}$ given in model theory textbooks \cite{Chang-Model-Theory,marker-model-theory}.
The following definition represents its concretization over the \sparql{} theory signature $\mathcal{L}$ corresponding to queries $\kod{Q}_1$ and $\kod{Q}_2$.

\begin{defn}[$\mathcal{L}$-structure $\mathfrak{D}$ corresponding to the queries $\kod{Q}_1$, $\kod{Q}_2$ and a dataset $\dataq$]
	\label{defn:l_structure}
	An $\mathcal{L}$-structure corresponding to a dataset $\dataq$ over the \sparql{} theory signature $\mathcal{L}$ corresponding to the queries $\kod{Q}_1$ and $\kod{Q}_2$, named $\mathfrak{D}$, is a pair $(\mathcal{D},\intrprt^{\mathcal{D}})$ where:
	\begin{itemize}[noitemsep,nolistsep,leftmargin=4mm]		
		\item $\mathcal{D}$ is a nonempty domain equal to $\ible{} \cup \siginv(\mathcal{C})$, where \kod{I}, \kod{B}, \kod{L} are sets of \iri{}s, blank nodes and literals respectively, appearing in the dataset $\dataq$,
		\item $\intrprt^{\mathcal{D}}$ is a function that maps the non-logical terms of the signature in the following way:		
		\begin{itemize}[noitemsep,nolistsep,leftmargin=2mm]			
			\item The interpretation $\intrprt^{\mathcal{D}}$ of constants from $\mathcal{C}$ are from $\ibl{}\kod{e}$, and are defined as $\intrprt^{\mathcal{D}}(c) \eqdef \siginv(c)$, for $c \in \mathcal{C}$. 		
			
			\item The interpretation $\intrprt^{\mathcal{D}}$ of the function symbol $datatype$ is a function $\kod{DATATYPE}: \ibl{} \rightarrow \kod{IBLe}$, which is defined in the following way:
			$$\kod{DATATYPE}(\kod{t}) \eqdef 
			\begin{cases}
				\kod{dt}(\kod{t}), & \kod{t} \in \kod{L}\\
				\error, & \text{otherwise}
			\end{cases}
			$$
			
			\item The interpretation $\intrprt^{\mathcal{D}}$ of the predicate symbol $isliteral$ is a function $\kod{ISLITERAL}: \ibl{} \rightarrow \bool$, which is defined in the following way:
			$$\kod{ISLITERAL}(\kod{t}) = \top \text{ if and only if } \kod{t} \in \kod{L}$$
			
			\item The interpretation $\intrprt^{\mathcal{D}}$ of the predicate symbol $\ppp$ is a function $\pppi: \kod{IB} \times \kod{I} \times \ibl{} \rightarrow \bool$, which is defined in the following way:
			$$\pppi(\kod{s}, \kod{p}, \kod{o}) = \top \text{ if and only if } (\kod{s}, \kod{p}, \kod{o}) \in \mathit{df}(\dataq)$$
			
			\item The interpretation $\intrprt^{\mathcal{D}}$ of the predicate symbol $\pp$ is a function $\ppi: \kod{IB} \times \kod{I} \times \ibl{} \times \kod{I} \rightarrow \bool$, which is defined in the following way:
			$$\ppi(\kod{s}, \kod{p}, \kod{o}, \kod{i}) = \top \text{ if and only if } (\kod{s}, \kod{p}, \kod{o}) \in gr_\kod{\,$\dataq$}(\kod{i})$$
		\end{itemize}	
	\end{itemize}
\end{defn}

Note that $\sigvibl$ is bijective by Definition \ref{def:sigmat}, therefore $\siginv$ is well defined.
As a direct consequence of this definition, if $\ppi(\kod{s}, \kod{p}, \kod{o}, \kod{i}) = \top$ then $\kod{i} \in names(\dataq)$. Otherwise, $gr_\kod{\,$\dataq$}(\kod{i})$ would be equal to $\kod{\gr}_{\varnothing}$ and $(\kod{s}, \kod{p}, \kod{o})$ does not belong to $\kod{\gr}_{\varnothing}$.
%\notin gr_\kod{\,$\dataq$}(\kod{i})$.
Note also that the domain $\mathcal{D}$ is at most countable.

A valuation of $\mathcal{L}$-formulas within $\mathcal{L}$-structure $\mathfrak{D}$, usually denoted by $\nu$, is a partial function from the set of variables $\vars$ to $\ibl{}$.

For an $\mathcal{L}$-structure $\mathfrak{M}$, valuation $\nu$, and formula $\Upsilon$ over the given \sparql{} theory signature $\mathcal{L}$, we use standard notation $(\mathfrak{M}, \nu) \models \Upsilon$ denoting that $\mathfrak{M}$ with valuation $\nu$ is a model of formula $\Upsilon$. If formula $\Upsilon$ is a sentence, i.e.~if it does not contain free variables, we use the common notation $\mathfrak{M} \models \Upsilon$.

\begin{defn}[Model $\intrprt_\nu$]
	\label{defn:model}
	Let $\Upsilon$ be a formula over the given \sparql{} theory signature $\mathcal{L}$, and let $(\mathfrak{D}, \nu)$ be a model of it, i.e.~$(\mathfrak{D}, \nu) \models \Upsilon$. For interpretation function of $(\mathfrak{D}, \nu)$ defined in the standard way for \fol{}, we use a notation $\intrprt_\nu$, i.e.~$\intrprt_\nu(\Upsilon)$ is true.
\end{defn}

\noindent If a formula $\Upsilon$ contains variables $\overline{x}$, and a valuation $\nu$ maps them to $\overline{c}$, then $(\mathfrak{D}, \nu) \models \Upsilon(\overline{c})$ denotes that $\intrprt_\nu(\Upsilon(\overline{x}))$ is true.

There are standard definitions of embeddings, isomorphisms, and substructures given in model theory textbooks \cite{Chang-Model-Theory,marker-model-theory}.
The following definition represents their concretization over the presented signature $\mathcal{L}$ corresponding to $\kod{Q}_1$ and $\kod{Q}_2$.

\begin{defn}[$\mathcal{L}$-embedding, $\mathcal{L}$-isomorphism, Relation $\cong$, $\mathcal{L}$-substructure]
	\label{def:isomorfic}
	Let $\mathcal{M}$ and $\mathcal{N}$ be nonempty domains.
	Let $\mathfrak{M} = (\mathcal{M}, \intrprt^{\mathcal{M}})$ and $\mathfrak{N} = (\mathcal{N}, \intrprt^{\mathcal{N}})$ be $\mathcal{L}$-structures.
	\begin{description}
		\item[$\mathcal{L}$-embedding] An \emph{$\mathcal{L}$-embedding} $\theta: \mathfrak{M} \rightarrow \mathfrak{N}$ is a function $\theta: \mathcal{M} \rightarrow \mathcal{N}$ with the following properties:
		\begin{itemize}[noitemsep,nolistsep,leftmargin=2mm]
			\item[(i)] $\theta$ is injective,
			
			\item[(ii)] for relation symbols $\ppp$ and $\pp$ in signature $\mathcal{L}$, and all $s, p, o, i \in \mathcal{M}$
			\begin{center}
				$(s, p, o) \in \intrprt^{\mathcal{M}}(\ppp)$ if and only if $(\theta(s), \theta(p), \theta(o)) \in \intrprt^{\mathcal{N}}(\ppp)$, and\\
				$(s, p, o, i) \in \intrprt^{\mathcal{M}}(\pp)$ if and only if $(\theta(s), \theta(p), \theta(o), \theta(i)) \in \intrprt^{\mathcal{N}}(\pp)$,
			\end{center}
			
			\item[(iii)] for any constant symbol $c$ in the signature $\mathcal{L}$
			\begin{center}
				$\theta(\intrprt^{\mathcal{M}}(c)) = \intrprt^{\mathcal{N}}(c)$.
			\end{center}
		\end{itemize}
		
		\item[$\mathcal{L}$-isomorphism] An \emph{$\mathcal{L}$-isomorphism} from $\mathfrak{M}$ to $\mathfrak{N}$ is a bijective $\mathcal{L}$-embedding from $\mathfrak{M}$ to $\mathfrak{N}$.
		
		\item[Relation $\cong$] $\mathfrak{M}$ and $\mathfrak{N}$ are \emph{isomorphic}, denoted by $\mathfrak{M} \cong \mathfrak{N}$, if there is an $\mathcal{L}$-isomorphism $\theta: \mathfrak{M} \rightarrow \mathfrak{N}$.
		
		\item[$\mathcal{L}$-substructure] $\mathfrak{M}$ is an \emph{$\mathcal{L}$-substructure} of $\mathfrak{N}$ if $\mathcal{M} \subseteq \mathcal{N}$ and the inclusion map $\iota : \mathcal{M} \rightarrow \mathcal{N}$, such that $\iota(a) = a$ for all $a \in \mathcal{M}$, is an $\mathcal{L}$-embedding.
	\end{description}
\end{defn}

Note that the definition of $\mathcal{L}$-embedding should provide the congruence of all relation symbols from $\preds$.
The property (ii) should hold not only for $\ppp$ and $\pp$, but also for the equality symbol, but it can be proven from property (i), i.e.~from the injectiveness of $\theta$:
\begin{itemize}[noitemsep,nolistsep,leftmargin=4mm]
	\item[-] for all $m_1, m_2 \in \mathcal{M}$, if $m_1 = m_2$ then $\theta(m_1) = \theta(m_2)$, as $\theta$ is a function;
	\item[-] the converse holds because $\theta$ is injective.
\end{itemize}
From the other side, if we include the property (ii) for the equality symbol, the property (i) is not necessary, as it is a corollary of property (ii) for the equality symbol.

\begin{lemma}
	\label{thm:model_model} 
	Let $\Upsilon$ be a sentence over a given signature $\mathcal{L}$,
	and let $\mathfrak{D'}$ and $\mathfrak{D''}$ be $\mathcal{L}$-structures such that $\mathfrak{D'} \cong \mathfrak{D''}$.
	Then:
	\begin{center}
		$\mathfrak{D'} \models \Upsilon$ ~~~if and only if~~~ $\mathfrak{D''} \models \Upsilon$
	\end{center}
	\begin{proof}
		A proof of this lemma can be found in \cite{marker-model-theory}.
	\end{proof}
\end{lemma}

\begin{lemma}
	\label{thm:model}
	Let $\mathfrak{D'} = (\mathcal{D'}, \intrprt^{\mathcal{D'}})$ be an $\mathcal{L}$-structure over the \sparql{} theory signature $\mathcal{L}$ that corresponds to $\kod{Q}_1$ and $\kod{Q}_2$.
	Let $\Upsilon$ be a sentence over a given signature $\mathcal{L}$, such that
	$\mathfrak{D'} \models \Upsilon$.
	Then, there exists a dataset $\dataq$ and the corresponding $\mathcal{L}$-structure $\mathfrak{D} = (\mathcal{D}, \intrprt^{\mathcal{D}})$ over the same signature, such that
	$\mathfrak{D} \models \Upsilon$.
	\begin{proof} 
		%\ref{proof:model} is given in Appendix \ref{sec:appendix}. It uses downward L\"{o}wenheim-Skolem Theorem and it constructs such dataset.
		
		According to Definition \ref{def:isomorfic}, $\mathfrak{D'}$ has a nonempty domain $\mathcal{D'}$, possibly infinite.
		By the downward L\"{o}wenheim-Skolem Theorem, there exists an $\mathcal{L}$-structure $$\mathfrak{D''} = (\mathcal{D''}, \intrprt^{\mathcal{D''}})$$ over the same signature, such that
		its domain $\mathcal{D''}$ is countable and $\mathfrak{D}'' \models \Upsilon$.
		
		Let us construct an \rdfs{} dataset $\dataq$ and the corresponding $\mathcal{L}$-structure $\mathfrak{D} = (\mathcal{D}, \intrprt^{\mathcal{D}})$, such that it holds $\mathfrak{D''} \cong \mathfrak{D}$.
		First, we define a function $\theta: \mathcal{D''} \rightarrow \ible{}$ as follows:
		\begin{enumerate}[noitemsep,nolistsep]
			
			\item[$\circ$] for all $c'' \in \mathcal{D''}$ such that $c'' = \intrprt^{\mathcal{D''}}(c)$ for some constant from $\mathcal{C}$, $\theta(c'')$ is equal to $\siginv(c)$. Note that the restriction of $\theta$ to the set of such $c''$ is injective, due to Definition \ref{def:sigmat}.
			
			\item[$\circ$] for all other $x$ from $\mathcal{D''}$, $\theta(x)$ is equal to an arbitrary element from $\ibl{}$, but taking into account that function $\theta$ should stay injective.
			This is possible, as $\ibl{}$ and $\mathcal{D''}$ are countable sets.
		\end{enumerate}
		Then, we construct an \rdfs{} dataset $\dataq$ as follows:
		\begin{itemize}[noitemsep,nolistsep]
			\item[-] for each constant $i$ from $\mathcal{D''}$ we define graph $\kod{G}_i$, and function $gr_\kod{\,$\dataq$}$:
			\begin{align*}
				\kod{G}_i &:= \{ \ \big( \,\theta(s), \theta(p), \theta(o)\,\big) \ | \ s, p, o \in \mathcal{D''} \\
				&\hspace*{10mm} \text{ and } \intrprt^\mathcal{D''}(\pp)(s, p, o, i) \text{ is true}\}\\
				gr_\kod{\,$\dataq$}(\theta(i)) &:= \kod{G}_i
			\end{align*}
			\item[-] we define default graph $\dg$:
			\begin{align*}
				\dg &:= \{ \ \big( \,\theta(s), \theta(p), \theta(o)\,\big) \ | \ s, p, o \in \mathcal{D''} \\
				& \hspace*{10mm} \text{ and } \intrprt^\mathcal{D''}(\ppp)(s, p, o) \text{ is true}\}
			\end{align*}
		\end{itemize}
		Then, dataset $\dataq$ contains default graph $\dg$ and named graphs $\kod{G}_i$ with the corresponding names $\theta(i)$.
		Note that according to the modeling of queries where formula contains predicates $\ppp$ and/or $\pp$ and the existence of its model in $\mathfrak{D''}$, these graphs are not empty.\\
		The function $\theta$ is an $\mathcal{L}$-embedding, according to Definition \ref{def:isomorfic}, because:
		\begin{enumerate}
			\item[(i)] $\theta$ is injective, by its construction;
			\item[(ii)] for all $s, p, o, i \in \mathcal{D''}$, it holds
			\begin{align*}
				& \hspace*{-3mm} \intrprt^\mathcal{D''}(\ppp)(s, p, o) \hspace*{-4mm} & \text{ iff } \\
				& \text{(by constr.)} & \text{ iff } &( \,\theta(s), \theta(p), \theta(o)\,) \hspace*{-1mm} \in \hspace*{-1mm} \dg \\
				& \text{(by Def \ref{defn:l_structure})}& \text{ iff } &\pppi(\theta(s), \theta(p), \theta(o))  \\
				& \text{(by Def \ref{defn:l_structure})}& \text{ iff }  &\intrprt^\mathcal{D}(\ppp)(\theta(s), \theta(p), \theta(o)), \\
				&\text{and}\\
				& \hspace*{-3mm} \intrprt^\mathcal{D''}(\pp)(s, p, o, i) \hspace*{-4mm} & \text{ iff } \\
				& \text{(by constr.)} & \text{ iff } &( \,\theta(s), \theta(p), \theta(o)\,) \hspace*{-1mm} \in \hspace*{-1mm} gr_\kod{\,$\dataq$}(\theta(i))\\
				& \text{(by Def \ref{defn:l_structure})} & \text{ iff } &\ppi(\theta(s), \theta(p), \theta(o), \theta(i)) \\
				& \text{(by Def \ref{defn:l_structure})} & \text{ iff } &\intrprt^\mathcal{D}(\pp)(\theta(s), \theta(p), \theta(o), \theta(i));
			\end{align*}
			\item[(iii)] for each constant symbol $c$ of the signature $\mathcal{L}$, $\intrprt^{\mathcal{D''}}\hspace*{-2pt}(c) \hspace*{-2pt} \in \hspace*{-2pt} \mathcal{D''}$, and $\theta(\intrprt^{\mathcal{D''}}(c)) = \intrprt^\mathcal{D}(c)$ by construction.
		\end{enumerate}
		Function $\theta$ is surjective, because for each $\kod{x} \in \mathcal{D}$, there is an element $x \in \mathcal{D''}$ such that $\theta(x) = \kod{x}$:
		\begin{itemize}[noitemsep,nolistsep]
			\item[-] if $\kod{x}$ is the interpretation of a constant from $\mathcal{C}$, according to Definition \ref{defn:l_structure}, construction of $\theta$ and the fact that $\mathfrak{D''}$ and $\mathfrak{D}$ are corresponding $\mathcal{L}$-structures, there is an element $x$ such that $\theta(x) = \kod{x}$;
			\item[-] otherwise, each other element from $\mathcal{D}$ is an image of an element from $\mathcal{D''}$ by the construction of $\theta$ and the dataset $\dataq$.
		\end{itemize}
		Therefore, $\mathcal{L}$-embedding $\theta$ is bijective, and by Definition \ref{def:isomorfic}, it holds $\mathfrak{D''} \cong \mathfrak{D}$.
		Then, by Lemma \ref{thm:model_model}, it holds
		$$\mathfrak{D} \models \Upsilon.$$
	\end{proof}
\end{lemma}

\begin{defn}[Relations $\definisemini, \definisenimi, \povezano$]
	\label{def:povezano}
	Let $\mu {: \kod{VB} \rightarrow \ibl{}}$ be a mapping and $\nu {: \vars \rightarrow \ibl{}}$ be a valuation.
	\begin{itemize}[noitemsep,nolistsep,leftmargin=4mm]
		\item[-]	Mapping $\mu$ \emph{defines} a valuation $\nu$, denoted $\mu \definisemini \nu$, ~if $\sigvibl(dom(\mu)) \subseteq dom(\nu)$ and for each $\kod{x} \in dom(\mu)$ it holds
		$\nu (\sigvibl(\kod{x})) = \mu (\kod{x}).$
		\item[-]	Mapping $\mu$ \emph{is defined by} a valuation $\nu$, denoted $\mu \definisenimi \nu$, if $dom(\mu) \supseteq \sigvibl^{-1}(dom(\nu))$ and for each  ${x} \in dom(\nu)$ it holds
		$\mu(\sigvibl^{-1}(x)) = \nu(x).$
		\item[-]	Mapping $\mu$ \emph{corresponds to} valuation $\nu$, denoted $\mu \povezano \nu$, if $\mu \definisemini \nu$ and $\mu \definisenimi \nu$.
	\end{itemize}
\end{defn}
\noindent
These relations are well defined as $\sigvibl$ is bijective.
Note that if $\mu \povezano \nu$ holds, $dom(\mu) = \sigvibl^{-1}(dom(\nu))$ and $dom(\nu) = \sigvibl(dom(\mu))$.
%Definition \ref{def:povezano} provides the possibility to construct a valuation $\nu$ based on a mapping $\mu$ such that $\mu \povezano \nu$, as its domain and definition are explicitly given depending on the mapping $\mu$. Note that it is also possible to construct a mapping $\mu$ based on a valuation $\nu$, such that $\mu \povezano \nu$, i.e.~if domain and definition of $\nu$ are given, we can construct the corresponding mapping $\mu$. In such case, the domain $dom(\mu)$ is $\{ \sigvibl^{-1}(x) \ | \ x \in dom(\nu) \}$, while $\mu$ is defined as $\mu(\sigvibl^{-1}(x)) = \nu(x)$. This holds as a direct consequence of the fact that the restriction $\sigvibl$ of $\sigma$ on $\vibl{}$ is bijective, thus the inverse exists. 

Similarly to the notation $\mu_{\kod{x} \rightarrow \kod{c}}$, we introduce $\nu_{x \rightarrow \kod{c}}$ denoting a valuation such that $dom(\nu_{x \rightarrow \kod{c}}) \eqdef \{x\}$ and $\nu_{x \rightarrow \kod{c}}(x) \eqdef \kod{c}$.
Note that, by Definition \ref{def:povezano}, $\mu_{\kod{x} \rightarrow \kod{c}} \povezano \nu_{x \rightarrow \kod{c}}$, as $dom(\nu) = \{x\} = \{\sigma(\kod{x})\} = \sigma(\{\kod{x}\}) = \sigma(dom(\mu))$, and $\nu(x) = \kod{c} = \mu(\kod{x})$.

Definition \ref{def:compatible_mappings} can be used in the context of valuations as well: $\nu_1$ and $\nu_2$ are compatible, denoted by $\nu_1 \comp \nu_2$, if for each ${x}$ such that $x \in dom(\nu_1) \cap dom(\nu_2)$ it holds that $\nu_1(x) = \nu_2(x)$. 

\begin{lemma}
	\label{thm:mu12_nu12_compatibility}
	Let ${\mu}_{1}, \mu_{2} : \kod{VB} \rightarrow \ibl{}$ be mappings, and let ${\nu}_{1}, {\nu}_{2} {: \vars \rightarrow \ibl{}}$ be valuations such that ${\mu}_{1} \povezano {\nu}_{1}$ and ${\mu}_{2} \povezano {\nu}_{2}$.
	Then:
	\begin{center}
		$\mu_1 \comp \mu_2$ \\
		if and only if \\
		$\nu_1 \comp \nu_2$.
	\end{center}
	\begin{proof}
		%\ref{proof:mu12_nu12_compatibility} is given in Appendix \ref{sec:appendix}. It uses Definitions \ref{def:compatible_mappings} and \ref{def:povezano}.
		
		($\Rightarrow$)
		If $dom(\nu_1) \cap dom(\nu_2)$ is an empty set, by Definition \ref{def:compatible_mappings}, it holds $\nu_1 \comp \nu_2$.
		Otherwise, let $x$ be an element of $dom(\nu_1) \cap dom(\nu_2)$, i.e.~$x \in dom(\nu_1)$ and $x \in dom(\nu_2)$.
		From $\mu_1 \povezano \nu_1$ and $\mu_2 \povezano \nu_2$, by Definition \ref{def:povezano}, it holds 
		$x \in \{ \sigvibl(\kod{x}) \ | \ \kod{x} \in dom(\mu_1) \}$ and $x \in \{ \sigvibl(\kod{x}) \ | \ \kod{x} \in dom(\mu_2) \}$.
		Therefore, since $\sigvibl$ is a bijective function, there exists a unique $\kod{x}$, such that 
		$$x = \sigvibl(\kod{x}), \ \kod{x} \in dom(\mu_1) \text{ and } \kod{x} \in dom(\mu_2)$$ i.e.~$\kod{x} \in dom(\mu_1) \cap dom(\mu_2)$. 
		As it holds $\mu_1 \comp \mu_2$ (Definition \ref{def:compatible_mappings}), it holds 
		$$\mu_1(\kod{x})=\mu_2(\kod{x})$$ 
		and by the assumptions $\mu_1 \povezano \nu_1$ and $\mu_2 \povezano \nu_2$, it holds 
		$$\nu_1(\sigvibl(\kod{x}))=\mu_1(\kod{x}) \text{ and } \mu_2(\kod{x})=\nu_2(\sigvibl(\kod{x})).$$ 
		Therefore, it holds 
		$$\nu_1(\sigvibl(\kod{x}))=\nu_2(\sigvibl(\kod{x})), \text{ i.e. } \nu_1(x)=\nu_2(x).$$ 
		By Definition \ref{def:compatible_mappings}, it holds $\nu_1 \comp \nu_2$.
		
		($\Leftarrow$)
		If $dom(\mu_1) \cap dom(\mu_2)$ is an empty set, by Definition \ref{def:compatible_mappings}, it holds $\mu_1 \comp \mu_2$.
		Otherwise, let $\kod{x}$ be an element of $dom(\mu_1) \cap dom(\mu_2)$, i.e.~$\kod{x} \in dom(\mu_1)$ and $x \in dom(\mu_2)$.
		From $\mu_1 \povezano \nu_1$ and $\mu_2 \povezano \nu_2$, by Definition \ref{def:povezano}, it holds 
		$\kod{x} \in \{ \siginv(x) \ | \ x \in dom(\nu_1) \}$ and $\kod{x} \in \{ \siginv(x) \ | \ x \in dom(\nu_2) \}$. 
		Therefore, since $\siginv$ is a bijective function, there exists a unique $x$, such that 
		$$\kod{x} = \siginv(x), \ x \in dom(\nu_1) \text{ and } x \in dom(\nu_2)$$ i.e.~$x \in dom(\nu_1) \cap dom(\nu_2)$. 
		As it holds $\nu_1 \comp \nu_2$ (Definition \ref{def:compatible_mappings}), it holds 
		$$\nu_1(x)=\nu_2(x)$$ 
		and by the assumptions $\mu_1 \povezano \nu_1$ and $\mu_2 \povezano \nu_2$, it holds 
		$$\mu_1(\siginv(x))=\nu_1(x) \text{ and } \nu_2(x)=\mu_2(\siginv(x)).$$ 
		Therefore, it holds 
		$$\mu_1(\siginv(x))=\mu_2(\siginv(x)), \text{ i.e. } \mu_1(\kod{x})=\mu_2(\kod{x}).$$ 
		By Definition \ref{def:compatible_mappings}, it holds $\mu_1 \comp \mu_2$.
	\end{proof}
\end{lemma}

\begin{lemma}
	\label{thm:mu_1_2_nu_1_2}
	Let $\mu, \mu_1, \mu_2 : \kod{VB} \rightarrow \ibl{}$ be mappings, and $\nu, \nu_1, \nu_2 : \vars \rightarrow \ibl{}$ be valuations such that $\mu \povezano \nu$, $\mu_1 \povezano \nu_1$ and $\mu_2 \povezano \nu_2$.
	Then:
	\begin{center}
		$\mu_1 \comp \mu_2$ ~and~ $\mu = \mu_1 \cup \mu_2$\\
		if and only if\\
		$\nu_1 \comp \nu_2$ ~and~ $\nu = \nu_1 \cup \nu_2$.
	\end{center}
	\begin{proof}
		%\ref{proof:mu_1_2_nu_1_2} is given in Appendix \ref{sec:appendix}. It uses Definitions \ref{def:compatible_mappings}, \ref{def:function_over_sets} and \ref{def:povezano} and Lemma \ref{thm:mu12_nu12_compatibility}.
		
		($\Rightarrow$)
		%$\mu = \mu_1 \cup \mu_2$ implies that $\mu_1$ is compatible with $\mu_2$.
		From $\mu_1 \povezano \nu_1$, $\mu_2 \povezano \nu_2$ and  $\mu_1 \comp \mu_2$, by Lema \ref{thm:mu12_nu12_compatibility} it holds that 
		$$	\nu_1 \comp \nu_2.$$  
		Therefore, $\nu_1 \cup \nu_2$ is a well defined valuation.
		
		From $\mu \povezano \nu$, $\mu_1 \povezano \nu_1$ and $\mu \comp \mu_1$ (as $\mu$ is an extension of $\mu_1$), by Lemma \ref{thm:mu12_nu12_compatibility} it holds that $$\nu \comp \nu_1.$$
		
		Similarly, from $\mu \povezano \nu$, $\mu_2 \povezano \nu_1$, and $\mu \comp \mu_2$ (as $\mu$ is an extension of $\mu_2$),  by Lemma \ref{thm:mu12_nu12_compatibility}, it holds that $$\nu \comp \nu_2.$$	
		Also, 
		\begin{align*}
			&\hspace*{20mm} dom(\nu) \hspace*{-20mm} & = \;\\
			&\text{(by Def \ref{def:povezano} as } \mu \povezano \nu) \hspace*{-3mm} & = \; & \sigvibl(dom(\mu)) \\
			&\text{(as } \mu = \mu_1 \cup \mu_2 \text{)} & = \; & \sigvibl(dom(\mu_1) \cup dom(\mu_2)) \\
			& \text{(by Def \ref{def:function_over_sets})} & = \; &  \sigvibl(dom(\mu_1)) \cup \sigvibl(dom(\mu_2)) \\
			& \text{(by Def \ref{def:povezano}, as } \mu_1 \povezano \nu_1 \text{ and } \mu_2 \povezano \nu_2) \hspace*{-30mm}\\
			& & = \; & dom(\nu_1) \cup dom(\nu_2)\\
			&\text{ (} \nu_1 \text{ and } \nu_2 \text{ are comp.)} \hspace*{-8mm} & = \; & dom(\nu_1 \cup \nu_2)
		\end{align*}
		As $dom(\nu) = dom(\nu_1) \cup dom(\nu_2)$, for $x \in dom(\nu)$, it holds $$x \in dom(\nu_1) \text{ or } x \in dom(\nu_2).$$ 
		From $\nu \comp \nu_1$ and $\nu \comp \nu_2$, by Definition \ref{def:compatible_mappings}, it holds $$\nu(x) = \nu_1(x) \text{ or } \nu(x) = \nu_2(x).$$
		Therefore, as $\nu_1 \comp \nu_2$,
		$$x \in dom(\nu) \text{ implies } \nu(x) = (\nu_1 \cup \nu_2)(x) \text{.}$$
		Finally, from $dom(\nu) = dom(\nu_1 \cup \nu_2)$ and the last equation, it holds 
		$$\nu = \nu_1 \cup \nu_2.$$
		
		($\Leftarrow$)
		%$\nu = \nu_1 \cup \nu_2$ implies that $\nu_1$ is compatible with $\nu_2$.
		From $\mu_1 \povezano \nu_1$, $\mu_2 \povezano \nu_2$ and $\nu_1 \comp \nu_2$, by Lema \ref{thm:mu12_nu12_compatibility} it holds that 
		$$\mu_1 \comp \mu_2.$$  
		Therefore, $\mu_1 \cup \mu_2$ is a well defined mapping.
		
		From $\mu \povezano \nu$, $\mu_1 \povezano \nu_1$ and $\nu \comp \nu_1$ (as $\nu$ is an extension of $\nu_1$), by Lemma \ref{thm:mu12_nu12_compatibility} it holds that $$\mu \comp \mu_1.$$
		
		Similarly, from $\mu \povezano \nu$, $\mu_2 \povezano \nu_1$, and $\nu \comp \nu_2$ (as $\nu$ is an extension of $\nu_2$),  by Lemma \ref{thm:mu12_nu12_compatibility}, it holds that $$\mu \comp \mu_2.$$	
		Also, 
		\begin{align*}
			&\hspace*{20mm} dom(\mu) \hspace*{-20mm} & = \;\\
			&\text{(by Def \ref{def:povezano} as } \mu \povezano \nu) \hspace*{-4mm} &=\;& \siginv(dom(\nu)) \\
			& \text{ (as } \nu = \nu_1 \cup \nu_2 \text{)}&=\;&  \siginv(dom(\nu_1) \cup dom(\nu_2))\\
			& \text{(by Def \ref{def:function_over_sets})} &=\;&  \siginv(dom(\nu_1)) \ \cup\\
			&&& \siginv(dom(\nu_2)) & \\
			& \text{(by Def \ref{def:povezano}, as }\mu_1 \povezano \nu_1 \text{ and } \mu_2 \povezano \nu_2 \text{)} \hspace*{-40mm}\\
			&&=\;& dom(\mu_1) \cup dom(\mu_2)\\
			&\text{ (} \mu_1 \text{ and } \mu_2 \text{ are comp.)} \hspace*{-3mm} &=\;& dom(\mu_1 \cup \mu_2)
		\end{align*}	
		As $dom(\mu) = dom(\mu_1) \cup dom(\mu_2)$, for  $\kod{x} \in dom(\mu)$, it holds 
		$$\kod{x} \in dom(\mu_1) \text{ or } \kod{x} \in dom(\mu_2).$$
		From $\mu \comp \mu_1$ and $\mu \comp \mu_2$, by Definition \ref{def:compatible_mappings}, it holds $$\mu(\kod{x}) = \mu_1(\kod{x}) \text{ or } \mu(\kod{x}) = \mu_2(x).$$
		Therefore, as $\mu_1 \comp  \mu_2$,
		$$\kod{x} \in dom(\mu) \text{ implies } \mu(\kod{x}) = (\mu_1 \cup \mu_2)(\kod{x}) \text{.}$$
		Finally, from $dom(\mu) = dom(\mu_1 \cup \mu_2)$ and the last equation, it holds 
		$$\mu = \mu_1 \cup \mu_2.$$
	\end{proof}
\end{lemma}

%The following definition extends the standard definition of a term value in the $\mathcal{L}$-structure $\mathfrak{D}$ according to the valuation $\nu$ with the $\error$ value in codomain.

\begin{defn}[Notation $\valuation{\cdot}{\nu}$]
	\label{defn:expr_in_nu}
	Let $\nu$ be a valuation, and $t$ a term over \sparql{} theory signature $\mathcal{L}$.
	A value of the term $t$, according to the valuation $\nu$, in notation $\valuation{t}{\nu}$ is a value from $\ible$, defined in the following way:
	\begin{align*}
		\valuation{t}{\nu} & \eqdef \ 
		\begin{cases}
			\error,& t \in \vars \text{ and } t \notin dom(\nu)\\
			\error,& t \text{ is } datatype(t_1) \text{ and } \valuation{t_1}{\nu}=\error\\
			\error,& t \text{ is } isliteral(t_1) \text{ and } \valuation{t_1}{\nu}=\error\\
			\intrprt_\nu(t) ,& \text{otherwise}.\\
		\end{cases}
	\end{align*}
\end{defn}

\begin{lemma}
	\label{thm:expr_mu_nu}
	Let $\kod{E}$ be an expression.
	Let $\mu : \kod{VB} \rightarrow \ibl{}$ be a mapping and $\nu {: \vars \rightarrow \ibl{}}$ be a valuation such that $\mu \povezano \nu$.
	Then:
	\begin{center}
		$\valuation{\kod{E}}{\mu} = \valuation{\sigma(\kod{E})}{\nu}$.
	\end{center}
	\begin{proof}
		%\ref{proof:expr_mu_nu} is given in Appendix \ref{sec:appendix}. It is done by induction on the expression $\kod{E}$.
		
		The lemma is proved by induction on an expression $\kod{E}$.
		\begin{description}
			
			\item \kod{E}  is  \kod{c}, $\kod{c} \in \kod{IL}$		
			\begin{align*}
				&\hspace*{22mm} \valuation{\kod{c}}{\mu} \hspace*{-20mm} & = \;\\
				& \text{(by Def \ref{defn:expr_in_mu})} & = \; & \kod{c} \\
				& \text{(by Def \ref{def:sigmat})} & = \; & \siginv(\sigvibl(\kod{c})) \\
				& \text{(by Def \ref{defn:l_structure})} & = \; & \intrprt^{\mathcal{D}}(\sigvibl(\kod{c})) \\
				& \text{(by Def \ref{def:sigma})} & = \; &\intrprt^{\mathcal{D}}(\sigma(\kod{c})) \\
				& \text{(by Def \ref{defn:expr_in_nu})} & = \; &\valuation{\sigma(\kod{c})}{\nu} \\
			\end{align*}
			
			\item \kod{E}  is  $\kod{v}$, $\kod{v} \in \kod{V}$
			\begin{itemize}
				\item $\kod{v} \in dom(\mu)$\\
				As $\mu \povezano \nu$ and $\kod{v} \in dom(\mu)$, by Definition \ref{def:povezano}, it holds $\sigma(\kod{v}) \in dom(\nu)$.
				\begin{align*}
					&\hspace*{37mm} \valuation{\kod{v}}{\mu} \hspace*{-20mm} & = \;\\
					& \text{(by Def \ref{defn:expr_in_mu}, as } \kod{v} \in dom(\mu) \text{)} & =\;  &\mu(\kod{v}) \\
					& \text{(by Def \ref{def:povezano} as $\mu \povezano \nu$)} & =\;  &\nu(\sigvibl(\kod{v}))\\ &\text{(by Def \ref{def:sigma})} & =\;  &\nu(\sigma(\kod{v})) \\
					& \text{(by Def \ref{defn:model})}  & =\;  &\intrprt_\nu(\sigma(\kod{v}))\\
					& \text{(by Def \ref{defn:expr_in_nu}, as } \sigma(\kod{v}) \hspace*{-1mm} \in \hspace*{-1mm} dom(\nu) \text{)}  \hspace*{-3mm} & =\;  &\valuation{\sigma(\kod{v})}{\nu}
				\end{align*}
				
				\item $\kod{v} \notin dom(\mu)$\\
				As $\mu \povezano \nu$ and $\kod{v} \notin dom(\mu)$, by Definition \ref{def:povezano}, it holds $\sigma(\kod{v}) \notin dom(\nu)$.
				\begin{align*}
					& \hspace*{37mm} \valuation{\kod{v}}{\mu} \hspace*{-20mm} & = \;\\
					& \text{(by Def \ref{defn:expr_in_mu}, as } \kod{v} \notin dom(\mu) \text{)} & =\; & \error \\
					& \text{(by Def \ref{defn:expr_in_nu}, as } \sigma(\kod{v}) \hspace*{-1mm} \notin \hspace*{-1mm} dom(\nu) \text{)} \hspace*{-3mm} & =\; & \valuation{\sigma(\kod{v})}{\nu}
				\end{align*}
			\end{itemize}
			
			\item \kod{E}  is  $\kod{datatype(E}_1\kod{)}$
			\begin{itemize}
				\item $\valuation{\kod{E}_1}{\mu} = \error$\\
				By induction hypothesis, it holds $\valuation{\kod{E}_1}{\mu} = \error$ iff $\valuation{\sigma(\kod{E}_1)}{\nu} = \error$.
				\begin{align*}
					& \hspace*{-6mm} \valuation{\kod{datatype(E}_1\kod{)}}{\mu} \hspace*{-3mm} & = \\
					& \text{(by Def \ref{defn:expr_in_mu})} \hspace*{-3mm} & = &\error \\
					& \text{(by induc. hyp.)} \hspace*{-3mm}& = &\valuation{\sigma(\kod{E}_1)}{\nu} \\
					& \text{(by Def \ref{defn:expr_in_nu})} \hspace*{-3mm}& = &\valuation{datatype(\sigma(\kod{E}_1))}{\nu} \\
					& \text{(by Def \ref{def:sigma})} \hspace*{-3mm}& = &\valuation{\sigma(\kod{datatype(E}_1\kod{)})}{\nu} 
				\end{align*}
				
				\item $\valuation{\kod{E}_1}{\mu} \not= \error$ and $\valuation{\kod{E}_1}{\mu} \notin \kod{L}$: \\
				By induc. hypothesis, it holds $\valuation{\sigma(\kod{E}_1)}{\nu} \hspace*{-1mm} \notin \hspace*{-1mm} \kod{L}$.
				\begin{align*}
					&\hspace*{-6mm} \valuation{\kod{datatype(E}_1\kod{)}}{\mu} \hspace*{-3mm} & = \\
					& \text{(by Def \ref{defn:expr_in_mu})} \hspace*{-3mm}& =  &\error\\
					& \text{(by Def \ref{defn:l_structure})} \hspace*{-3mm}& = &\kod{DATATYPE}(\valuation{\sigma(\kod{E}_1)}{\nu}) \\
					& \text{(by Def \ref{defn:l_structure})} \hspace*{-3mm}& = &\intrprt^{\mathcal{D}}(datatype)(\valuation{\sigma(\kod{E}_1)}{\nu}) \\
					& \text{(by Def \ref{defn:expr_in_nu})} \hspace*{-3mm}& = &\valuation{datatype(\sigma(\kod{E}_1))}{\nu} \\
					& \text{(by Def \ref{def:sigma})} \hspace*{-3mm}& = &\valuation{\sigma(\kod{datatype(E}_1\kod{)})}{\nu}
				\end{align*}
				
				\item $\valuation{\kod{E}_1}{\mu} \not= \error$ and $\valuation{\kod{E}_1}{\mu} \in \kod{L}$: \\
				By induc. hypothesis, it holds $\valuation{\kod{E}_1}{\mu} \not= \error$ iff $\valuation{\sigma(\kod{E}_1)}{\nu} \not= \error$.
				\begin{align*}
					&\hspace*{-9mm} \valuation{\kod{datatype(E}_1\kod{)}}{\mu} \hspace*{-3mm} & = \\
					&\hspace*{-3mm} \text{(by Def \ref{defn:expr_in_mu})} \hspace*{-3mm} & =  &\kod{dt}(\valuation{(\kod{E}_1)}{\mu})\\
					&\hspace*{-3mm} \text{(by induc. hyp.)} \hspace*{-3mm} & = &\kod{dt}(\valuation{\sigma(\kod{E}_1)}{\nu}) \\
					&\hspace*{-3mm} \text{(by Def \ref{defn:l_structure})} \hspace*{-3mm} & = &\kod{DATATYPE}(\valuation{\sigma(\kod{E}_1)}{\nu}) \\
					&\hspace*{-3mm} \text{(by Def \ref{defn:l_structure})} \hspace*{-3mm} & = &\intrprt^{\mathcal{D}}(datatype)(\valuation{\sigma(\kod{E}_1)}{\nu}) \\
					&\hspace*{-3mm} \text{(by Def \ref{defn:expr_in_nu})} \hspace*{-3mm} & = &\valuation{datatype(\sigma(\kod{E}_1))}{\nu} \\
					&\hspace*{-3mm} \text{(by Def \ref{def:sigma})} \hspace*{-3mm} & = &\valuation{\sigma(\kod{datatype(E}_1\kod{)})}{\nu}
				\end{align*}
			\end{itemize}	
		\end{description}
	\end{proof}	
\end{lemma}

\begin{lemma}
	\label{thm:thm78}
	Let $\kod{E}$ be an expression, and $\kod{v}$ a variable.
	Let $\mu : \kod{VB} \rightarrow \ibl{}$ be a mapping and $\nu {: \vars \rightarrow \ibl{}}$ be a valuation such that $\mu \povezano \nu$.
	Then:
	\begin{center}
		$\mu(\kod{v}) = \valuation{\kod{E}}{\mu}$ ~~~if and only if~~~ $\model{\nu} \models \sigma(\kod{v})=\sigma(\kod{E})$.
	\end{center}
	\begin{proof}
		%\ref{proof:thm78} is given in Appendix \ref{sec:appendix}. It uses Definitions \ref{defn:model}, \ref{def:povezano}, \ref{defn:expr_in_nu} and Lemma \ref{thm:expr_mu_nu}.
		
		\begin{align*}
			& \hspace*{16mm} \mu(\kod{v}) = \valuation{\kod{E}}{\mu} \hspace*{-10mm} & \text{ iff } \\
			& \text{(by Lemma \ref{thm:expr_mu_nu})} & \text{ iff } & \mu(\kod{v}) = \valuation{\sigma(\kod{E})}{\nu} \\
			& \text{(by Def \ref{defn:expr_in_nu})} & \text{ iff } & \mu(\kod{v}) = \intrprt_\nu(\sigma(\kod{E})) \\
			& \text{(by Def \ref{def:povezano} as $\mu \povezano \nu$)} & \text{ iff } & \nu(\sigma(\kod{v})) = \intrprt_\nu(\sigma(\kod{E})) \\
			& \text{(by Def \ref{defn:model})} & \text{ iff } & \model{\nu} \models \sigma(\kod{v})=\sigma(\kod{E})
		\end{align*}
	\end{proof}	
\end{lemma}

\begin{lemma}
	\label{thm:thm56}
	Let $\mu : \kod{VB} \rightarrow \ibl{}$ be a mapping and $\nu {: \vars \rightarrow \ibl{}}$ be a valuation such that $\mu \povezano \nu$.
	Then:
	\begin{center}
		$\mu \zadovoljava \kod{R}$ ~~~if and only if~~~ $\model{\nu} \models \sigma(\kod{R})$.
	\end{center}
	\begin{proof}
		%\ref{proof:thm56} is given in Appendix \ref{sec:appendix}. It is done by induction on the condition $\kod{R}$.
		
		The lemma is proved by induction on a condition $\kod{R}$. 
		\begin{description}
			
			\item \kod{R}  is  $\kod{E}_1$ = $\kod{E}_2$
			\begin{align*}
				& \hspace*{10mm} \mu \zadovoljava \kod{E}_1 = \kod{E}_2 \hspace*{-3mm} & \text{ iff } \\
				& \text{(by Def \ref{defn:definition_filter})} & \text{ iff } &\valuation{\kod{E}_1}{\mu} \not= \error \text{ and }\\
				&&& \valuation{\kod{E}_2}{\mu} \not= \error \text{ and } \\
				&&& \valuation{\kod{E}_1}{\mu} = \valuation{\kod{E}_2}{\mu} \\
				& \text{(by Lemma \ref{thm:expr_mu_nu})} &\text{ iff } &\valuation{\sigma(\kod{E}_1)}{\nu} \not= \error \text{ and } \\
				&&& \valuation{\sigma(\kod{E}_2)}{\nu} \not= \error \text{ and } \\
				&&& \valuation{\sigma(\kod{E}_1)}{\nu} = \valuation{\sigma(\kod{E}_2)}{\nu} \\
				& \text{(by Def \ref{defn:model} and \ref{defn:expr_in_nu})} \hspace*{-3mm} &\text{ iff } &\model{\nu} \models \sigma(\kod{E}_1) = \sigma(\kod{E}_2)  \\
				& \text{(by Def \ref{def:sigma})} & \text{ iff } & \model{\nu} \models \sigma(\kod{E}_1 = \kod{E}_2) & 
			\end{align*}
			
			\item \kod{R}  is  $\kod{!R}_1$
			\begin{align*}
				& \hspace*{17mm} \mu \zadovoljava \kod{!R}_1 \hspace*{-10mm} & \text{ iff } \\
				& \text{(by Def \ref{defn:definition_filter})} & \text{ iff } & \text{not } \mu \zadovoljava \kod{R}_1 \\
				& \text{(by induc. hyp.)} & \text{ iff } & \text{not } \model{\nu} \models \sigma(\kod{R}_1) \\
				& \text{(by Def \ref{defn:model})} & \text{ iff } & \model{\nu} \models \neg \sigma(\kod{R}_1)\\
				& \text{(by Def \ref{def:sigma})} & \text{ iff } & \model{\nu} \models \sigma(\kod{!R}_1) &
			\end{align*}
			
			\item \kod{R}  is  $\kod{R}_1 \kod{\&\&} \kod{R}_2$
			\begin{align*}
				& \hspace*{10mm} \mu \zadovoljava \kod{R}_1 \kod{\&\&} \kod{R}_2 \hspace*{-3mm} & \text{ iff } \\
				& \text{(by Def \ref{defn:definition_filter})} & \text{ iff } & \mu \zadovoljava \kod{R}_1 \text{ and } \mu \zadovoljava \kod{R}_2 \\
				& \text{(by induc. hyp.)} & \text{ iff } & \model{\nu} \models \sigma(\kod{R}_1) \text{ and } \\
				&&& \model{\nu} \models \sigma(\kod{R}_2) \\
				& \text{(by Def \ref{defn:model})} & \text{ iff } & \model{\nu} \models \sigma(\kod{R}_1) \wedge \sigma(\kod{R}_2) \\
				& \text{(by Def \ref{def:sigma})} & \text{ iff } & \model{\nu} \models \sigma(\kod{R}_1 \kod{\&\&} \kod{R}_2) &
			\end{align*}
			
			\item $\kod{R} \text{ is } \kod{R}_1 \kod{||} \kod{R}_2$\\
			Proof is analogous to the previous case.
			
			\item $\kod{R} \text{ is } \kod{(R}_1\kod{)}$\\
			Proof is analogous to the case $\kod{R} \text{ is } \kod{!R}_1$.
			
			\item \kod{R}  is  $\kod{isliteral(E)}$
			\begin{align*}
				& \hspace*{-5mm} \mu \zadovoljava \kod{isliteral(E)} \hspace*{-5mm} & \text{ iff } \\
				& \text{(by Def \ref{defn:definition_filter})} &\text{ iff }  &\valuation{\kod{E}}{\mu} \not= \error \text{ and }\\
				&&& \valuation{\kod{E}}{\mu} \in \kod{L}\\
				& \text{(by Lemma \ref{thm:expr_mu_nu})} &\text{ iff }  & \valuation{\sigma(\kod{E})}{\nu} \not= \error \text{ and }\\
				&&& \valuation{\sigma(\kod{E})}{\nu} \in \kod{L}\\
				& \text{(by Def \ref{defn:l_structure} } & \text{ iff }  & \kod{ISLITERAL}(\valuation{\sigma(\kod{E})}{\nu})\\
				& \text{(by Def \ref{defn:l_structure} } & \text{ iff } & 	\intrprt^{\mathcal{D}}(isliteral)(\valuation{\sigma(\kod{E})}{\nu}) \\
				& \text{(by Def \ref{defn:model})} & \text{ iff } & \model{\nu} \models isliteral(\sigma(\kod{E})) \\
				& \text{(by Def \ref{def:sigma})} & \text{ iff }  & \model{\nu} \models \sigma(\kod{isliteral(E)}) & 
			\end{align*}
		\end{description}
	\end{proof}	
\end{lemma}

\begin{lemma}
	\label{thm:thm34}
	Let $\ag$ be a query dataset, \kod{\gr} a graph within $\ag$, and \kod{\pat} a %\kod{union}-free 
	graph pattern.
	Let $\mu : \kod{VB} \rightarrow \ibl{}$ be a mapping and $\nu {: \vars \rightarrow \ibl{}}$ be a valuation such that $\mu \povezano \nu$.
	Then:
	\begin{center}
		$\mu \in \semdg{\kod{\pat}}{{\ag}}{\kod{\gr}}$\\
		if and only if\\
		$\model{\nu} \models \sigma^{\kod{\gr}}(\kod{\pat})$ ~and~ $dom(\nu) = var(\sigma^{\kod{\gr}}(\kod{\pat}))$.
	\end{center}
	\begin{proof}
		%\ref{proof:thm34} is given in Appendix \ref{sec:appendix}. It is done by induction over the graph pattern $\kod{\pat}$.
		
		By Lemma \ref{thm:pattern_domain}, from $\mu \in \semdg{\kod{\pat}}{{\ag}}{\kod{G}}$, it holds $dom(\mu) = \kod{var}(\kod{\pat})$.
		Therefore, by Definitions \ref{def:sigma} and \ref{def:povezano}, as $\mu \povezano \nu$, it holds $dom(\nu) = \sigma(dom(\mu)) = \sigma(\kod{var}(\kod{\pat}))$.
		Then, by Lemma \ref{thm:dom_mu_nu1}, it holds $dom(\nu) = var(\sigma(\kod{\pat}))$.
		The rest of specified equivalence is proved by induction over graph pattern $\kod{\pat}$.
		\begin{description}
			\item \kod{\pat{}} is \kod{\tpat} where \kod{\tpat{}} is \kod{s p o}\\
			($\Rightarrow$) 
			By Definition \ref{definition_sem}, 
			$$\mu \in \semdg{\kod{s p o}}{{\ag}}{\kod{G}} \text{ implies } \valuation{\kod{s p o}}{\mu} \in \kod{G},$$
			i.e. by Definition \ref{defn:expr_in_mu} $$\valuation{\kod{s}}{\mu} \kod{ } \valuation{\kod{p}}{\mu} \kod{ } \valuation{\kod{o}}{\mu} \in \kod{G}.$$
			By Definition \ref{defn:query_dataset}, $\kod{\gr}$ is equal to $\dg$ ($\dg = \mathit{df}(\dataq)$) or to an \rdfs{} merge of a nonempty set of named graphs, i.e.~$merge(\kod{\gr}_{k_1}, \ldots, \kod{\gr}_{k_m}), m > 0$, or a named graph $\kod{\gr}_{l_j}$, $j \in \{1, n\}$ within dataset $\ag$.
			Therefore, one option of the following three holds:
			\begin{equation}
				\label{thm34:eq1}
				\valuation{\kod{s}}{\mu} \kod{\,} \valuation{\kod{p}}{\mu} \kod{\,} \valuation{\kod{o}}{\mu} \hspace*{-2pt} \in \hspace*{-2pt} \dg
			\end{equation}
			\vspace*{-5mm}
			\begin{equation}
				\label{thm34:eq2}
				\valuation{\kod{s}}{\mu} \kod{\,} \valuation{\kod{p}}{\mu} \kod{\,} \valuation{\kod{o}}{\mu} \hspace*{-2pt} \in \hspace*{-2pt} \kod{\gr}_{k_j}\text{, for some }j \hspace*{-2pt} \in \hspace*{-2pt} \{1,...,m\} \hspace*{-1pt}
			\end{equation}
			\vspace*{-5mm}
			\begin{equation}
				\label{thm34:eq3}
				\valuation{\kod{s}}{\mu} \kod{\,} \valuation{\kod{p}}{\mu} \kod{\,} \valuation{\kod{o}}{\mu} \hspace*{-2pt} \in \hspace*{-2pt} \kod{\gr}_{l_j}\text{, for some }j \hspace*{-2pt} \in \hspace*{-2pt} \{1,..., n \}.
			\end{equation}
			From option (\ref{thm34:eq1}), by Definitions \ref{defn:l_structure}, it holds
			$$\pppi(\valuation{\kod{s}}{\mu}, \valuation{\kod{p}}{\mu}, \valuation{\kod{o}}{\mu}) = \top \text{. }$$
			Therefore, by Lemma \ref{thm:expr_mu_nu}, from $\mu \povezano \nu$, it holds
			$$\pppi(\valuation{\sigma(\kod{s})}{\nu}, \valuation{\sigma(\kod{p})}{\nu}, \valuation{\sigma(\kod{o})}{\nu}) = \top \text{, }$$
			i.e. by Definitions \ref{defn:l_structure}, \ref{defn:model} and \ref{defn:expr_in_nu}:
			$$\model{\nu} \models \ppp(\sigma(\kod{s}), \sigma(\kod{p}), \sigma(\kod{o})).$$
			From option (\ref{thm34:eq2}), by Definition \ref{def:context}, the set $\context(\kod{\gr})$ contains \iri{} constants from $\mathcal{C}$ corresponding to graphs that form $\kod{G}$. Therefore, there exists $i \in \context(\kod{\gr})$, such that $i$ corresponds to a graph that forms $\kod{G}$, i.e.~$i$ is equal to an $i_{k_j}$ from $\context(\kod{\gr})$, corresponding to the named graph $\kod{\gr}_{k_j}$, i.e.~$gr_\kod{\,D}(\kod{i}) = \kod{\gr}_{k_j}$.
			By Definition \ref{defn:l_structure}, from 
			$\valuation{\kod{s}}{\mu} \kod{ } \valuation{\kod{p}}{\mu} \kod{ } \valuation{\kod{o}}{\mu} \in \kod{\gr}_{k_j}$,
			it holds
			$$\ppi(\valuation{\kod{s}}{\mu}, \valuation{\kod{p}}{\mu}, \valuation{\kod{o}}{\mu}, i) = \top.$$
			From option (\ref{thm34:eq3}), by Definition \ref{def:context}, there is a constant $i = i_{l_j}$, corresponding to the named graph $\kod{\gr}_{l_j}$, i.e. $gr_\kod{\,D}(\kod{i}) = \kod{\gr}_{l_j}$. 
			Note that $i \in \context(\kod{\gr})$.
			By Definition \ref{defn:l_structure}, from  $\valuation{\kod{s}}{\mu} \kod{\,} \valuation{\kod{p}}{\mu} \kod{\,} \valuation{\kod{o}}{\mu} \in \kod{\gr}_{l_j}$,
			it holds (the same as in the previous option):
			$$\ppi(\valuation{\kod{s}}{\mu}, \valuation{\kod{p}}{\mu}, \valuation{\kod{o}}{\mu}, i) = \top \text{. }$$
			Therefore, from $\mu \povezano \nu$, by Lemma \ref{thm:expr_mu_nu}, it holds
			$$\ppi(\valuation{\sigma(\kod{s})}{\nu}, \valuation{\sigma(\kod{p})}{\nu}, \valuation{\sigma(\kod{o})}{\nu}, i) = \top.$$
			By Definitions \ref{defn:l_structure}, \ref{defn:model} and \ref{defn:expr_in_nu}, it holds
			$$\model{\nu} \models \pp(\sigma(\kod{s}), \sigma(\kod{p}), \sigma(\kod{o}), i),$$ i.e.
			$$\model{\nu} \models \underset{i_j \in \context(\kod{\gr})} \bigvee \pp(\sigma(\kod{s}), \sigma(\kod{p}), \sigma(\kod{o}), i_j).$$
			Putting all options together, by Definition \ref{def:sigma}, it holds
			$$\model{\nu} \models \sigma^\kod{\gr}(\kod{s p o}).$$
			
			($\Leftarrow$) By Lemma \ref{thm:dom_mu_nu1}, from $dom(\nu) = var(\sigma^{\kod{\gr}}(\kod{\tpat}))$, it holds $dom(\nu) = \sigma(\kod{var}(\kod{\tpat}))$.
			By Definitions \ref{def:povezano}, as $\mu \povezano \nu$, it holds $$dom(\mu) \hspace*{-3pt} = \hspace*{-3pt} \siginv(dom(\nu)) \hspace*{-2pt} = \hspace*{-2pt} \siginv(\sigma(\kod{var}(\kod{\tpat}))) \hspace*{-2pt} = \hspace*{-2pt} \kod{var}(\kod{\tpat})$$
			From $\model{\nu} \models \sigma^\kod{\gr}(\kod{s p o})$,
			by Definition \ref{def:sigma}, one option of the following two holds:
			\begin{multline}
				\label{thm34:eq4}
				\model{\nu} \models  \ppp(\sigma(\kod{s}), \sigma(\kod{p}), \sigma(\kod{o})),\\ \text{if } \kod{\gr} \notin dom(\context)
			\end{multline}
			\vspace*{-5mm}
			\begin{multline}
				\label{thm34:eq5}
				\model{\nu} \models \underset{i_j \in \context(\kod{\gr})} \bigvee \pp(\sigma(\kod{s}), \sigma(\kod{p}), \sigma(\kod{o}), i_j), \\ \text{if } \kod{\gr} \in dom(\context)
			\end{multline}
			From option (\ref{thm34:eq4}), by Definitions \ref{defn:l_structure}, \ref{defn:model} and \ref{defn:expr_in_nu}, it holds 
			$$\pppi(\valuation{\sigma(\kod{s})}{\nu}, \valuation{\sigma(\kod{p})}{\nu}, \valuation{\sigma(\kod{o})}{\nu}) = \top \text{. }$$
			Therefore, by Lemma \ref{thm:expr_mu_nu}, from $\mu \povezano \nu$, it holds
			$$\pppi(\valuation{\kod{s}}{\mu}, \valuation{\kod{p}}{\mu}, \valuation{\kod{o}}{\mu}) = \top \text{. }$$
			Then, by Definition \ref{defn:l_structure}, it holds 
			$$\valuation{\kod{s}}{\mu} \kod{\,} \valuation{\kod{p}}{\mu} \kod{\,} \valuation{\kod{o}}{\mu} \in \mathit{df}(\dataq),$$
			i.e.~as $\kod{\gr} \notin dom(\context)$,
			$$\valuation{\kod{s}}{\mu} \kod{\,} \valuation{\kod{p}}{\mu} \kod{\,} \valuation{\kod{o}}{\mu} \in \kod{\gr}.$$
			From option (\ref{thm34:eq5}), there exists $i_j \in \context(\kod{\gr})$, $gr_{\kod{\,D}}(\kod{i}_j)=\kod{\gr}_j$, such that $$\model{\nu} \models \pp(\sigma(\kod{s}), \sigma(\kod{p}), \sigma(\kod{o}), i_j).$$
			By Definitions \ref{defn:l_structure}, \ref{defn:model} and \ref{defn:expr_in_nu}, it holds:
			$$\ppi(\valuation{\sigma(\kod{s})}{\nu}, \valuation{\sigma(\kod{p})}{\nu}, \valuation{\sigma(\kod{o})}{\nu}, i_j) = \top.$$
			Therefore, by Lemma \ref{thm:expr_mu_nu}, from $\mu \povezano \nu$, it holds
			$$\ppi(\valuation{\kod{s}}{\mu}, \valuation{\kod{p}}{\mu}, \valuation{\kod{o}}{\mu}, i_j) = \top.$$
			By Definition \ref{defn:l_structure}, it holds
			$$\valuation{\kod{s}}{\mu} \kod{ } \valuation{\kod{p}}{\mu} \kod{ } \valuation{\kod{o}}{\mu} \in gr_{\kod{\,D}}(\siginv(i_j)),$$
			i.e.
			$$\valuation{\kod{s}}{\mu} \kod{ } \valuation{\kod{p}}{\mu} \kod{ } \valuation{\kod{o}}{\mu} \in \kod{\gr}_j.$$
			By Definition \ref{defn:expr_in_mu}, it holds $\valuation{\kod{s}  \kod{ p }  \kod{o}}{\mu} \in \kod{\gr}_j$ for some $\kod{\gr}_j$ that forms $\kod{G}$, i.e.~$\valuation{\kod{s}  \kod{ p }  \kod{o}}{\mu} \in \kod{\gr}$.
			Therefore, putting all options together, by Definition \ref{definition_sem}, as $dom(\mu) = \kod{var}(\kod{\tpat})$, it holds
			$$\mu \in \semdg{\kod{s p o}}{\ag}{\kod{G}}.$$		
			
			\item $\kod{\pat}$ is $\kod{\pat}_1 . \kod{\pat}_2$\\
			($\Rightarrow$) By Definition \ref{definition_sem}, 
			$$\mu \in \semdg{\kod{\pat}_1 . \kod{\pat}_2}{\ag}{\kod{G}} \text{ implies } \mu \in \semdg{\kod{\pat}_1}{\ag}{\kod{G}} \bowtie \semdg{\kod{\pat}_2}{\ag}{\kod{G}}.$$
			By Definition \ref{def:union_join}, $\mu = \mu_1 \cup \mu_2$, where $\mu_1 \in \semdg{\kod{\pat}_1}{\ag}{\kod{G}}$, $\mu_2 \in \semdg{\kod{\pat}_2}{\ag}{\kod{G}}$ and $\mu_1 \comp \mu_2$.
			By induction hypotheses on $\kod{\pat}_1$ and $\kod{\pat}_2$, mappings $\mu_1$ and $\mu_2$, and their corresponding valuations $\nu_1$ and $\nu_2$, where $\mu_1 \povezano \nu_1$ and $\mu_2 \povezano \nu_2$, it holds:\footnote{Until the end of this proof, we use notation $\sigma$ instead of $\sigma^{\kod{\gr}}$, for brevity.}
			$$\model{\nu_1} \models \sigma(\kod{\pat}_1) \text{ and } \model{\nu_2} \models \sigma(\kod{\pat}_2).$$
			By Lemma \ref{thm:mu_1_2_nu_1_2}, from $\mu_1\comp \mu_2$, $\mu = \mu_1 \cup \mu_2$, $\mu_1 \povezano \nu_1$, $\mu_2 \povezano \nu_2$ and $\mu \povezano \nu$, it holds $\nu_1 \comp \nu_2$ and $\nu = \nu_1 \cup \nu_2$.
			Therefore, $\nu \succeq \nu_1$ and $\nu \succeq \nu_2$, and it holds
			$$\model{\nu} \models \sigma(\kod{\pat}_1) \text{ and } \model{\nu} \models \sigma(\kod{\pat}_2),$$
			i.e. by Definition \ref{defn:model},
			$$\model{\nu} \models \sigma(\kod{\pat}_1) \wedge \sigma(\kod{\pat}_2).$$
			Then, by Definition \ref{def:sigma}, it holds
			$$\model{\nu} \models \sigma(\kod{\pat}_1 . \kod{\pat}_2).$$
			
			($\Leftarrow$) 
			By Definition \ref{def:sigma}, from $\model{\nu} \models \sigma(\kod{\pat}_1 . \kod{\pat}_2)$, it holds 
			$$\model{\nu} \models \sigma(\kod{\pat}_1) \wedge \sigma(\kod{\pat}_2),$$
			i.e. by Definition \ref{defn:model},
			$$\model{\nu} \models \sigma(\kod{\pat}_1) \text{ and } \model{\nu} \models \sigma(\kod{\pat}_2).$$
			All free variables appearing in these two formulas belong to $var(\sigma(\kod{\pat}_1))$ and $var(\sigma(\kod{\pat}_2))$, while restrictions of $\nu$ to these domains are denoted $\nu_1$ and $\nu_2$, respectively.
			Note that it holds $$dom(\nu) = dom(\nu_1) \cup dom(\nu_2) \text{ and } \nu = \nu_1 \cup \nu_2.$$
			Therefore, models $\model{\nu_1}$ and $\model{\nu_2}$ can be used in the previous formulas:
			$$\model{\nu_1} \models \sigma(\kod{\pat}_1) \text{ and } \model{\nu_2} \models \sigma(\kod{\pat}_2).$$
			By induction hypotheses on $\kod{\pat}_1$ and $\kod{\pat}_2$, valuations $\nu_1$ and $\nu_2$, and their corresponding mappings $\mu_1$ and $\mu_2$, such that $\mu_1 \povezano \nu_1$ and $\mu_1 \povezano \nu_1$, as $dom(\nu_1) = var(\sigma(\kod{\pat}_1))$ and $dom(\nu_2) = var(\sigma(\kod{\pat}_2))$, it holds $\mu_1 \in \semdg{\kod{\pat}_1}{\ag}{\kod{G}}$ and $\mu_2 \in \semdg{\kod{\pat}_2}{\ag}{\kod{G}}$.
			By Lemma \ref{thm:mu_1_2_nu_1_2}, from $\nu_1 \comp \nu_2$, $\nu = \nu_1 \cup \nu_2$, $\mu \povezano \nu$, $\mu_1 \povezano \nu_1$ and $\mu_2 \povezano \nu_2$, it holds $\mu_1 \comp \mu_2$ and $\mu = \mu_1 \cup \mu_2$.
			Therefore, by Definition \ref{def:union_join} of $\bowtie$, it holds 
			$$\mu \in \semdg{\kod{\pat}_1}{\ag}{\kod{G}} \bowtie \semdg{\kod{\pat}_2}{\ag}{\kod{G}}.$$
			Then, by Definition \ref{definition_sem}, it holds $$\mu \in \semdg{\kod{\pat}_1.\kod{\pat}_2}{\ag}{\kod{G}}.$$ 
			
			\item $\kod{\pat}$ is $\kod{\pat}_1 \kod{ filter } \kod{R}$
			\begin{align*}
				& \mu \in \semdg{\kod{\pat}_1 \kod{ filter } \kod{R}}{\ag}{\kod{G}} \text{ iff } \hspace*{-18mm} \\
				& \text{(by Def \ref{definition_sem})} & \text{ iff }  & \mu \in \semdg{\kod{\pat}_1}{\ag}{\kod{G}} \text{ and } \\
				&&& \mu \zadovoljava \kod{R} \\
				& \text{(by induc.hyp.)} \hspace*{-3mm} & \text{ iff }  & \model{\nu} \models \sigma(\kod{\pat}_1) \text{ and } \\
				&&& dom(\nu)=var(\sigma(\kod{\pat}_1))  \\
				&&& \text{ and } \mu \zadovoljava \kod{R} \\
				& \text{(by Lemma \ref{thm:thm56})} \hspace*{-3mm} & \text{ iff }  & \model{\nu} \models \sigma(\kod{\pat}_1) \text{ and } \\
				&&& dom(\nu)=var(\sigma(\kod{\pat}_1))  \\
				&&& \text{ and } \model{\nu} \models \sigma(\kod{R})\\
				& \text{(by Def \ref{defn:model})} & \text{ iff }  & \model{\nu} \models \sigma(\kod{\pat}_1) \wedge \sigma(\kod{R}) \text{ and }\\
				&&& dom(\nu)=var(\sigma(\kod{\pat}_1)) \\
				& \text{(by Def \ref{def:var}, as} & \text{ iff }  & \model{\nu} \models \sigma(\kod{\pat}_1) \wedge \sigma(\kod{R}) \text{ and }\\
				& ~\kod{var}(\kod{R}) \hspace*{-1mm}\subseteq\hspace*{-1mm} \kod{var}(\kod{\pat}_1) \text{)} \hspace*{-7mm} && dom(\nu)=var(\sigma(\kod{\pat})) \\
				& \text{(by Def \ref{def:sigma})} & \text{ iff }  & \model{\nu} \models \sigma(\kod{\pat}_1 \kod{ filter } \kod{R})\\
				&&& dom(\nu)=var(\sigma(\kod{\pat})) 
			\end{align*}
			
			\item $\kod{\pat}$ is $\kod{\{\pat}_1\kod{\}}$
			\begin{align*}
				& \hspace*{5mm} \mu \in \semdg{\kod{\{\pat}_1\kod{\}}}{\ag}{\kod{G}} \hspace*{-3mm} & \text{ iff } \\
				& \text{(by Def \ref{definition_sem})} & \text{ iff }  & \mu \in \semdg{\kod{\pat}_1}{\ag}{\kod{G}} \\
				& \text{(by induc.hyp.)} & \text{ iff }  & \model{\nu} \models \sigma(\kod{\pat}_1) \text{ and }\\
				&&& dom(\nu) = var(\sigma(\kod{\pat}_1)) \\
				& \text{(by Def \ref{def:var})} & \text{ iff }  & \model{\nu} \models \sigma(\kod{\pat}_1) \text{ and }\\
				&&& dom(\nu) = var(\sigma(\kod{\{\pat}_1\kod{\}})) \\
				& \text{(by Def \ref{def:sigma})} & \text{ iff }  & \model{\nu} \models \sigma(\kod{\{\pat}_1\kod{\}}) \text{ and } \\
				&&& dom(\nu) = var(\sigma(\kod{\{\pat}_1\kod{\}})) \\
			\end{align*}
			
			\item $\kod{\pat}$ is $\kod{\pat}_1 \kod{ diff } \kod{\pat}_2$\\
			($\Rightarrow$) 
			By Definiton \ref{definition_sem}, from
			$$\mu \in \semdg{\kod{\pat}_1 \  \kod{ diff } \ \kod{\pat}_2}{\ag}{\kod{\gr}}$$
			it holds $\mu \in \semdg{\kod{\pat}_1}{\ag}{\kod{\gr}} \setminus \semdg{\kod{\pat}_2}{\ag}{\kod{\gr}}.$
			By Definition \ref{def:union_join}, it holds
			$\mu \in \semdg{\kod{\pat}_1}{\ag}{\kod{\gr}}$
			and there is no $\mu_2$ such that 
			$\mu_2 \in \semdg{\kod{\pat}_2}{\ag}{\kod{\gr}}$ and $\mu \comp \mu_2$.
			By induction hypothesis, from $\mu \in \semdg{\kod{\pat}_1}{\ag}{\kod{\gr}}$, for mapping $\mu$ and a corresponding valuation $\nu$ ($\mu \povezano \nu$), it holds 
			$$\model{\nu} \models \sigma(\kod{\pat}_1) \text{ and } dom(\nu) = var(\sigma(\kod{\pat}_1)).$$
			Let us prove by contraposition that it holds
			$\model{\nu} \models \forall \overline{x} \ \neg \sigma(\kod{\pat}_2)$,
			i.e.~let us assume
			$$\model{\nu} \models \exists \overline{x} \ \sigma(\kod{\pat}_2).$$
			This means that we can extend the valuation $\nu$ to all the variables from $var(\sigma(\kod{\pat}_2))$ (by adding values to the variables from $\overline{x}$), and then restrict it only to the variables from $var(\sigma(\kod{\pat}_2))$ (by removing variables from $var(\sigma(\kod{\pat}_1))$ that are not in $var(\sigma(\kod{\pat}_2))$).
			Let $\nu_2$ denote such valuation.
			Then, it holds
			$$\model{\nu_2} \models \sigma(\kod{\pat}_2).$$
			Note that it holds $\nu_2 \comp \nu$ by construction.
			From the induction hypothesis applied on $\nu_2$ and a mapping $\mu_2$ such that $\mu_2 \povezano \nu_2$, it holds 
			$\mu_2 \in \semdg{\kod{\pat}_2}{\ag}{\kod{\gr}}$.
			By Lemma \ref{thm:mu12_nu12_compatibility}, from $\mu \povezano \nu$, $\mu_2 \povezano \nu_2$ and $\nu\comp \nu_2$, it holds $\mu \comp \mu_2$.
			This is a contradiction with nonexistence of a mapping compatible with $\mu$ and from $\semdg{\kod{\pat}_2}{\ag}{\kod{\gr}}$.
			Therefore, it holds:
			$$\model{\nu} \models \forall \overline{x} \ \neg \sigma(\kod{\pat}_2),$$
			and then by Definition \ref{defn:model}
			$$\model{\nu} \models \sigma(\kod{\pat}_1) \wedge \forall \overline{x} \ \neg \sigma(\kod{\pat}_2).$$
			By Definition \ref{def:sigma}, it holds
			$$\model{\nu} \models \sigma(\kod{\pat}_1 \kod{ diff } \kod{\pat}_2).$$
			By Lemma \ref{thm:dom_mu_nu1}, from $dom(\nu) = var(\sigma(\kod{\pat}_1))$, it holds $dom(\nu) = \sigma(\kod{var}(\kod{\pat}_1))$, and then by Definition \ref{def:var} it holds $dom(\nu) = \sigma(\kod{var}(\kod{\pat}_1 \kod{ diff } \kod{\pat}_2))$.
			Therefore, by Lemma \ref{thm:dom_mu_nu1}, it holds
			$dom(\nu) = var(\sigma(\kod{\pat}_1 \kod{ diff } \kod{\pat}_2))$.
			
			($\Leftarrow$) 
			By Definition \ref{def:sigma}, from
			$$dom(\nu) = var(\sigma(\kod{\pat}_1  \kod{ diff } \kod{\pat}_2)), \text{ and }$$
			$$\model{\nu} \models \sigma(\kod{\pat}_1  \kod{ diff } \kod{\pat}_2)$$
			it holds
			$$dom(\nu) = var(\sigma(\kod{\pat}_1) \wedge \forall \overline{x} \ \neg \sigma(\kod{\pat}_2)), \text{ and }$$
			$$\model{\nu} \models \sigma(\kod{\pat}_1) \wedge \forall \overline{x} \ \neg \sigma(\kod{\pat}_2),$$
			i.e.~by Definition \ref{defn:model},
			$$\model{\nu} \models \sigma(\kod{\pat}_1) \text{ and } \model{\nu} \models \forall \overline{x} \ \neg \sigma(\kod{\pat}_2).$$
			By Definition \ref{def:var_sigma}, as $\overline{x}$ denotes all variables that appear in $\sigma(\kod{\pat}_2)$, but not in $\sigma(\kod{\pat}_1)$, it holds $dom(\nu) = var(\sigma(\kod{\pat}_1))$.
			Applying the induction hypothes on the valuation $\nu$ and its corresponding mapping $\mu$ ($\mu \povezano \nu$), it holds $\mu \in \semdg{\kod{\pat}_1}{\ag}{\kod{\gr}}$.
			
			It should also be proved that there does not exist a $\mu_2$ such that $\mu_2 \in \semdg{\kod{\pat}_2}{\ag}{\kod{\gr}}$ and $\mu_2 \comp \mu$.
			This can be proved by contraposition, i.e.~let us assume the existence of such a mapping $\mu_2$.
			By induction hypothesis, applied on the mapping $\mu_2$ and a valuation $\nu_2$ such that $\mu_2 \povezano \nu_2$, it holds
			$$\model{\nu_2} \models \sigma(\kod{\pat}_2).$$
			By Lemma \ref{thm:mu12_nu12_compatibility}, from $\mu_2 \comp \mu$, $\mu \povezano \nu$ and $\mu_2 \povezano \nu_2$, it holds  $\nu_2 \comp \nu$.
			Then, $\nu \cup \nu_2$ is a well defined valuation, denoted by $\nu'$.
			As $\nu'$ is an extension of $\nu_2$, from $\model{\nu_2} \models \sigma(\kod{\pat}_2)$, it holds
			$$\model{\nu'} \models \sigma(\kod{\pat}_2).$$
			As $\nu'$ is an extension of $\nu$, from $\model{\nu} \models \forall \overline{x} \ \neg \sigma(\kod{\pat}_2)$, it holds
			$$\model{\nu'} \models \forall \overline{x} \ \neg \sigma(\kod{\pat}_2).$$
			Therefore, there is an extension $\nu''$ of $\nu'$, such that
			$$\model{\nu''} \models \neg \sigma(\kod{\pat}_2).$$
			Also, as $\nu''$ is an extension of $\nu'$, from $\model{\nu'} \models \sigma(\kod{\pat}_2)$, it holds
			$$\model{\nu''} \models \sigma(\kod{\pat}_2).$$
			Therefore, $\model{\nu''}$ is a model of formula $\sigma(\kod{\pat}_2)$ and its negation as well, which is a contradiction.
			
			As it holds $\mu \in \semdg{\kod{\pat}_1}{\ag}{\kod{\gr}}$ and it holds that there does not exist $\mu_2$ such that $\mu_2 \in \semdg{\kod{\pat}_2}{\ag}{\kod{\gr}}$ and $\mu_2 \comp \mu$, then, by Definition \ref{def:union_join} $$\mu \in \semdg{\kod{\pat}_1}{\ag}{\kod{\gr}} \setminus \semdg{\kod{\pat}_2}{\ag}{\kod{\gr}}.$$
			Then, by Definition \ref{definition_sem}, it holds $$\mu \in \semdg{\kod{\pat}_1 \  \kod{ diff } \ \kod{\pat}_2}{\ag}{\kod{\gr}}.$$
			
			\item $\kod{\pat}$ is $\kod{\pat}_1 \kod{ minus } \kod{\pat}_2$\\
			
			If $\kod{var}(\kod{\pat}_1) \cap \kod{var}(\kod{\pat}_2) = \varnothing$ then:
			
			($\Rightarrow$) 
			By Definiton \ref{definition_sem}, from
			$$\mu \in \semdg{\kod{\pat}_1 \  \kod{ minus } \ \kod{\pat}_2}{\ag}{\kod{\gr}}$$
			it holds $\mu \in \semdg{\kod{\pat}_1}{\ag}{\kod{\gr}} - \semdg{\kod{\pat}_2}{\ag}{\kod{\gr}}.$
			By Definition \ref{def:union_join}, it holds
			$\mu \in \semdg{\kod{\pat}_1}{\ag}{\kod{\gr}}$.
			%and there is no $\mu_2$ such that 
			%$\mu_2 \in \semdg{\kod{\pat}_2}{\ag}{\kod{\gr}}$, $\mu \comp \mu_2$ and $dom(\mu) \cap dom(\mu_2) \not= \varnothing$.
			By induction hypothesis, from $\mu \in \semdg{\kod{\pat}_1}{\ag}{\kod{\gr}}$, for mapping $\mu$ and a corresponding valuation $\nu$ ($\mu \povezano \nu$), it holds 
			$$\model{\nu} \models \sigma(\kod{\pat}_1) \text{ and } dom(\nu) = var(\sigma(\kod{\pat}_1)).$$
			By Definition \ref{def:sigma}, it holds
			$$\model{\nu} \models \sigma(\kod{\pat}_1 \kod{ minus } \kod{\pat}_2).$$
			By Lemma \ref{thm:dom_mu_nu1}, from $dom(\nu) = var(\sigma(\kod{\pat}_1))$, it holds $dom(\nu) = \sigma(\kod{var}(\kod{\pat}_1))$, and then by Definition \ref{def:var} it holds $dom(\nu) = \sigma(\kod{var}(\kod{\pat}_1 \kod{ minus } \kod{\pat}_2))$.
			Therefore, by Lemma \ref{thm:dom_mu_nu1}, it holds
			$dom(\nu) = var(\sigma(\kod{\pat}_1 \kod{ minus } \kod{\pat}_2))$.
			
			($\Leftarrow$) 
			By Definition \ref{def:sigma}, from
			$$dom(\nu) = var(\sigma(\kod{\pat}_1  \kod{ minus } \kod{\pat}_2)), \text{ and }$$
			$$\model{\nu} \models \sigma(\kod{\pat}_1  \kod{ minus } \kod{\pat}_2)$$
			it holds
			$$dom(\nu) = var(\sigma(\kod{\pat}_1)) \text{ and } \model{\nu} \models \sigma(\kod{\pat}_1).$$
			Applying the induction hypothesis on the valuation $\nu$ and its corresponding mapping $\mu$ ($\mu \povezano \nu$), it holds $\mu \hspace*{2pt} \in \hspace*{2pt} \semdg{\kod{\pat}_1}{\ag}{\kod{\gr}}$ .
			Therefore, by Lemma \ref{thm:dom_mu_nu1}, it holds $dom(\mu) = \kod{var}(\kod{\pat}_1)$.
			For all $\mu_2 \in \semdg{\kod{\pat}_2}{\ag}{\kod{\gr}}$, by Lemma \ref{thm:dom_mu_nu1}, it holds $dom(\mu_2) = \kod{var}(\kod{\pat}_2)$.
			Therefore, from $\kod{var}(\kod{\pat}_1) \cap \kod{var}(\kod{\pat}_2) = \varnothing$, for all $\mu_2 \in \semdg{\kod{\pat}_2}{\ag}{\kod{\gr}}$, it holds
			$$dom(\mu) \cap dom(\mu_2) = \varnothing.$$
			As it holds $\mu \in \semdg{\kod{\pat}_1}{\ag}{\kod{\gr}}$ and for all $\mu_2 \in \semdg{\kod{\pat}_2}{\ag}{\kod{\gr}}$ it holds $dom(\mu) \cap dom(\mu_2) = \varnothing$, then, by Definition \ref{def:union_join} $$\mu \in \semdg{\kod{\pat}_1}{\ag}{\kod{\gr}} - \semdg{\kod{\pat}_2}{\ag}{\kod{\gr}}.$$
			Then, by Definition \ref{definition_sem}, it holds $$\mu \in \semdg{\kod{\pat}_1 \  \kod{ minus } \ \kod{\pat}_2}{\ag}{\kod{\gr}}.$$
			
			Otherwise:
			
			($\Rightarrow$) 
			By Definiton \ref{definition_sem}, from
			$$\mu \in \semdg{\kod{\pat}_1 \  \kod{ minus } \ \kod{\pat}_2}{\ag}{\kod{\gr}}$$
			it holds $\mu \in \semdg{\kod{\pat}_1}{\ag}{\kod{\gr}} - \semdg{\kod{\pat}_2}{\ag}{\kod{\gr}}.$
			By Definition \ref{def:union_join}, it holds
			$\mu \in \semdg{\kod{\pat}_1}{\ag}{\kod{\gr}}$
			and there is no $\mu_2$ such that 
			$\mu_2 \in \semdg{\kod{\pat}_2}{\ag}{\kod{\gr}}$, $\mu \comp \mu_2$ and $dom(\mu) \cap dom(\mu_2) \not= \varnothing$.
			By induction hypothesis, from $\mu \in \semdg{\kod{\pat}_1}{\ag}{\kod{\gr}}$, for mapping $\mu$ and a corresponding valuation $\nu$ ($\mu \povezano \nu$), it holds 
			$$\model{\nu} \models \sigma(\kod{\pat}_1) \text{ and } dom(\nu) = var(\sigma(\kod{\pat}_1)).$$
			Let us prove by contraposition that it holds
			$\model{\nu} \models \forall \overline{x} \ \neg \sigma(\kod{\pat}_2)$,
			i.e.~let us assume
			$$\model{\nu} \models \exists \overline{x} \ \sigma(\kod{\pat}_2).$$
			This means that we can extend the valuation $\nu$ to all the variables from $var(\sigma(\kod{\pat}_2))$ (by adding values to the variables from $\overline{x}$), and then restrict it only to the variables from $var(\sigma(\kod{\pat}_2))$ (by removing variables from $var(\sigma(\kod{\pat}_1))$ that are not in $var(\sigma(\kod{\pat}_2))$).
			Let $\nu_2$ denote such valuation.
			Then, it holds
			$$\model{\nu_2} \models \sigma(\kod{\pat}_2).$$
			Note that it holds $\nu_2 \comp \nu$ by construction.
			From the induction hypothesis applied on $\nu_2$ and a mapping $\mu_2$ such that $\mu_2 \povezano \nu_2$, it holds 
			$\mu_2 \in \semdg{\kod{\pat}_2}{\ag}{\kod{\gr}}$.
			By Lemma \ref{thm:mu12_nu12_compatibility}, from $\mu \povezano \nu$, $\mu_2 \povezano \nu_2$ and $\nu\comp \nu_2$, it holds $\mu \comp \mu_2$.
			By Lemma \ref{thm:pattern_domain}, from $\mu \in \semdg{\kod{\pat}_1}{\ag}{\kod{\gr}}$ and $\mu_2 \in \semdg{\kod{\pat}_2}{\ag}{\kod{\gr}}$, it holds $dom(\mu_1) = \kod{var}(\kod{\pat}_1)$ and $dom(\mu_2) = \kod{var}(\kod{\pat}_2)$.
			From $\kod{var}(\kod{\pat}_1) \cap \kod{var}(\kod{\pat}_2) \not= \varnothing$, it holds $dom(\mu_1) \cap dom(\mu_2) \not= \varnothing$.
			This is a contradiction with nonexistence of a mapping compatible with $\mu$, from $\semdg{\kod{\pat}_2}{\ag}{\kod{\gr}}$ such that $dom(\mu_1) \cap dom(\mu_2) \not= \varnothing$.
			Therefore, it holds:
			$$\model{\nu} \models \forall \overline{x} \ \neg \sigma(\kod{\pat}_2),$$
			and then by Definition \ref{defn:model}
			$$\model{\nu} \models \sigma(\kod{\pat}_1) \wedge \forall \overline{x} \ \neg \sigma(\kod{\pat}_2).$$
			By Definition \ref{def:sigma}, it holds
			$$\model{\nu} \models \sigma(\kod{\pat}_1 \kod{ minus } \kod{\pat}_2).$$
			By Lemma \ref{thm:dom_mu_nu1}, from $dom(\nu) = var(\sigma(\kod{\pat}_1))$, it holds $dom(\nu) = \sigma(\kod{var}(\kod{\pat}_1))$, and then by Definition \ref{def:var} it holds $dom(\nu) = \sigma(\kod{var}(\kod{\pat}_1 \kod{ minus } \kod{\pat}_2))$.
			Therefore, by Lemma \ref{thm:dom_mu_nu1}, it holds
			$dom(\nu) = var(\sigma(\kod{\pat}_1 \kod{ minus } \kod{\pat}_2))$.
			
			($\Leftarrow$) 
			By Definition \ref{def:sigma}, from
			$$dom(\nu) = var(\sigma(\kod{\pat}_1  \kod{ minus } \kod{\pat}_2)), \text{ and }$$
			$$\model{\nu} \models \sigma(\kod{\pat}_1  \kod{ minus } \kod{\pat}_2)$$
			it holds
			$$dom(\nu) = var(\sigma(\kod{\pat}_1) \wedge \forall \overline{x} \ \neg \sigma(\kod{\pat}_2)), \text{ and }$$
			$$\model{\nu} \models \sigma(\kod{\pat}_1) \wedge \forall \overline{x} \ \neg \sigma(\kod{\pat}_2),$$
			i.e.~by Definition \ref{defn:model},
			$$\model{\nu} \models \sigma(\kod{\pat}_1) \text{ and } \model{\nu} \models \forall \overline{x} \ \neg \sigma(\kod{\pat}_2).$$
			By Definition \ref{def:var_sigma}, as $\overline{x}$ denotes all variables that appear in $\sigma(\kod{\pat}_2)$, but not in $\sigma(\kod{\pat}_1)$, it holds $dom(\nu) = var(\sigma(\kod{\pat}_1))$.
			Applying the induction hypothes on the valuation $\nu$ and its corresponding mapping $\mu$ ($\mu \povezano \nu$), it holds $\mu \in \semdg{\kod{\pat}_1}{\ag}{\kod{\gr}}$.
			
			It should also be proved that there does not exist a $\mu_2$ such that $\mu_2 \in \semdg{\kod{\pat}_2}{\ag}{\kod{\gr}}$, $\mu_2 \comp \mu$ and $dom(\mu_2) \cap dom(\mu) \not= \varnothing$.
			This can be proved by contraposition, i.e.~let us assume the existence of such a mapping $\mu_2$.
			By induction hypothesis, applied on the mapping $\mu_2$ and a valuation $\nu_2$ such that $\mu_2 \povezano \nu_2$, it holds
			$$\model{\nu_2} \models \sigma(\kod{\pat}_2).$$
			By Lemma \ref{thm:mu12_nu12_compatibility}, from $\mu_2 \comp \mu$, $\mu \povezano \nu$ and $\mu_2 \povezano \nu_2$, it holds  $\nu_2 \comp \nu$.
			Then, $\nu \cup \nu_2$ is a well defined valuation, denoted by $\nu'$.
			As $\nu'$ is an extension of $\nu_2$, from $\model{\nu_2} \models \sigma(\kod{\pat}_2)$, it holds
			$$\model{\nu'} \models \sigma(\kod{\pat}_2).$$
			As $\nu'$ is an extension of $\nu$, from $\model{\nu} \models \forall \overline{x} \ \neg \sigma(\kod{\pat}_2)$, it holds
			$$\model{\nu'} \models \forall \overline{x} \ \neg \sigma(\kod{\pat}_2).$$
			Therefore, there is an extension $\nu''$ of $\nu'$, such that
			$$\model{\nu''} \models \neg \sigma(\kod{\pat}_2).$$
			Also, as $\nu''$ is an extension of $\nu'$, from $\model{\nu'} \models \sigma(\kod{\pat}_2)$, it holds
			$$\model{\nu''} \models \sigma(\kod{\pat}_2).$$
			Therefore, $\model{\nu''}$ is a model of formula $\sigma(\kod{\pat}_2)$ and its negation as well, which is a contradiction.
			
			As it holds $\mu \in \semdg{\kod{\pat}_1}{\ag}{\kod{\gr}}$ and it holds that there does not exist $\mu_2$ such that $\mu_2 \in \semdg{\kod{\pat}_2}{\ag}{\kod{\gr}}$ and $\mu_2 \comp \mu$, then, by Definition \ref{def:union_join} $$\mu \in \semdg{\kod{\pat}_1}{\ag}{\kod{\gr}} - \semdg{\kod{\pat}_2}{\ag}{\kod{\gr}}.$$
			Then, by Definition \ref{definition_sem}, it holds $$\mu \in \semdg{\kod{\pat}_1 \  \kod{ minus } \ \kod{\pat}_2}{\ag}{\kod{\gr}}.$$
			
			\item $\kod{\pat}$ is $\kod{graph x \{\pat$_1$\}}$\\
			
			$(\Rightarrow)$	By Definition \ref{definition_sem},
			from $$\mu \in \semdg{\kod{ graph x \{\pat$_1$\}}}{\ag}{\kod{\gr}},$$
			it holds
			$$\mu \in \underset{\kod{i} \in names(\ag)} \bigcup \left(\ \semdg{\kod{\pat}_1}{\ag}{gr_{\,\ag}(\kod{i})} \bowtie \{ \mu_{\kod{x} \rightarrow \kod{i}} \} \ \right).$$
			Therefore, $\agfn \not= \varnothing$ and there is an IRI $\kod{i}$ from $names(\ag)$ such that
			$$\mu \in \semdg{\kod{\pat}_1}{\ag}{gr_{\,\ag}(\kod{i})} \bowtie \{ \mu_{\kod{x} \rightarrow \kod{i}} \}.$$
			By Definition \ref{def:union_join}, there exists mapping $\mu_1$ such that $\mu = \mu_1 \cup \mu_{\kod{x} \rightarrow \kod{i}}$, $\mu_1 \in \semdg{\kod{\pat}_1}{\ag}{gr_{\,\ag}(\kod{i})}$ and $\mu_1 \comp\mu_{\kod{x} \rightarrow \kod{i}}$.
			Applying the induction hypothesis on the mapping $\mu_1$ and a valuation $\nu_1$ such that $\mu_1 \povezano \nu_1$, it holds 
			$$\model{\nu_1} \models \sigma^{gr_{\,\ag}(\kod{i})}(\kod{\pat}_1).$$
			By Lemma \ref{thm:mu_1_2_nu_1_2}, from $\mu = \mu_1 \cup \mu_{\kod{x} \rightarrow \kod{i}}$, $\mu \povezano \nu$, $\mu_1 \povezano \nu_1$ and $\mu_{\kod{x} \rightarrow \kod{i}} \povezano \nu_{x \rightarrow \kod{i}}$, it holds $\nu = \nu_1 \cup \nu_{x \rightarrow \kod{i}}$.
			Therefore, $\nu \succeq \nu_1$, and it holds
			$$\model{\nu} \models \sigma^{gr_{\,\ag}(\kod{i})}(\kod{\pat}_1).$$
			From $\mu = \mu_1 \cup \mu_{\kod{x} \rightarrow \kod{i}}$, it holds $\mu(\kod{x}) = \kod{i}$.
			As $\kod{i} \in \kod{I}$, by Definition \ref{defn:expr_in_mu}, it holds $\kod{i} = \valuation{\kod{i}}{\mu}$, and $\mu(\kod{x}) = \valuation{\kod{i}}{\mu}$.
			By Lemma \ref{thm:thm78}, from the last equation and $\mu \povezano \nu$, it holds
			$$\model{\nu} \models \sigma(\kod{x}) = \sigma(\kod{i}).$$
			Then, by Definition \ref{defn:model}, it holds:
			$$\model{\nu} \models \sigma^{gr_{\,\ag}(\kod{i})}(\kod{\pat}_1) \wedge \sigma(\kod{x}) = \sigma(\kod{i}).$$
			By Definition \ref{def:sigma}, from $\kod{i} \in names(\ag)$, for $i = \siginv(\kod{i})$, it holds $i \in \agfn$.
			Therefore, it holds
			$$\model{\nu} \models \underset{i \in \agfn}\bigvee \Big( \sigma^{gr_{\,\ag}(\siginv({i}))}(\kod{\pat}_1) \wedge \sigma(\kod{x}) = i \Big),$$
			and, by Definition \ref{def:sigma},
			$$\model{\nu} \models \sigma(\kod{graph x \{\pat$_1$\}}).$$
			
			$(\Leftarrow)$ From
			$$dom(\nu) = var(\sigma(\kod{graph x \{\pat$_1$\}})) \text{ and }$$
			$$\model{\nu} \models \sigma(\kod{graph x \{\pat$_1$\}}),$$
			by Definition \ref{def:sigma}, it holds
			$$dom(\nu) = var(\sigma(\kod{\pat}_1)) \cup \{\sigma(\kod{x})\} \text{ and }$$
			$$\model{\nu} \models \underset{i \in \agfn}\bigvee \Big( \sigma^{gr_{\,\ag}(\siginv({i}))}(\kod{\pat}_1) \wedge \sigma(\kod{x}) = i \Big).$$
			Therefore, there is a constant $i$ from $\agfn$, such that $i = \sigma(\kod{i})$ and 
			$$\model{\nu} \models \sigma^{gr_{\,\ag}(\kod{i})}(\kod{\pat}_1) \wedge \sigma(\kod{x}) = \sigma(\kod{i}).$$
			%By construction of set $\agfn$, $i \in \agfn$ implies that there is an IRI $\kod{i}$, such that $\kod{i} \in names(\ag)$, $i = \sigma(\kod{i})$ and $\context(gr_{\,\ag}(\kod{i})) = \{i\}$.
			Then, by Definition \ref{defn:model}, it holds
			$$\model{\nu} \models \sigma^{gr_{\,\ag}(\kod{i})}(\kod{\pat}_1) \text{ and } \model{\nu} \models \sigma(\kod{x}) = \sigma(\kod{i}).$$
			By Lemma \ref{thm:thm78}, from the second satisfiability and $\mu \povezano \nu$, it holds $\mu(x) = \valuation{\kod{i}}{\mu},$ i.e.~by Definition \ref{defn:expr_in_mu}, it holds
			$\mu(x) = \kod{i}$.
			We consider two cases:
			\begin{itemize}
				\item If $\sigma(\kod{x}) \in var(\sigma(\kod{\pat}_1))$, i.e.~$dom(\nu) = var(\sigma(\kod{\pat}_1))$.
				By induction hypothesis, from the first satisfiability and $\mu \povezano \nu$, it holds
				$\mu \in \semdg{\kod{\pat}_1}{\ag}{gr_{\,\ag}(\kod{i})}$.
				Therefore, $\mu = \mu \cup \mu_{\kod{x} \rightarrow \kod{i}}$.
				\item If $\sigma(\kod{x}) \notin var(\sigma(\kod{\pat}_1))$, let $\nu'$ denotes a restriction of $\nu$ to the set $var(\sigma(\kod{\pat}_1))$.
				Then, it holds $\model{\nu'} \models \sigma^{gr_{\,\ag}(\kod{i})}(\kod{\pat}_1)$.
				Then, for a mapping $\mu'$ such that $\mu' \povezano \nu'$, by induction hypothesis, it holds $\mu' \in \semdg{\kod{\pat}_1}{\ag}{gr_{\,\ag}(\kod{i})}$.
				By Definition \ref{def:povezano}, from $\mu \povezano \nu$, $\mu' \povezano \nu'$ and $\nu' \preceq \nu$, it holds $\mu' \preceq \mu$.
				Also, it holds 
				\begin{align*}
					& \hspace*{20mm} dom(\mu)& = &\\
					& \text{(by Def \ref{def:povezano}, as } \mu \povezano \nu \text{)} \hspace*{-4mm} & = & \siginv(dom(\nu)) \\
					&  & = & \siginv(var(\sigma(\kod{\pat}_1)) \\
					&&& \cup \{\sigma(\kod{x})\}) \\
					& \text{(by Def \ref{def:function_over_sets})} & = & \siginv(var(\sigma(\kod{\pat}_1))) \\
					&&&\cup \siginv(\{\sigma(\kod{x})\}) \\
					& \text{(by Lemma \ref{thm:dom_mu_nu1})} & = & \siginv(\sigma(var(\kod{\pat}_1))) \\
					&&&\cup \siginv(\{\sigma(\kod{x})\}) \\
					& \text{(by Def \ref{def:sigmat}, \ref{def:sigma})} & = & \kod{var}(\kod{\pat}_1) \cup \{\kod{x}\}
				\end{align*}
				By Lemma \ref{thm:pattern_domain}, as $\mu' \hspace*{1pt} \in \hspace*{1pt} \semdg{\kod{\pat}_1}{\ag}{gr_{\,\ag}(\kod{i})}$, it holds $dom(\mu') = \kod{var}(\kod{\pat}_1)$.
				Therefore, it holds $\mu = \mu' \cup \mu_{\kod{x} \rightarrow \kod{i}}$.
			\end{itemize}
			Putting both cases together, by Definition \ref{def:union_join} it holds
			$$\mu \in \semdg{\kod{\pat}_1}{\ag}{gr_{\,\ag}(\kod{i})} \bowtie \{ \mu_{\kod{x} \rightarrow \kod{i}} \},$$
			i.e.
			$$\mu \in \underset{\kod{i} \in names(\ag)} \bigcup \left(\ \semdg{\kod{\pat}_1}{\ag}{gr_{\,\ag}(\kod{i})} \bowtie \{ \mu_{\kod{x} \rightarrow \kod{i}} \} \ \right).$$
			Finally, by Definition \ref{definition_sem} it holds that
			$$\mu \in \semdg{\kod{ graph x \{\pat$_1$\}}}{\ag}{\kod{\gr}}.$$
			
			\item $\kod{\pat}$ is $\kod{graph i \{\pat$_1$\}}$
			\begin{align*}
				& \mu \in \semdg{\kod{graph i \{\pat$_1$\}}}{\ag}{\kod{\gr}} \hspace*{2mm} \text{ iff } \hspace*{-30mm} \\
				& \text{(by Def \ref{definition_sem})} \hspace*{-4mm} & \text{ iff }  & \mu \in \semdg{\kod{\pat}_1}{\ag}{gr_{\,\ag}(\kod{i})} \\
				& \text{(by induc.hyp.)} \hspace*{-4mm} & \text{ iff }  & \model{\nu} \hspace*{-1mm} \models \hspace*{-1mm} \sigma^{gr_{\,\ag}(\kod{i})}(\kod{\pat}_1) \text{ and } \\
				&&& dom(\nu) = var(\sigma^{gr_{\,\ag}(\kod{i})}(\kod{\pat}_1)) \\
				& \text{(by Def \ref{def:var})} \hspace*{-4mm} & \text{ iff }  & \model{\nu} \hspace*{-1mm} \models \hspace*{-1mm} \sigma^{gr_{\,\ag}(\kod{i})}(\kod{\pat}_1) \text{ and } \\
				&&& dom(\nu) = var(\sigma(\kod{\pat})) \\
				& \text{(by Def \ref{def:sigma})} \hspace*{-4mm} & \text{ iff }  & \model{\nu} \hspace*{-1mm} \models \hspace*{-1mm} \sigma(\kod{graph\,i\,\{\pat$_1$\}}) \text{ and } \\
				&&& dom(\nu) = var(\sigma(\kod{\pat}))
			\end{align*}
		\end{description}
	\end{proof}	
\end{lemma}

\begin{lemma}
	\label{thm:thm12}
	Let $\dataq$ be a dataset. Let \kod{Q} be a query, $\overline{\kod{rv}}$ a set of its relevant variables and $\Phi(\overline{rv})$ its corresponding formula.
	Let $\mu : \kod{VB} \rightarrow \ibl{}$ be a mapping and $\nu : \vars \rightarrow \ibl{}$ be a valuation such that $\mu \povezano \nu$.
	Then:
	\begin{center}
		$\mu \in \semd{\kod{Q}}{\dataq}$\\
		if and only if\\
		$\model{\nu} \models \Phi(\overline{rv})$ ~and~ $dom(\nu) = var(\Phi(\overline{rv}))$.
	\end{center}
	\begin{proof}
		%\ref{proof:thm12} is given in Appendix \ref{sec:appendix}. It uses Lemma \ref{thm:thm34}.
		
		Let $\ag$ be a query dataset of \kod{Q}, $\kod{qpat}$ its query pattern and $\kod{dv}$ a set of its distinguished variables.
		
		($\Rightarrow$)
		By Definition \ref{def:rv}, from $\mu \hspace*{-2pt} \in \hspace*{-2pt} \semd{\kod{Q}}{\dataq}$, it holds $dom(\mu) \hspace*{-2pt} = \hspace*{-2pt} \overline{\kod{rv}}$.
		By Definition \ref{def:povezano}, as $\mu \povezano \nu$, it holds $dom(\nu) = \overline{rv}$.
		By Definitions \ref{def:projection} and \ref{defn:query_evaluation}, from $\mu \in \semd{\kod{Q}}{\dataq}$, we conclude that there exists an extension $\mu'$ of $\mu$, such that $\mu' \in \semdg{\kod{qpat}}{\ag}{\mathit{df}(\ag)}$.
		%By Definition \ref{def:extension_restriction}, $\mu'$ is an extension of $\mu$ ($\mu' \succeq \mu$), such that $dom(\mu) = dom(\mu') \cap \overline{\kod{dv}}$.
		By Lemma \ref{thm:thm34}, for a valuation $\nu'$ such that $\mu' \povezano \nu'$, it holds
		$$\model{\nu'} \models \sigma^{\mathit{df}(\ag)}(\kod{qpat}) \text{ ~and~ }$$
		$$dom(\nu') = var(\sigma^{\mathit{df}(\ag)}(\kod{qpat})).$$
		Therefore, it holds
		$$\model{\nu'} \models \exists \overline{ov} \  \sigma^{\mathit{df}(\ag)}(\kod{qpat}),$$
		i.e.~by Definition \ref{def:phi},
		$$\model{\nu'} \models \Phi(\overline{rv}).$$
		By Definitions \ref{def:extension_restriction} and \ref{def:povezano}, from $\mu \povezano \nu$, $\mu' \povezano \nu'$ and $\mu \preceq \mu'$, it holds $\nu \preceq \nu'$.
		By Lemma \ref{thm:var_phi}, it holds $dom(\nu) = var(\Phi(\overline{rv})$.
		Therefore, $\nu$ is a restriction of $\nu'$ defined on all free variables from $\Phi(\overline{rv})$, and then it holds
		$$\model{\nu} \models \Phi(\overline{rv}).$$
		
		($\Leftarrow$) By Definition \ref{def:phi}, from $\model{\nu} \models \Phi(\overline{rv})$, it holds
		$$\model{\nu} \models \exists \overline{ov} \ \sigma^{\mathit{df}(\ag)}(\kod{qpat}).$$
		Then, there exists an extension $\nu'$ of $\nu$ defined on $dom(\nu) \cup \overline{ov}$, such that
		$$\model{\nu'} \models \sigma^{\mathit{df}(\ag)}(\kod{qpat}).$$
		Therefore, from $dom(\nu) = var(\exists \overline{ov} \ \sigma^{\mathit{df}(\ag)}(\kod{qpat}))$, it holds
		$$dom(\nu') = var(\sigma^{\mathit{df}(\ag)}(\kod{qpat})).$$
		By Lemma \ref{thm:thm34}, for a mapping $\mu'$ such that $\mu' \povezano \nu'$, it holds that $\mu' \in \semdg{\kod{qpat}}{\ag}{\mathit{df}(\ag)}$.
		By Definitions \ref{def:extension_restriction} and \ref{def:povezano}, from $\mu \povezano \nu$, $\mu' \povezano \nu'$ and $\nu \preceq \nu'$, it holds $\mu \preceq \mu'$.
		Therefore, $\mu$ is a restriction of $\mu'$ with the following domain:
		\begin{align*}
			& \hspace*{27mm} dom(\mu) \hspace*{-10mm} &=\; \\
			&\text{(by Def \ref{def:povezano}, as $\mu \povezano \nu$)}&=\;& \siginv(dom(\nu)) \\
			& \text{(by Lemma \ref{thm:var_phi})} &=\;& \siginv(\overline{rv})\\
			& &=\;& \overline{\kod{rv}} \\
			& \text{(by Lemma \ref{thm:query_domain})} &=\;& \kod{var}(\kod{qpat}) \cap \overline{\kod{dv}} \\
			& \text{(by Lemma \ref{thm:pattern_domain})}&=\;& dom(\mu') \cap \overline{\kod{dv}}
		\end{align*}
		Therefore, by Definition \ref{def:extension_restriction}, $\mu = \mu'_{\overline{\kod{dv}}}$, and by Definitions \ref{def:projection} and \ref{defn:query_evaluation}
		$$\mu \in \semd{\kod{Q}}{\dataq}.$$
	\end{proof}
\end{lemma}

Note that a mapping $\mu$ is always considered as a mapping from \kod{VB} to \ibl{}. However, in Lemma \ref{thm:thm12}, the actual $dom(\mu)$ is a subset of \kod{V} and it does not contain blank nodes (according to Definition \ref{defn:query_evaluation}).

\begin{lemma}
	\label{thm:theta}
	$\Theta$ is valid if and only if $\kod{Q}_1$ is unsatisfiable.
	\begin{proof} 
		%\ref{proof:theta} is given in Appendix \ref{sec:appendix}. The proofs of both implications are done by contraposition.
		
		($\Rightarrow$) Assume that $\kod{Q}_1$ is satisfiable.
		By Definition \ref{def:satisfiable}, there exist a dataset $\dataq$ and a mapping $\mu$ such that $\mu \in \semdq{\kod{Q}}$.
		Then, by Lemma \ref{thm:thm12}, for a valuation $\nu$ such that $\mu \povezano \nu$, it holds
		$$\model{\nu} \models \Phi_1(\overline{rv_1}),$$
		where $\overline{\kod{rv}_1}$ is a set of relevant variables of $\kod{Q}_1$ and formula $\Phi_1(\overline{rv_1})$ corresponds to $\kod{Q}_1$.
		Therefore, it holds
		$$\model{\nu} \models \exists \overline{rv_1} \ \Phi_1(\overline{rv_1}).$$
		By Definition \ref{def:theta}, from the validity of $\Theta$, it holds
		$$\model{\nu} \models \neg \big(\exists \overline{rv_1} \ \Phi_1(\overline{rv_1})\big).$$
		Therefore, $\model{\nu}$ is a model of a formula and its negation, that is not possible, i.e.~$\kod{Q}_1$ is unsatisfiable.
		
		($\Leftarrow$) Assume that $\Theta$ is not valid.
		Therefore, there exists an $\mathcal{L}$-structure $\mathfrak{D'} = (\mathcal{D'}, \intrprt^\mathcal{D'})$, such that $\mathfrak{D'} \models \neg \Theta$.
		By Definition \ref{def:theta}, $\neg \Theta$ is equal to
		$$\exists \overline{rv_1} \ \Phi_1(\overline{rv_1}).$$	
		Note that this formula is a sentence (all variables are quantified). %, thus the valuation within $\mathfrak{D'}$ matter.
		By Lemma \ref{thm:model}, there exists a dataset $\dataq$ and the corresponding $\mathcal{L}$-structure $\mathfrak{D} = (\mathcal{D}, \intrprt^{\mathcal{D}})$, such that
		$$\mathfrak{D} \models \exists \overline{rv_1} \ \Phi_1(\overline{rv_1}).$$
		Therefore, there exists a valuation $\nu$ defined on $\overline{rv_1}$, such that 
		$$(\mathfrak{D}, \nu) \models \Phi_1(\overline{rv_1}),$$
		i.e.~by Definition \ref{defn:model},
		$$\model{\nu} \models \Phi_1(\overline{rv_1}).$$
		By Lemma \ref{thm:var_phi}, it holds $dom(\nu) = var(\Phi_1(\overline{rv_1}))$.
		Therefore, by Lemma \ref{thm:thm12}, for a mapping $\mu$ such that $\mu \povezano \nu$, it holds
		$$\mu \in \semd{\kod{Q}_1}{\dataq}.$$
		This is a contradiction with the unsatisfiability of $\kod{Q}_1$.
		Therefore, $\Theta$ is valid.
	\end{proof}
\end{lemma}

\subsection{Soundness}
\label{subsec:soundness}

The following soundness theorem states that if the Procedure \ref{thm:main} claims that two queries are in the containment relation, they really are.

\begin{thm}[Soundness]
	\label{thm:soundness}
	Let $\kod{Q}_1$ and $\kod{Q}_2$ be queries, $\Theta$ be a formula generated from $\kod{Q}_1$ (Def \ref{def:theta}) and $\Psi$ be a formula generated from $\kod{Q}_1$ and $\kod{Q}_2$ (Def \ref{def:psi}).
	It holds $\kod{Q}_1 \sqsubseteq \kod{Q}_2$ if one of the following conditions is satisfied:
	\begin{itemize}[noitemsep,topsep=0pt]
		\item[(1)] $\Theta$ is valid, or
		\item[(2)] $\kod{Q}_1 \seleq \kod{Q}_2$ holds and $\Psi$ is valid.
	\end{itemize}
	\begin{proof} 
		%\ref{proof:soundness} is given in Appendix \ref{sec:appendix}. It uses Lemmas \ref{thm:theta} and \ref{thm:thm12}.
		
		Case (1):  $\Theta$ is valid.\\
		By Lemma \ref{thm:theta}, $\kod{Q}_1$ is unsatisfiable.
		By Definition \ref{def:satisfiable}, $\semd{\kod{Q}_1}{\dataq}$ is an empty set for any dataset $\dataq$.
		Therefore, it holds $\semd{\kod{Q}_1}{\dataq} \subseteq \semd{\kod{Q}_2}{\dataq}$, i.e.~by Definition \ref{defn:query_containment}, $\kod{Q}_1 \sqsubseteq \kod{Q}_2$.
		
		Case (2): It holds $\kod{Q}_1 \seleq \kod{Q}_2$ and $\Psi$ is valid.\\
		Let $\dataq$ be any dataset, and $\mu$ a mapping such that
		$$\mu \in \semd{\kod{Q}_1}{\dataq}.$$
		By Lemma \ref{thm:thm12}, for a valuation $\nu$ such that $\mu \povezano \nu$, it holds
		$$\model{\nu} \models \Phi_1(\overline{rv_1}) \text{ and } dom(\nu) = var(\Phi_1(\overline{rv_1})).$$
		By Lemma \ref{thm:var_phi}, it holds $dom(\nu) = \overline{rv_1}$.
		By Lemma \ref{thm:set_equality}, as $\kod{Q}_1 \seleq \kod{Q}_2$ and $\kod{Q}_2$ does not contain projections, it holds 
		$$\overline{\kod{rv}_1} = \kod{var}(\kod{qpat}_2),$$
		i.e.
		$$\sigma(\overline{\kod{rv}_1}) = \sigma(\kod{var}(\kod{qpat}_2)).$$
		Then, by Lemma \ref{thm:dom_mu_nu1}, it holds
		$$\overline{rv_1} = var(\sigma(\kod{qpat}_2)).$$
		Therefore, $dom(\nu)$ is equal to $var(\sigma(\kod{qpat}_2))$, as well.
		By Lemma \ref{thm:psi}, as $\kod{Q}_1 \seleq \kod{Q}_2$, $\kod{Q}_2$ does not contain projections and $\Psi$ is valid, it holds
		$$\model{\nu} \models \forall \overline{rv_1} \ (\Phi_1(\overline{rv_1}) \Rightarrow \sigma^{\mathit{df}({\ag_2})}(\kod{qpat}_2)).$$
		As $\nu$ is defined on all free variables from formula $\Phi_1(\overline{rv_1}) \Rightarrow \sigma^{\mathit{df}({\ag_2})}(\kod{qpat}_2)$, it holds:
		$$\model{\nu} \models \Phi_1(\overline{rv_1}) \Rightarrow \sigma^{\mathit{df}({\ag_2})}(\kod{qpat}_2).$$
		From $\model{\nu} \models \Phi_1(\overline{rv_1})$, it holds
		$$\model{\nu} \models \sigma^{\mathit{df}({\ag_2})}(\kod{qpat}_2).$$
		Then, by Lemma \ref{thm:thm34}, as $\mu \povezano \nu$ and $dom(\nu) = var(\sigma(\kod{qpat}_2))$, it holds
		$$\mu \in \semdg{\kod{qpat}_2}{{\ag_2}}{\mathit{df}({\ag_2})},$$
		i.e.~by Definition \ref{defn:query_evaluation}, as $\kod{Q}_2$ does not contain projections, it holds
		$$\mu \in \semd{\kod{Q}_2}{\dataq}.$$
		Therefore, each element of the set $\semd{\kod{Q}_1}{\dataq}$ belongs to the set $\semd{\kod{Q}_2}{\dataq}$ for any dataset $\dataq$, i.e.~$\semd{\kod{Q}_1}{\dataq} \subseteq \semd{\kod{Q}_2}{\dataq}$.
		By Definition \ref{defn:query_containment}, it holds $\kod{Q}_1 \sqsubseteq \kod{Q}_2$.
	\end{proof}
\end{thm}

Soundness theorem for subsumption (dual to Theorem \ref{thm:soundness}) can be proved as well.
It validates the results of Procedure \ref{thm:main_subsumption}.

\begin{thm}[Soundness for subsumption]
	\label{thm:soundness_subsumption}
	Let $\kod{Q}_1$ and $\kod{Q}_2$ be queries, $\Theta$ be a formula generated from $\kod{Q}_1$ (Def \ref{def:theta}) and $\Psi$ be a formula generated from $\kod{Q}_1$ and $\kod{Q}_2$ (Def \ref{def:psi}).
	It holds $\kod{Q}_1 \dot{\sqsubseteq} \kod{Q}_2$ if one of the following conditions is satisfied:
	\begin{itemize}[noitemsep,topsep=0pt]
		\item[(1)] $\Theta$ is valid, or
		\item[(2)] $\kod{Q}_1 \dot{\seleq} \kod{Q}_2$ holds and $\Psi$ is valid.
	\end{itemize}
	\begin{proof} 
		%\ref{proof:soundness_subsumtion} is given in Appendix \ref{sec:appendix}. It is done by changing the proof of Theorem \ref{thm:soundness}.
		
		This proof is equal to the proof of Theorem \ref{thm:soundness}, from beginning to the applying Lemma \ref{thm:set_equality}.
		Instead of it, in the subsumption case, we apply its dual lemma, i.e.~Lemma \ref{thm:set_equality_subsumption}, and from $\kod{Q}_1 \dot{\seleq} \kod{Q}_2$ it holds
		$$\overline{\kod{rv}_1} \subseteq \kod{var}(\kod{qpat}_2) \cap \overline{\kod{dv}_2},$$
		i.e.~
		$$\sigma(\overline{\kod{rv}_1}) \subseteq \sigma(\kod{var}(\kod{qpat}_2) \cap \overline{\kod{dv}_2}),$$
		and, by Definition \ref{def:function_over_sets}, it holds
		$$\sigma(\overline{\kod{rv}_1}) \subseteq \sigma(\kod{var}(\kod{qpat}_2)) \cap \sigma(\overline{\kod{dv}_2}).$$
		Therefore, by Lemma \ref{thm:dom_mu_nu1}, it holds
		\begin{equation}
			\label{eq:subsumpt1}
			\overline{rv_1} \subseteq var(\sigma(\kod{qpat}_2)) \cap \overline{dv_2},
		\end{equation}
		The rest of the proof follows.
		
		Let $\nu'$ denote any extension of $\nu$, such that $dom(\nu') = dom(\nu) \cup (var(\sigma(\kod{qpat}_2)) \setminus \overline{ov_2})$.
		By Definitions \ref{def:phi} and \ref{def:psi}, as $\Psi$ is valid, it holds
		$$\model{\nu'} \models \forall \overline{rv_1} \ (\Phi_1(\overline{rv_1}) \Rightarrow \exists \overline{ov_2} \ \sigma^{\mathit{df}({\ag_2})}(\kod{qpat}_2)).$$
		As $\nu'$ is defined on all free variables from formula $\Phi_1(\overline{rv_1}) \Rightarrow \exists \overline{ov_2} \ \sigma^{\mathit{df}({\ag_2})}(\kod{qpat}_2)$, it holds:
		$$\model{\nu'} \models \Phi_1(\overline{rv_1}) \Rightarrow \exists \overline{ov_2} \ \sigma^{\mathit{df}({\ag_2})}(\kod{qpat}_2).$$
		From $\model{\nu} \models \Phi_1(\overline{rv_1})$, as $\nu' \succeq \nu$ it holds $\model{\nu'} \models \Phi_1(\overline{rv_1})$.
		Therefore, it holds
		$$\model{\nu'} \models \exists \overline{ov_2} \ \sigma^{\mathit{df}({\ag_2})}(\kod{qpat}_2).$$
		Then, there exists $\nu''$, an extension of $\nu'$, such that $dom(\nu'') = dom(\nu') \cup \overline{ov_2}$ and 
		$$\model{\nu''} \models \sigma^{\mathit{df}({\ag_2})}(\kod{qpat}_2).$$
		\begin{align*}
			& \text{Also: } dom(\nu'') \hspace*{-3mm}& = & \\
			& \text{(by constr.)} & = & dom(\nu) \cup (var(\sigma(\kod{qpat}_2)) \setminus \overline{ov_2}) \cup \overline{ov_2} \\
			&  & = & dom(\nu) \cup var(\sigma(\kod{qpat}_2)) \cup \overline{ov_2}\\
			& \text{(by Def \ref{def:phi})} & = & dom(\nu) \cup var(\sigma(\kod{qpat}_2)) \cup \\
			&&& (var(\sigma(\kod{qpat}_2)) \setminus \overline{rv}_1)\\
			&  & = & dom(\nu) \cup var(\sigma(\kod{qpat}_2)) \\
			& \text{(} dom(\nu) = \overline{rv_1}\text{)} \hspace*{-3mm} & = & \overline{rv_1} \cup var(\sigma(\kod{qpat}_2))\\
			& \text{(by (\ref{eq:subsumpt1}))} & = & var(\sigma(\kod{qpat}_2)).
		\end{align*}
		Therefore, by Lemma \ref{thm:thm34}, for a mapping $\mu''$ such that $\mu'' \povezano \nu''$, it holds
		$$\mu'' \in \semdg{\kod{qpat}_2}{{\ag_2}}{\mathit{df}({\ag_2})}.$$
		By Lemma \ref{thm:pattern_domain}, $dom(\mu'') = \kod{var}(\kod{qpat}_2)$.
		By Definition \ref{defn:query_evaluation}, it holds $\mu''_{\overline{\kod{dv}_2}} \in \semd{\kod{Q}_2}{\dataq}$, while
		$$dom(\mu''_{\overline{\kod{dv}_2}}) = \kod{var}(\kod{qpat}_2) \cap \overline{\kod{dv}_2}.$$
		By Definitions \ref{def:extension_restriction} and \ref{def:povezano}, from $\mu \povezano \nu$, $\mu'' \povezano \nu''$ and $\nu \preceq \nu''$, it holds $\mu \preceq \mu''$.
		Therefore, $\mu$ and $\mu''_{\overline{\kod{dv}_2}}$ are restrictions of $\mu''$.
		By Definition \ref{def:set_equality}, as $\kod{Q}_1 \dot{\seleq} \kod{Q}_2$, it holds $\overline{\kod{rv}_1} \subseteq \overline{\kod{rv}_2}$.
		Then, by Lemma \ref{thm:query_domain}, it holds $\kod{var}(\kod{qpat}_1) \cap \overline{\kod{dv}_1} \subseteq \kod{var}(\kod{qpat}_2) \cap \overline{\kod{dv}_2}$, i.e.~$dom(\mu) \subseteq dom(\mu''_{\overline{\kod{dv}_2}})$ and $\mu \preceq \mu''_{\overline{\kod{dv}_2}}$.
		Therefore, for each element of the set $\semd{\kod{Q}_1}{\dataq}$ there exists an extension that belongs to the set $\semd{\kod{Q}_2}{\dataq}$ for any dataset $\dataq$.
		By Definition \ref{defn:query_subsumption}, it holds $\kod{Q}_1 \dot{\sqsubseteq} \kod{Q}_2$.
	\end{proof}
\end{thm}

\subsection{Completeness}
\label{subsec:completeness}

The following completeness theorem states that for each pair of queries satisfying the containment relation, the Procedure \ref{thm:main} will confirm that.

\begin{thm}[Completeness]
	\label{thm:completeness}
	Let $\kod{Q}_1$ and $\kod{Q}_2$ be queries, $\Theta$ be a formula generated from $\kod{Q}_1$ (Def \ref{def:theta}) and $\Psi$ be a formula generated from $\kod{Q}_1$ and $\kod{Q}_2$ (Def \ref{def:psi}).
	If $\kod{Q}_1 \sqsubseteq \kod{Q}_2$ holds, then one of the following conditions is satisfied:
	\begin{itemize}[noitemsep,topsep=0pt]
		\item[(1)] $\Theta$ is valid, or
		\item[(2)] $\kod{Q}_1 \seleq \kod{Q}_2$ holds and $\Psi$ is valid.
	\end{itemize}
	\begin{proof} 
		%\ref{proof:completeness} is given in Appendix \ref{sec:appendix}. It is done using a contraposition.
		
			Case (1): $\kod{Q}_1$ is unsatisfiable.\\
		By Lemma \ref{thm:theta}, from the unsatisfiability of $\kod{Q}_1$, $\Theta$ is valid.
		
		Case (2): $\kod{Q}_1$ is satisfiable.\\
		We prove $\kod{Q}_1 \seleq \kod{Q}_2$ and the validity of $\Psi$.
		As query $\kod{Q}_1$ is satisfiable, there exists a dataset $\dataq$ and a mapping $\mu$ such that $\mu \in \semd{\kod{Q}_1}{\dataq}$.
		By Definition \ref{defn:query_containment}, from $\kod{Q}_1 \sqsubseteq \kod{Q}_2$, it holds $\semd{\kod{Q}_1}{\dataq} \subseteq \semd{\kod{Q}_2}{\dataq}$.
		Therefore, it holds $\mu \in \semd{\kod{Q}_2}{\dataq}$.
		By Definition \ref{def:rv}, it holds $dom(\mu) = \overline{\kod{rv}_1}$ and $dom(\mu) = \overline{\kod{rv}_2}$.
		%As query $\kod{Q}_2$ does not contain projection, it holds $\overline{\kod{dv}_2} = \kod{var}(\kod{qpat}_2)$.
		Therefore, it holds $\overline{\kod{rv}_1} = \overline{\kod{rv}_2}$, i.e.~by Definition \ref{def:set_equality}, $\kod{Q}_1 \seleq \kod{Q}_2$.
		
		Let us assume that $\Psi$ is not valid.
		Therefore, there exists an $\mathcal{L}$-structure $\mathfrak{D'} = (\mathcal{D'}, \intrprt^\mathcal{D'})$, such that $\mathfrak{D'} \models \neg \Psi$.
		By Lemma \ref{thm:psi}, as $\kod{Q}_1 \seleq \kod{Q}_2$ and $\kod{Q}_2$ does not contain projections, $\neg \Psi$ is equal to
		\begin{align}
			\label{eq:for_subsumption0}
			\neg \Big(\forall \overline{rv_1} \big( \Phi_1(\overline{rv}_1) \Rightarrow \sigma^{{\mathit{df}({\ag_2})}}(\kod{qpat}_2)\big)\Big),
		\end{align}	
		where $\ag_2$ is a query dataset of $\kod{Q}_2$, i.e.
		\begin{align}
			\label{eq:for_subsumption1}
			\exists \overline{rv_1} \ \big( \Phi_1(\overline{rv}_1) \wedge \neg \sigma^{{\mathit{df}({\ag_2})}}(\kod{qpat}_2) \big).
		\end{align}
		Note that this formula is a sentence (all variables are quantified). %, thus the valuation within $\mathfrak{D'}$ does not matter.
		By Lemma \ref{thm:model}, there exists a dataset $\dataq$ and the corresponding $\mathcal{L}$-structure $\mathfrak{D} = (\mathcal{D}, \intrprt^{\mathcal{D}})$, such that
		\begin{align}
			\label{eq:for_subsumption2}
			\mathfrak{D} \models \exists \overline{rv_1} \ \big( \Phi_1(\overline{rv}_1) \wedge \neg \sigma^{{\mathit{df}({\ag_2})}}(\kod{qpat}_2) \big).
		\end{align}
		Therefore, there exists a valuation $\nu$ defined on $\overline{rv_1}$, such that 
		\begin{align}
			\label{eq:for_subsumption3}
			(\mathfrak{D}, \nu) \models \Phi_1(\overline{rv}_1) \wedge \neg \sigma^{{\mathit{df}({\ag_2})}}(\kod{qpat}_2),
		\end{align}
		i.e.~by Definition \ref{defn:model},
		\begin{align}
			\label{eq:for_subsumption4}
			\model{\nu} \models \Phi_1(\overline{rv}_1) \text{~~~and~~~} \model{\nu} \models \neg \sigma^{{\mathit{df}({\ag_2})}}(\kod{qpat}_2).
		\end{align}
		%Let $\nu'$ denote a restriction of $\nu$ such that $dom(\nu') = var(\Phi_1(\overline{dv_1}))$.
		By Lemma \ref{thm:var_phi}, it holds $dom(\nu) = var(\Phi_1(\overline{rv_1}))$.
		Therefore, by Lemma \ref{thm:thm12}, for a mapping $\mu$ such that $\mu \povezano \nu$, it holds
		$$\mu \in \semd{\kod{Q}_1}{\dataq}.$$
		%By Definitions \ref{def:var_sigma} and \ref{def:phi}, it holds $dom(\nu') = var(\sigma(\kod{qpat}_1)) \setminus \overline{ov_1} = var(\sigma(\kod{qpat}_1)) \cap \overline{dv_1}$.
		By Lemma \ref{thm:set_equality}, as $\kod{Q}_1 \seleq \kod{Q}_2$ and $\kod{Q}_2$ does not contain projection, it holds
		$$\overline{\kod{rv}_1} = \kod{var}(\kod{qpat}_2).$$
		Therefore, it holds
		$$\sigma(\overline{\kod{rv}_1}) = \sigma(\kod{var}(\kod{qpat}_2)),$$
		i.e.~by Lemma \ref{thm:dom_mu_nu1},
		$$\overline{rv_1} = var(\sigma(\kod{qpat}_2)).$$
		Therefore, $dom(\nu)$ is equal to $var(\sigma(\kod{qpat}_2))$, and then, by Lemma \ref{thm:thm34}, it holds
		$$\mu \notin \semdg{\kod{qpat}_2}{{\ag_2}}{{\mathit{df}({\ag_2})}}.$$
		Then, by Definition \ref{defn:query_evaluation}, as query $\kod{Q}_2$ does not have projections, it holds
		$$\mu \notin \semd{\kod{Q}_2}{{\dataq}}.$$
		We conclude that $\semd{\kod{Q}_1}{\dataq} \subseteq \semd{\kod{Q}_2}{\dataq}$ does not hold, i.e.~by Definition \ref{defn:query_containment}, it does not hold $\kod{Q}_1 \sqsubseteq \kod{Q}_2$.
		This is a contradiction with the theorem assumption.
		Therefore, $\Psi$ is valid.
	\end{proof}
\end{thm}

Completeness theorem for subsumption (dual to Theorem \ref{thm:completeness}) can be proved as well.

\begin{thm}[Completeness for subsumption]
	\label{thm:completeness_subsumption}
	Let $\kod{Q}_1$ and $\kod{Q}_2$ be queries, $\Theta$ be a formula generated from $\kod{Q}_1$ (Def \ref{def:theta}) and $\Psi$ be a formula generated from $\kod{Q}_1$ and $\kod{Q}_2$ (Def \ref{def:psi}).
	If $\kod{Q}_1 \dot{\sqsubseteq} \kod{Q}_2$ holds, then one of the following conditions is satisfied:
	\begin{itemize}[noitemsep,topsep=0pt]
		\item[(1)] $\Theta$ is valid, or
		\item[(2)] $\kod{Q}_1 \dot{\seleq} \kod{Q}_2$ holds and $\Psi$ is valid.
	\end{itemize}
	\begin{proof} 
		%\ref{proof:completeness_subsumtion} is given in Appendix \ref{sec:appendix}. It is done by changing the proof of Theorem \ref{thm:completeness}.
		
		In the case when $\kod{Q}_1$ is unsatisfiable, the proof is exactly the same as the proof of Theorem \ref{thm:completeness}.
		Otherwise, the proof of $\kod{Q}_1 \dot{\seleq} \kod{Q}_2$ is done in the similar way as in the proof of Theorem \ref{thm:completeness}:
		By Definition \ref{def:satisfiable}, as query $\kod{Q}_1$ is satisfiable, there exists a dataset $\dataq$ and a mapping $\mu$ such that $\mu \in \semd{\kod{Q}_1}{\dataq}$.
		By Definition \ref{defn:query_subsumption}, from $\kod{Q}_1 \dot{\sqsubseteq} \kod{Q}_2$, there exists an extension $\mu'$ of $\mu$ ($dom(\mu) \subseteq dom(\mu')$) such that $\mu' \in \semd{\kod{Q}_2}{\dataq}$.
		By Definition \ref{def:rv}, it holds $dom(\mu) = \overline{\kod{rv}_1}$ and $dom(\mu') = \overline{\kod{rv}_2}$.
		Therefore, it holds $\overline{\kod{rv}_1} \subseteq \overline{\kod{rv}_2}$, i.e.~by Definition \ref{def:set_equality}, $\kod{Q}_1 \dot{\seleq} \kod{Q}_2$.
		
		In the rest of the proof, until the applying Lemma \ref{thm:set_equality}, the following changes are necessary.
		Instead of satisfiabilities (\ref{eq:for_subsumption0}), (\ref{eq:for_subsumption1}), (\ref{eq:for_subsumption2}), (\ref{eq:for_subsumption3}) and (\ref{eq:for_subsumption4}), in the subsumption case, as $\kod{Q}_2$ can contain projections, it holds respectively:
		\begin{align*}
			\neg \Big(\forall \overline{rv_1} \big( \Phi_1(\overline{rv}_1) \Rightarrow \exists \overline{ov_2} \  \sigma^{{\mathit{df}({\ag_2})}}(\kod{qpat}_2)\big)\Big)\\
			\exists \overline{rv_1} \ \big( \Phi_1(\overline{rv}_1) \wedge \neg (\exists \overline{ov_2} \ \sigma^{{\mathit{df}({\ag_2})}}(\kod{qpat}_2)) \big)\\
			\mathfrak{D} \models \exists \overline{rv_1} \ \big( \Phi_1(\overline{rv}_1) \wedge \forall \overline{ov_2} \ \neg \sigma^{{\mathit{df}({\ag_2})}}(\kod{qpat}_2) \big)\\
			(\mathfrak{D}, \nu) \models \Phi_1(\overline{rv}_1) \wedge \forall \overline{ov_2} \ \neg \sigma^{{\mathit{df}({\ag_2})}}(\kod{qpat}_2)\\
			\model{\nu} \models \Phi_1(\overline{rv}_1) \text{~~~and~~~} \model{\nu} \models \forall \overline{ov_2} \ \neg \sigma^{{\mathit{df}({\ag_2})}}(\kod{qpat}_2)
		\end{align*}
		The rest of the proof follows.
		By Definition \ref{defn:query_subsumption}, from $\mu \in \semd{\kod{Q}_1}{\dataq}$ and $\kod{Q}_1 \dot{\sqsubseteq} \kod{Q}_2$, there exists a mapping $\mu'$ such that $\mu \preceq \mu'$ and $\mu' \in \semd{\kod{Q}_2}{\dataq}$.
		%By Definition \ref{defn:query_evaluation}, there exists a mapping $\mu''$ such that $\mu'' \in \semdg{\kod{qpat}_2}{{\ag_2}}{{\mathit{df}({\ag_2})}}$ and $\mu_2 = \mu''_{\overline{\kod{dv}_2}}$.
		By Lemma \ref{thm:thm12}, for a valuation $\nu'$, such that $\mu' \povezano \nu'$, it holds
		$$\model{\nu'} \models \Phi_2(\overline{rv_2}),$$
		i.e.~by Definition \ref{def:phi},
		$$\model{\nu'} \models \exists \dot{\overline{ov_2}} \ \sigma^{{\mathit{df}({\ag_2})}}(\kod{qpat}_2),$$
		where $\dot{\overline{ov_2}} = var(\sigma^{{\mathit{df}({\ag_2})}}(\kod{qpat}_2)) \setminus \overline{rv_2}$.
		Note that by Definitions \ref{def:extension_restriction} and \ref{def:povezano}, as $\mu \povezano \nu$, $\mu' \povezano \nu'$, and $\mu \preceq \mu'$, it holds $\nu \preceq \nu'$.
		There exists an extension $\nu''$ of $\nu'$ to the variables from $\dot{\overline{ov_2}}$, such that
		$$\model{\nu''} \models \sigma^{{\mathit{df}({\ag_2})}}(\kod{qpat}_2),$$
		but also
		$$\model{\nu''} \models \exists \overline{ov_2} \ \sigma^{{\mathit{df}({\ag_2})}}(\kod{qpat}_2).$$
		From $\nu \preceq \nu'$ and $\nu' \preceq \nu''$, $\nu$ is a restriction of $\nu''$.
		Therefore, it holds 
		$$\model{\nu''} \models \forall \overline{ov_2} \ \neg \sigma^{{\mathit{df}({\ag_2})}}(\kod{qpat}_2).$$
		This is a contradiction, as $\model{\nu''}$ is a model of both, a formula and its negation.
		Therefore, $\Psi$ is valid.
	\end{proof}
\end{thm}

%--------------------------------------------------------------------
\section{Dealing with Non-Conjunctive Queries} 
\label{sec:modeling_ncq}

The containment problem of non-conjunctive queries, or more precisely queries containing the operators \kod{union} (Section \ref{subsec:union}) and \kod{optional} (Section \ref{subsec:optional}), and containing subqueries (Section \ref{subsec:subqueries}) can be transformed and/or reduced to the containment problem over (a set of) conjunctive queries presented in Section \ref{sec:modeling}.
Having correctness proved for conjunctive queries (Section \ref{sec:correctness}), the query containment problem for non-conjunctive queries presented in this section is reduced to a sound and complete approach.

\subsection{Operator \kod{union}} 
\label{subsec:union}

Dealing with queries containing the \kod{union} operator requires a transformation of their graph patterns into a special equivalent form where the \kod{union} operator can appear only outside the scope of other operators.
This condition is presented in the following definition.

\begin{defn}[Simple normal form]
A \textit{simple normal form} of a graph pattern is a graph pattern containing $n$ \kod{union}-free graph patterns $\kod{\pat}^\kod{i} \; \; (1 \le \kod{i} \le \kod{n})$ connected with $n-1$ \kod{union} operators: 	
$$\kod{\pat}^1 \ \kod{union} \ \kod{\pat}^2  \ \dots \ \kod{union} \ \kod{\pat}^\kod{n}$$
\end{defn}
\noindent Each graph pattern can be reduced to an equivalent pattern in a simple normal form following a standard DNF-style expansion \cite{perez2006}. 

\begin{lemma}
	\label{thm:union}
	For queries $\kod{Q}_\kod{1}$ and $\kod{Q}_\kod{2}$ of the form $\kod{select} ~ \overline{\kod{dv}} \ \kod{qpat}_\kod{1}$ and $\kod{select} \ \kod{*} \ \kod{qpat}_\kod{2}$ respectively, where query patterns $\kod{qpat}_\kod{1}$ and $\kod{qpat}_\kod{2}$ are in simple normal form consisting of union-free  graph patterns $\kod{\pat}_\kod{1}^\kod{i}$ ($1 \le \kod{i} \le \kod{m}$) and $\kod{\pat}_\kod{2}^\kod{j}$ ($1 \le \kod{j} \le \kod{n}$) respectively, queries $\kod{Q}_\kod{1}^\kod{i}$ ($1 \le \kod{i} \le \kod{m}$) and 
	$\kod{Q}_\kod{2}^\kod{j}$ ($1 \le \kod{j} \le \kod{n}$) are defined as:
	\begin{center}
		$\kod{Q}_\kod{1}^\kod{i} = \kod{select} ~ \overline{\kod{dv}} \ \kod{\{} \kod{\pat}_\kod{1}^\kod{i} \kod{\}}$, \\
		$\kod{Q}_\kod{2}^\kod{j} = \kod{select} \ \kod{*}  \kod{\{} \kod{\pat}_\kod{2}^\kod{j} \kod{\}}$,
	\end{center}
	where \kod{from} and \kod{from named} clauses of queries $\kod{Q}_\kod{1}$ and $\kod{Q}_\kod{2}$ are propagated to the queries $\kod{Q}_\kod{1}^\kod{i}$ and $\kod{Q}_\kod{2}^\kod{j}$, respectively.
	It holds:
	\begin{center}
		$\kod{Q}_\kod{1} \sqsubseteq \kod{Q}_\kod{2}$\\
		if and only if\\
		for each $i$, $1 \le \kod{i} \le \kod{m}$\\ 
		there exists $j$, $1 \le \kod{j} \le \kod{n}$, \\
		such that  $\kod{Q}_\kod{1}^\kod{i} \sqsubseteq \kod{Q}_\kod{2}^\kod{j}$.
	\end{center}
	\begin{proof}
		%\ref{proof:union} is given in Appendix \ref{sec:appendix}. It uses Definitions \ref{def:projection}, \ref{definition_sem} and \ref{defn:query_evaluation}.
		
		$(\Rightarrow)$
		In order to prove this direction of lemma,
		for any $\kod{i}=\kod{1}..\kod{m}$, there should be $\kod{j}=\kod{1}..\kod{n}$ such that $\kod{Q}_\kod{1}^\kod{i} \sqsubseteq \kod{Q}_\kod{2}^\kod{j}$.
		Assume that
		$$\mu \in \semd{\kod{Q}_\kod{1}^\kod{i}}{\dataq},$$
		for some $\kod{i}$, $\kod{i} \in \{\kod{1}, \dots, \kod{m}\}$ and some dataset $\dataq$.
		By Definitions \ref{def:projection} and \ref{defn:query_evaluation}, there exists a mapping $\mu'$, such that $\mu = \mu'_{\overline{\kod{dv}}}$, and $\mu' \in \semdg{\kod{\pat}_\kod{1}^\kod{i}}{\ag_1}{\mathit{df}(\ag_1)}$, where $\ag_1$ is a query dataset specified by the \kod{from} and \kod{from named} clauses of query $\kod{Q}_\kod{1}^\kod{i}$, and also of query $\kod{Q}_\kod{1}$.
		By Definition \ref{definition_sem}, it holds
		$$\mu' \in \semdg{\kod{\pat}_\kod{1}^\kod{1} \ \kod{union} \ \dots \ \kod{union} \ \kod{\pat}_\kod{1}^\kod{m}}{\ag_1}{\mathit{df}(\ag_1)}.$$
		By Definitions \ref{def:projection} and \ref{defn:query_evaluation}, it holds
		$$\mu \in \semd{\kod{Q}_\kod{1}}{\dataq},$$
		and also, as $\kod{Q}_\kod{1} \sqsubseteq \kod{Q}_\kod{2}$,
		$$\mu \in \semd{\kod{Q}_\kod{2}}{\dataq}.$$
		By Definitions \ref{def:projection} and \ref{defn:query_evaluation}, as there is no projections in $\kod{Q}_\kod{2}$,
		$$\mu \in \semdg{\kod{\pat}_\kod{2}^\kod{1} \ \kod{union} \ \dots \ \kod{union} \ \kod{\pat}_\kod{2}^\kod{n}}{\ag_2}{\mathit{df}(\ag_2)},$$
		where $\ag_2$ is a query dataset specified by the \kod{from} and \kod{from named} clauses of query $\kod{Q}_\kod{2}$, and also of every query $\kod{Q}_\kod{2}^\kod{j}, \kod{j}=\kod{1}..\kod{n}$.
		By Definition \ref{definition_sem}, it holds
		$$\mu \in \semdg{\kod{\pat}_\kod{2}^\kod{j}}{\ag_2}{\mathit{df}(\ag_2)},$$
		for some index $\kod{j}$, $\kod{j} \in \{\kod{1}, \dots ,\kod{n}\}$.
		By Definitions \ref{def:projection} and \ref{defn:query_evaluation}, as there is no projections in $\kod{Q}_\kod{2}^\kod{j}$, it holds
		$$\mu \in \semd{\kod{Q}_\kod{2}^\kod{j}}{\dataq}.$$
		
		$(\Leftarrow)$
		Let $\mu$ be a mapping, and $\dataq$ a dataset.
		Assume that
		$$\mu \in \semd{\kod{Q}_\kod{1}}{\dataq}.$$
		By Definitions \ref{def:projection} and \ref{defn:query_evaluation}, there exists a mapping $\mu'$, such that $\mu = \mu'_{\overline{\kod{dv}}}$, and
		$$\mu' \in \semdg{\kod{\pat}_\kod{1}^\kod{1} \ \kod{union} \ \dots \ \kod{union} \ \kod{\pat}_\kod{1}^\kod{m}}{\ag_1}{\mathit{df}(\ag_1)},$$
		where $\ag_1$ is a query dataset specified by the \kod{from} and \kod{from} \kod{named} clauses of query $\kod{Q}_\kod{1}$, and also of every query $\kod{Q}_\kod{1}^\kod{i}, \kod{i}=\kod{1}..\kod{m}$.
		By Definition \ref{definition_sem}, there exists an index $\kod{i}$, $\kod{i} \in \{\kod{1}, \dots, \kod{m}\}$, such that $\mu' \in \semdg{\kod{\pat}_\kod{1}^\kod{i}}{\ag_1}{\mathit{df}(\ag_1)}$.
		By Definitions \ref{def:projection} and \ref{defn:query_evaluation},
		it holds
		$$\mu \in \semd{\kod{Q}_\kod{1}^\kod{i}}{\dataq}.$$
		By the assumption, there exists an index $\kod{j}$, $\kod{j} \in \{\kod{1}, \dots, \kod{n}\}$, such that
		$$\mu \in \semd{\kod{Q}_\kod{2}^\kod{j}}{\dataq}.$$
		By Definitions \ref{def:projection} and \ref{defn:query_evaluation}, as there is no projections in $\kod{Q}_\kod{2}^\kod{j}$,
		$$\mu \in \semdg{\kod{\pat}_\kod{2}^\kod{j}}{\ag_2}{\mathit{df}(\ag_2)},$$
		where $\ag_2$ is a query dataset specified by the \kod{from} and \kod{from} \kod{named} clauses of query $\kod{Q}_\kod{2}^\kod{j}$, and also of query $\kod{Q}_\kod{2}$. 
		By Definition \ref{definition_sem}, it holds
		$$\mu \in \semdg{\kod{\pat}_\kod{2}^\kod{1} \ \kod{union} \ \dots \ \kod{union} \ \kod{\pat}_\kod{2}^\kod{n}}{\ag_2}{\mathit{df}(\ag_2)},$$
		i.e.~by Definitions \ref{def:projection} and \ref{defn:query_evaluation},
		$$\mu \in \semd{\kod{Q}_\kod{2}}{\dataq}.$$
	\end{proof}
\end{lemma}

Following Lemma \ref{thm:union}, the query containment problem of queries $\kod{Q}_\kod{1}$ and $\kod{Q}_\kod{2}$ can be reduced to the checks if for each $i$ ($1 \le \kod{i} \le \kod{m}$) there exists $j$ ($1 \le \kod{j} \le \kod{n}$) such that $\kod{Q}_\kod{1}^\kod{i} \sqsubseteq \kod{Q}_\kod{2}^\kod{j}$.
The soundness and the completeness of the proposed modeling are a direct consequence of this lemma.

For the query subsumption problem, a weaker form of Lemma \ref{thm:union} is needed, considering relation $\dot{\sqsubseteq}$ instead of $\sqsubseteq$, whose proof is analogous.

\subsection{Operator \kod{optional}} 
\label{subsec:optional}

All graph patterns explained so far demand to be matched completely. However, it is very useful to have possibility to add an information to the solution where the information is available, but do not reject the solution because some part of the query pattern does not match. Optional matching provides this facility.

For efficient usage of operator \kod{optional} in practice, it is necessary to consider only \textit{well-designed} graph patterns \cite{Pichler2014,Perez:2009,perez2006}.
\begin{defn}[Well-designed pattern]
\label{def:well-designed}
	A pattern $\kod{pat}$ is \emph{well-designed} if for all its subpatterns of the form $\kod{\pat}_1 \; \kod{optional} \; \kod{\pat}_2$, it holds that all variables appearing in $\kod{\pat}_2$, but not in $\kod{\pat}_1$, cannot appear in $\kod{pat}$ anywhere else except in $\kod{\pat}_2$. 
\end{defn}
\noindent The same restriction is also necessary for operators \kod{minus} and \kod{diff}.

Operator \kod{optional} can be reduced to a union of a conjunction and a negation, as proved in Lemma \ref{thm:optional}.
\begin{lemma}
	\label{thm:optional}
	Let $\ag$ be an \rdfs{} dataset, \kod{\gr} a graph within $\ag$, $\kod{\pat}_1$ and $\kod{\pat}_2$ graph patterns.
	Then:
	\begin{center}
		$\semdg{\kod{\pat}_1 \kod{ optional } \kod{\pat}_2}{\ag}{\kod{\gr}} = \semdg{\kod{\pat}_1 \, \kod{.} \,  \kod{\pat}_2 \ \kod{ union } \  \kod{\pat}_1 \; \kod{diff} \; \kod{\pat}_2}{\ag}{\kod{\gr}}$.
	\end{center}
	\begin{proof}
		This is a direct corollary of Definitions \ref{def:union_join} and \ref{definition_sem}.
	\end{proof}
\end{lemma}

\noindent Note that a graph pattern being introduced by a previously described reduction of a well-designed \kod{optional} pattern is, by  Definition \ref{def:well-designed}, also a well-designed pattern.

\subsection{Subqueries as Graph Patterns} 
\label{subsec:subqueries}

Subqueries allow embedding queries within other queries in order to facilitate a composition of new queries and a reuse of existing queries.
Subqueries are simpler than queries because they cannot contain \kod{from} and \kod{from named} clauses, and are used in a place of a graph pattern \cite{Angles:Subqueries}.
During the evaluation of a query, the inner most subquery is evaluated first (according to Definition \ref{definition_sem}), its results are projected up to the outer query, and only distinguished variables are visible (in scope) to the outer query (according to Definition \ref{def:var}).

In this section, we present a subquery elimination procedure, i.e.~reduction of a query to an equivalent query without a subquery. We prove the correctness of this procedure.
In order to simplify the proofs, we assume that queries do not contain \kod{graph}, \kod{diff}, and \kod{minus} operators. If these operators are present, the subquery elimination is also possible, as the expressive power of the \sparql{} language with and without subqueries is the same \cite{kaminski}.

Definition \ref{def:filter_nf} presents a special form of a graph pattern that simplifies reducing a graph pattern with a subquery into an equivalent query (by Definition \ref{defn:query_equivalence}) without the subquery.
%It requires that all \kod{filter} clauses within graph patterns are shifted to the right, at the very end of the pattern and if there exist more than one \kod{filter} clause, they are combined into a single one.
%\todo{prethodna reccenica je kao neka procedura koja govori kako se od obicnog dobija filter normal form - to svakako ne moze ovako da se napise jer niti definicija moze nesto da zahteva niti je jasno da je u pitanju procedura transformacije - ako zelimo proceduru transformacije, onda to treba da ide nakon definicije i da bude jasno da je u pitanju procedura.}

%\todo{U ovoj definiciji i narednih par lemi treba videti kako se uklapaju graf obrasci diff, minus i graph.}

\begin{defn}[\kod{filter}-normal form]
	\label{def:filter_nf}
	A graph pattern $\kod{\pat}$ is in \emph{\kod{filter}-normal form} if:
	\begin{itemize}
		\item[(i)] $\kod{\pat}$ is a triple pattern,
		\item[(ii)] $\kod{\pat}$ is $\kod{\{Q$_\kod{sq}$\}}$ and the query pattern $\kod{qpat}_\kod{sq}$ of $\kod{Q$_\kod{sq}$}$ is in \kod{filter}-normal form,
		\item[(iii)] $\kod{\pat}$ is $\kod{\pat}_1 \kod{.} \kod{\pat}_2$, where $\kod{\pat}_1$ and $\kod{\pat}_2$ are in \kod{filter}-normal form (i), (ii) or (iii),
		\item[(iv)] $\kod{\pat}$ is $\kod{\pat}_1 \kod{ filter R}$ and $\kod{\pat}_1$ is in \kod{filter}-normal form (i), (ii) or (iii).
	\end{itemize}
\end{defn}

The following lemma shows that any graph pattern (considered so far) can be reduced to a graph pattern in \kod{filter}-normal form.

\begin{lemma}
	\label{thm:filter_nf}
	Let $\ag$ be an \rdfs{} dataset and \kod{\gr} a graph within $\ag$.
	For any graph pattern $\kod{\pat}$ there exists a graph pattern $\kod{\pat}'$ in \kod{filter}-normal form such that:
	\begin{center}
		$\semdg{\kod{\pat}}{\ag}{\kod{G}} = \semdg{\kod{\pat}'}{\ag}{\kod{G}}.$
	\end{center}
	\begin{proof}
		%\ref{proof:filter_nf} is given in Appendix \ref{sec:appendix}. It is done by induction over the graph pattern $\kod{\pat}$.
		
		The lemma is proved by induction over graph pattern $\kod{\pat}$.
		\begin{description}
			\item \kod{\pat{}} is \kod{\tpat}\\
			By Definition \ref{def:filter_nf}, $\kod{\pat}$ is in \kod{filter}-normal form.
			
			\item $\kod{\pat}$ is $\kod{\pat}_1 . \kod{\pat}_2$\\
			By Definition \ref{definition_sem}, it holds
			$$\semdg{\kod{\pat}}{\ag}{\kod{G}} = \semdg{\kod{\pat}_1}{\ag}{\kod{G}} \bowtie \semdg{\kod{\pat}_2}{\ag}{\kod{G}}.$$
			By induction hypothesis, for $\kod{\pat}_1$ and $\kod{\pat}_2$ there exist graph patterns $\kod{\pat}_1'$ and $\kod{\pat}_2'$ in \kod{filter}-normal form such that $$\semdg{\kod{\pat}_1}{\ag}{\kod{G}} = \semdg{\kod{\pat}_1'}{\ag}{\kod{G}} \text{ and }$$
			$$\semdg{\kod{\pat}_2}{\ag}{\kod{G}} = \semdg{\kod{\pat}_2'}{\ag}{\kod{G}}.~~~~~~~$$
			\begin{itemize}
				\item[Case 1:] $\kod{\pat}_1'$ and $\kod{\pat}_2'$ are of the form (i), (ii) or (iii). \\
				Let $\kod{\pat}'$ denote $\kod{\pat}_1' . \kod{\pat}_2'$.
				Then, by Definition \ref{definition_sem}, it holds
				$$\semdg{\kod{\pat}'}{\ag}{\kod{G}} = \semdg{\kod{\pat}_1'}{\ag}{\kod{G}} \bowtie \semdg{\kod{\pat}_2'}{\ag}{\kod{G}},$$
				i.e.~
				$$\semdg{\kod{\pat}'}{\ag}{\kod{G}} = \semdg{\kod{\pat}_1}{\ag}{\kod{G}} \bowtie \semdg{\kod{\pat}_2}{\ag}{\kod{G}}.$$
				Therefore, it holds $\semdg{\kod{\pat}}{\ag}{\kod{G}} = \semdg{\kod{\pat}'}{\ag}{\kod{G}}$ and by Definition \ref{def:filter_nf}, $\kod{\pat}'$ is in \kod{filter}-normal form.
				
				\item[Case 2:] $\kod{\pat}_1'$ has the form (iv), i.e.\\
				$\kod{\pat}_1' = \kod{\pat}_1'' \kod{ filter R}_1$,\\ while $\kod{\pat}_2'$ is of the form (i), (ii) or (iii).\\
				Then, it holds
				$$\semdg{\kod{\pat}_1'}{\ag}{\kod{G}} \bowtie \semdg{\kod{\pat}_2'}{\ag}{\kod{G}} = \semdg{\kod{\pat}_1'' \kod{ filter R}_1}{\ag}{\kod{G}} \bowtie \semdg{\kod{\pat}_2'}{\ag}{\kod{G}},$$
				i.e.~by Definition \ref{definition_sem},
				$$\semdg{\kod{\pat}_1'}{\ag}{\kod{G}} \hspace*{-2pt} \bowtie \hspace*{-2pt} \semdg{\kod{\pat}_2'}{\ag}{\kod{G}} \hspace*{-2pt} = \hspace*{-2pt} \{ \mu \hspace*{-2pt} \in \hspace*{-2pt} \semdg{\kod{\pat}_1''}{\ag}{\kod{\gr}} \ | \ \mu \zadovoljava \kod{R}_1 \} \hspace*{-1pt} \bowtie \hspace*{-1pt} \semdg{\kod{\pat}_2'}{\ag}{\kod{G}}.$$
				By Definition \ref{defn:definition_filter}, if for a mapping $\mu \in \semdg{\kod{\pat}_1''}{\ag}{\kod{\gr}}$, it holds $\mu \zadovoljava \kod{R}_1$, then for its extension $\mu \cup \mu_2$, where $\mu_2 \in \semdg{\kod{\pat}_2'}{\ag}{\kod{G}}$ and $\mu_2 \comp \mu$, it holds $\mu \cup \mu_2 \zadovoljava \kod{R}_1$.
				Therefore, it holds:
				$$\semdg{\kod{\pat}_1'}{\ag}{\kod{G}} \bowtie \semdg{\kod{\pat}_2'}{\ag}{\kod{G}} = $$
				$$\{ \mu \cup \mu_2 \in \semdg{\kod{\pat}_1''}{\ag}{\kod{\gr}} \bowtie \semdg{\kod{\pat}_2'}{\ag}{\kod{\gr}} \ | \ \mu  \cup \mu_2 \zadovoljava \kod{R}_1 \},$$
				i.e.~by Definition \ref{definition_sem},
				$$\semdg{\kod{\pat}_1'}{\ag}{\kod{G}} \bowtie \semdg{\kod{\pat}_2'}{\ag}{\kod{G}} = \semdg{(\kod{\pat}_1'' \kod{.} \kod{\pat}_2') \kod{ filter R}_1}{\ag}{\kod{G}}.$$
				%By Definition \ref{def:filter_nf}, as $\kod{\pat}_1'$ is in \kod{filter}-normal form and $\kod{\pat}_1' = \kod{\pat}_3 \kod{ filter R}_3$, $\kod{\pat}_3$ is in \kod{filter}-normal form and not in form $\kod{\pat}_4 \kod{ filter R}_4$.
				By Definition \ref{def:filter_nf}, $(\kod{\pat}_1'' \kod{.} \kod{\pat}_2') \kod{ filter R}_1$ is in the \kod{filter}-normal form, and it holds
				$$\semdg{\kod{\pat}}{\ag}{\kod{G}} = \semdg{(\kod{\pat}_1'' \kod{.} \kod{\pat}_2') \kod{ filter R}_1}{\ag}{\kod{G}}.$$
				
				\item[Case 3:] $\kod{\pat}_2'$ is in the form (iv), while $\kod{\pat}_1'$ is in the form (i), (ii) or (iii).\\
				This case is reduced to the previous one because of the commutativity of the \sparql{} operator \kod{.} \cite{perez2006}, i.e.~
				$$\semdg{\kod{\pat}_1'}{\ag}{\kod{G}} \bowtie \semdg{\kod{\pat}_2'}{\ag}{\kod{G}} = \semdg{\kod{\pat}_2'}{\ag}{\kod{G}} \bowtie \semdg{\kod{\pat}_1'}{\ag}{\kod{G}}.$$
				
				\item[Case 4:] $\kod{\pat}_1'$ and $\kod{\pat}_2'$ are in the form (iv), i.e.~in the form $\kod{\pat}_1'' \kod{ filter R}_1$ and $\kod{\pat}_2'' \kod{ filter R}_2$, respectively.\\
				Applying construction of a normal form from the cases 2 and 3, we get:
				$$\semdg{\kod{\pat}}{\ag}{\kod{G}} = \semdg{(\kod{\pat}_1'' \kod{.} \kod{\pat}_2'') \kod{ filter R}_1 \kod{ filter R}_2}{\ag}{\kod{G}},$$
				i.e.~by Definition \ref{defn:definition_filter}, it holds
				$$\semdg{\kod{\pat}}{\ag}{\kod{G}} = \semdg{(\kod{\pat}_1'' \kod{.} \kod{\pat}_2'') \kod{ filter R}_1 \kod{\&\&R}_2}{\ag}{\kod{G}}.$$
				By Definition \ref{def:filter_nf}, $(\kod{\pat}_1'' \kod{.} \kod{\pat}_2'') \kod{ filter R}_1 \kod{\&\&R}_2$ is in the \kod{filter}-normal form.
			\end{itemize}
			Note that all the variables appearing in \kod{filter} clause (cases 2, 3 and 4) appear also in the graph pattern in front of it.
			This is a consequence of the same feature that holds for the initial graph patterns, while this transformation preserves it.
			
			\item $\kod{\pat}$ is $\kod{\pat}_1 \kod{ filter } \kod{R}$\\
			By Definition \ref{definition_sem}, it holds
			$$\semdg{\kod{\pat}}{\ag}{\kod{G}} = \{ \mu \in \semdg{\kod{\pat}_1}{\ag}{\kod{\gr}} \ | \ \mu \zadovoljava \kod{R} \}.$$
			By induction hypothesis, for $\kod{\pat}_1$ there exists a graph pattern $\kod{\pat}_1'$ in the \kod{filter}-normal form such that
			$$\semdg{\kod{\pat}_1}{\ag}{\kod{G}} = \semdg{\kod{\pat}_1'}{\ag}{\kod{G}}.$$
			Therefore, it holds
			$$\semdg{\kod{\pat}}{\ag}{\kod{G}} = \{ \mu \in \semdg{\kod{\pat}_1'}{\ag}{\kod{\gr}} \ | \ \mu \zadovoljava \kod{R} \},$$
			i.e.~by Definition \ref{definition_sem}, $$\semdg{\kod{\pat}}{\ag}{\kod{G}} = \semdg{\kod{\pat}_1' \kod{ filter } \kod{R}}{\ag}{\kod{G}}.$$
			If $\kod{\pat}_1'$ is in the form (i), (ii) or (iii) , let $\kod{\pat}'$ denote $\kod{\pat}_1' \kod{ filter } \kod{R}$.
			Then, it holds $\semdg{\kod{\pat}}{\ag}{\kod{G}} = \semdg{\kod{\pat}'}{\ag}{\kod{G}}$ and by Definition \ref{def:filter_nf}, $\kod{\pat}'$ is in the \kod{filter}-normal form.
			
			If $\kod{\pat}_1'$ is in the form (iv), i.e.~$\kod{\pat}_1'' \kod{ filter R}_1$, let $\kod{\pat}'$ denote $\kod{\pat}_1'' \kod{ filter } \kod{R\&\&R}_1$.
			By Definition \ref{defn:definition_filter}, it holds
			$$\semdg{\kod{\pat}_1'' \kod{ filter R}_1 \kod{ filter } \kod{R}}{\ag}{\kod{G}} =$$
			$$\semdg{\kod{\pat}_1'' \kod{ filter } \kod{R\&\&R}_1}{\ag}{\kod{G}}.$$
			Then, it holds $\semdg{\kod{\pat}}{\ag}{\kod{G}} = \semdg{\kod{\pat}'}{\ag}{\kod{G}}$ and by Definition \ref{def:filter_nf}, $\kod{\pat}'$ is in \kod{filter}-normal form.
			
			Note that it holds
			$$\kod{var}(\kod{R\&\&R}_1) \subseteq \kod{var}(\kod{\pat}_1''),$$
			i.e.~this feature holds for the transformed graph pattern as well.

			\item $\kod{\pat}$ is $\kod{\{\pat}_1\kod{\}}$\\
			By Definition \ref{definition_sem}, it holds
			$$\semdg{\kod{\{\pat}_1\kod{\}}}{\ag}{\kod{G}} = \semdg{\kod{\pat}_1}{\ag}{\kod{G}}.$$
			By induction hypothesis, for $\kod{\pat}_1$ there exists a graph pattern $\kod{\pat}_1'$ in \kod{filter}-normal form such that
			$$\semdg{\kod{\pat}_1}{\ag}{\kod{G}} = \semdg{\kod{\pat}_1'}{\ag}{\kod{G}}.$$
			Therefore, by Definition \ref{definition_sem}, $$\semdg{\kod{\pat}}{\ag}{\kod{G}} = \semdg{\kod{\pat}_1'}{\ag}{\kod{G}},$$
			where $\kod{\pat}_1'$ is in \kod{filter}-normal form.
			
			\item $\kod{\pat}$ is $\kod{\{Q$_\kod{sq}$\}}$\\
			By induction hypothesis, for $\kod{qpat}_\kod{sq}$ there exists a graph pattern $\kod{qpat}_\kod{sq}'$ in \kod{filter}-normal form such that
			$$\semdg{\kod{qpat}_\kod{sq}}{\ag}{\kod{G}} = \semdg{\kod{qpat}_\kod{sq}'}{\ag}{\kod{G}}.$$
			Let $\kod{Q}_\kod{sq}'$ be a query obtained from $\kod{Q}_\kod{sq}$ by switching the query pattern $\kod{qpat}_\kod{sq}$ with $\kod{qpat}_\kod{sq}'$.
			Therefore, by Definition \ref{defn:query_evaluation}, it holds
			$$\semd{\kod{Q}_\kod{sq}}{\ag} = \semd{\kod{Q}_\kod{sq}'}{\ag},$$
			i.e.~by Definition \ref{definition_sem},
			$$\semdg{\kod{qpat}}{\ag}{\kod{G}} = \semdg{\kod{\{Q$_\kod{sq}'$\}}}{\ag}{\kod{G}}.$$
			By Definition \ref{def:filter_nf}, $\kod{\{Q$_\kod{sq}'$\}}$ is in \kod{filter}-normal form.
		\end{description}
	\end{proof}
\end{lemma}

An example of a graph pattern $\kod{\pat}$ not satisfying \kod{filter}-normal form is given in Figure \ref{fig:filter_nf} (left), while its equivalent pattern $\kod{\pat}'$ in \kod{filter}-normal form is given in the same figure (right).
\begin{figure}[h]
	\startfig
	\begin{footnotesize}
		\begin{minipage}{0.005\textwidth}
			~
		\end{minipage}
		\begin{minipage}{0.46\textwidth}
			\startsubfig
			Graph pattern $\kod{\pat}$ \vspace*{-2mm}
			\begin{lstlisting}[style=sparql,numberstyle=\tiny\color{gray},numbers=left,numbersep=12pt,mathescape=true,escapeinside={(*}{*)}]
?y a ?t
filter (?t = :SoloArtist) .
?y :hometown ?z .
?z :name ?w
filter (?w = "Los Angeles")
			\end{lstlisting}
			\vspace*{7mm}
			\endsubfig
		\end{minipage}
		\begin{minipage}{0.03\textwidth}
		~
		\end{minipage}
		\begin{minipage}{0.48\textwidth}
		\startsubfig
		Graph pattern $\kod{\pat}'$  \vspace*{-2mm}
		\begin{lstlisting}[style=sparql,numberstyle=\tiny\color{gray},numbers=left,numbersep=12pt,mathescape=true,escapeinside={(*}{*)}]
{
	?y a ?t .
	?y :hometown ?z .
	?z :name ?w
}
filter (?t = :SoloArtist &&
         ?w = "Los Angeles")
	\end{lstlisting}
	\endsubfig
\end{minipage}
	\end{footnotesize}
	\vspace*{3px}
	\endfig
	\caption{Reducing the graph pattern $\kod{\pat}$ to its equivalent pattern $\kod{\pat}'$ in the \kod{filter}-normal form}
	\label{fig:filter_nf}
\end{figure}

\begin{lemma}
	\label{thm:subquery_form}
	Let $\ag$ be an \rdfs{} dataset and \kod{\gr} a graph within $\ag$.
	For a graph pattern $\kod{\pat}$ that contains a subquery $\kod{\{Q$_\kod{sq}$\}}$, there exists a graph pattern $\kod{\pat}'$ such that it holds 
	$\semdg{\kod{\pat}}{\ag}{\kod{G}} = \semdg{\kod{\pat}'}{\ag}{\kod{G}}$, and $\kod{\pat}'$ has one of the following forms: 
	\begin{itemize}
		\item $\kod{\{Q$_\kod{sq}'$\}}$, 
		\item $\kod{\{Q$_\kod{sq}'$\}} \; \kod{filter R}$, 
		\item $\kod{\pat}_1 \kod{.} \kod{\{Q$_\kod{sq}'$\}}$, 
		\item $\kod{\pat}_1 \kod{.} \kod{\{Q$_\kod{sq}'$\}} \; \kod{ filter R}$ 
	\end{itemize}
	where $\kod{\pat}_1$ is in \kod{filter}-normal form (i), (ii) or (iii) and $\kod{\{Q$_\kod{sq}'$\}}$ is in \kod{filter}-normal.
	\begin{proof}
		A proof of this lemma is a direct consequence of Definition \ref{def:filter_nf}, Lemma \ref{thm:filter_nf} and the associativity and commutativity of the SPARQL operator \kod{.} \cite{perez2006}.
		%		\ref{proof:subquery_form} is given in Appendix \ref{sec:appendix}.
	\end{proof}
\end{lemma}

Some variables in a query can be renamed, while the evaluation of the transformed query remains the same, as stated in the following lemma.

\begin{lemma}
	\label{thm:rename}
	Let $\dataq$ be an \rdfs{} dataset.
	Let $\kod{Q}$ be a query, $\kod{qpat}$ and $\overline{\kod{dv}}$ its query pattern and its set of distinguished variables, respectively.
	Let $\kod{qpat}'$ be a graph pattern obtained from $\kod{qpat}$ by renaming all the variables from $\kod{var}(\kod{qpat}) \setminus \overline{\kod{dv}}$, while introducing fresh new ones.
	Let $\kod{Q}'$ be a query obtained from $\kod{Q}$ by changing its query pattern $\kod{qpat}$ with $\kod{qpat}'$.
	Then it holds:
	\begin{center}
		$\semd{\kod{Q}}{\dataq} = \semd{\kod{Q}'}{\dataq}$.
	\end{center}
	\begin{proof}		
		%\ref{proof:rename} is given in Appendix \ref{sec:appendix}. It uses Definitions \ref{def:extension_restriction}, \ref{def:projection} and \ref{defn:query_evaluation}.
		
		By Definition \ref{defn:query_evaluation}, it holds
		$$\semd{\kod{Q}}{\dataq} = \PP_{\overline{\kod{dv}}}(\semdg{\kod{qpat}}{\ag}{\mathit{df}(\ag)}),$$
		i.e.~by Definition \ref{def:projection},
		$$\semd{\kod{Q}}{\dataq} = \{\mu_{\overline{\kod{dv}}} \;|\; \mu \in \semdg{\kod{qpat}}{\ag}{\mathit{df}(\ag)}\}.$$
		By the construction of $\kod{qpat}'$, it holds
		$$\{\mu_{\overline{\kod{dv}}} \;|\; \mu \in \semdg{\kod{qpat}}{\ag}{\mathit{df}(\ag)}\} = $$ $$\{\mu'_{\overline{\kod{dv}}} \;|\; \mu' \in \semdg{\kod{qpat}'}{\ag}{\mathit{df}(\ag)}\},$$
		as, by Definition \ref{def:extension_restriction}, mappings from $\semdg{\kod{qpat}}{\ag}{\mathit{df}(\ag)}$ and from $\semdg{\kod{qpat}'}{\ag}{\mathit{df}(\ag)}$ match on variables from $\overline{\kod{dv}}$.
		% (in the same time, they are not defined on them, or they are defined and their values are the same).
		Therefore, it holds
		$$\semd{\kod{Q}}{\dataq} = \{\mu'_{\overline{\kod{dv}}} \;|\; \mu' \in \semdg{\kod{qpat}'}{\ag}{\mathit{df}(\ag)}\}.$$
		By Definition \ref{def:projection} it holds
		$$\semd{\kod{Q}}{\dataq} = \PP_{\overline{\kod{dv}}}(\semdg{\kod{qpat}'}{\ag}{\mathit{df}(\ag)}).$$
		By Definition \ref{defn:query_evaluation} and the construction of the query $\kod{Q}'$, it holds
		$$\semd{\kod{Q}}{\dataq} = \semd{\kod{Q}'}{\dataq}.$$
	\end{proof}
\end{lemma}

Any query $\kod{Q}$ containing a subquery $\kod{Q}_\kod{sq}$ can be reduced to an equivalent query $\kod{Q}'$ that does not contain this subquery. The construction of such query $\kod{Q}'$ is given in the proof of the following lemma.

\begin{lemma}
	\label{thm:subquery}
	Let $\dataq$ be an \rdfs{} dataset.
	Let $\kod{Q}$ be a query such that its query pattern contains a subquery $\kod{\{Q$_\kod{sq}$\}}$.
	Then, there exists a graph pattern $\kod{\pat}_\kod{sq}$ such that it is not a subquery and if it is used within 
	\kod{Q} instead of $\kod{\{Q$_\kod{sq}$\}}$, forming a new query $\kod{Q}'$, it holds
	$\semd{\kod{Q}}{\dataq} = \semd{\kod{Q}'}{\dataq}$.
	
	\begin{proof}
		%\ref{proof:subquery} is given in Appendix \ref{sec:appendix}. It contains a construction of one such graph pattern, i.e.~$\kod{\pat}_\kod{sq}$ is constructed as a query pattern $\kod{qpat}_\kod{sq}$ of the subquery $\kod{Q}_\kod{sq}$, where all variables appearing only in the subquery and not in the outer query are renamed.
		
		According to Lemma \ref{thm:subquery_form}, there exists a query pattern that is in a form defined by the lemma, and that is equivalent to the original query pattern of the query \kod{Q}. Therefore, we can assume that the query \kod{Q} already contains a query pattern in such form. 
		
		We prove the case when the query pattern $\kod{qpat}$ of the query \kod{Q} is equal to $\kod{\pat}_1 \kod{.} \kod{\{Q$_\kod{sq}$\}} \; \kod{ filter R}$. All other cases are simpler and can be proved in analogous way.
		
		In the subquery $\kod{Q}_\kod{sq}$ of the query pattern $\kod{qpat}$, we can rename all the variables that are not distinguished ($\kod{var}(\kod{qpat}_\kod{sq}) \setminus \overline{\kod{dv}_\kod{sq}}$), by introducing fresh new variables that do not appear in the outer query \kod{Q} (anywhere else except in the subquery $\kod{Q}_\kod{sq}$). 
		This step is possible as a set of variables $\kod{V}$ is countable. 	After this renaming, by Lemma \ref{thm:rename}, it holds that the evaluation of the subquery $\kod{Q}'_\kod{sq}$ that contains renamed variables, has not been changed compared to the subquery $\kod{Q}_\kod{sq}$ . Let $\kod{qpat}_\kod{sq}'$ denote the query pattern of $\kod{Q}'_\kod{sq}$.
		
		The set of variables from the outer query \kod{Q} includes distinguished variables $\overline{\kod{dv}}$ and
		variables of the graph pattern $\kod{\pat}_1$, i.e.~$\kod{var}(\kod{\pat}_1)$.
		Therefore, it holds:
		$$\kod{var}(\kod{qpat}'_\kod{sq}) \cap (\kod{var}(\kod{\pat}_1) \cup \overline{\kod{dv}}) \subseteq \overline{\kod{dv}_\kod{sq}}$$
		i.e.~
		\begin{align}
			\kod{var}(\kod{qpat}'_\kod{sq}) \cap \kod{var}(\kod{\pat}_1) \subseteq \overline{\kod{dv}_\kod{sq}} \label{eq:subquery1}\\
			\text{and } \kod{var}(\kod{qpat}'_\kod{sq}) \cap \overline{\kod{dv}} \subseteq \overline{\kod{dv}_\kod{sq}} \label{eq:subquery2}
		\end{align}
		Note that it also holds
		$\kod{var}(\kod{R}) \subseteq \kod{var}(\kod{\pat}_1) \cup \kod{var}(\kod{\{Q}_\kod{sq}\kod{\}}).$
		i.e.~by Definition \ref{def:var}, %it holds
		\begin{align}
			\kod{var}(\kod{R}) \subseteq \kod{var}(\kod{\pat}_1) \cup \overline{\kod{dv}_\kod{sq}}.\label{eq:subquery0}
		\end{align}
		
		Let $\kod{qpat}'$ be a query pattern such that in query pattern $\kod{qpat}$, the subquery $\kod{Q}'_\kod{sq}$ is replaced with $\{\kod{qpat}'_\kod{sq}\}$.  Let $\kod{Q}'$ be a query such that in the query $\kod{Q}$, its query pattern \kod{qpat} is replaced with $\kod{qpat}'$. 
		Let us prove that $\semd{\kod{Q}}{\dataq} = \semd{\kod{Q}'}{\dataq}$.
		
		($\subseteq$)
		Let $\mu$ be a mapping such that $\mu \in \semd{\kod{Q}}{\dataq}$.
		By Definition \ref{defn:query_evaluation}, there exists a mapping $\mu^{\rm i}$, such that
		$$\mu = \mu^{\rm i}_{\overline{\kod{dv}}} \text{~~~and~~~} \mu^{\rm i} \in \semdg{\kod{qpat}}{\ag}{\mathit{df}(\ag)}.$$
		Therefore, by Definitions \ref{definition_sem} and \ref{definition_sem}, it holds
		$$\mu^{\rm i} \in \semdg{\kod{\pat}_1}{\ag}{\mathit{df}(\ag)} \bowtie \semd{\kod{Q}'_\kod{sq}}{\ag} \text{~~~and~~~}  \mu^{\rm i} \zadovoljava \kod{R}.$$
		By Definition \ref{def:union_join}, there exist compatible mappings $\mu_1$ and $\mu^{\rm ii}$ such that
		$$\mu^{\rm i} = \mu_1 \cup \mu^{\rm ii} \text{~~~and~~~} \mu_1 \in \semdg{\kod{\pat}_1}{\ag}{\mathit{df}(\ag)} \text{~~~and~~~} \mu^{\rm ii} \in \semd{\kod{Q}'_\kod{sq}}{\ag}.$$
		By Definition \ref{defn:query_evaluation}, there exists a mapping $\mu^{\rm iii}$ such that
		$$\mu^{\rm ii} = \mu^{\rm iii}_{\overline{\kod{dv}_\kod{sq}}} \text{~~~and~~~} \mu^{\rm iii} \in \semdg{\kod{qpat}'_\kod{sq}}{\ag}{\mathit{df}(\ag)}.$$
		By Definition \ref{def:compatible_mappings}, from $\mu_1 \comp \mu^{\rm ii}$, it holds $\mu_1 \comp \mu^{\rm iii}$, as there is no variable in $\kod{var}(\kod{qpat}'_\kod{sq})$ apart from $\overline{\kod{dv}_\kod{sq}}$ that can appear in $\kod{var}(\kod{\pat}_1)$ because of (\ref{eq:subquery1}).
		Let $\mu^{\rm iv}$ denote mapping $\mu_1 \cup \mu^{\rm iii}$.
		By Definition \ref{def:union_join}, it holds
		$$\mu^{\rm iv} \in \semdg{\kod{\pat}_1}{\ag}{\mathit{df}(\ag)} \bowtie \semdg{\kod{qpat}'_\kod{sq}}{\ag}{\mathit{df}(\ag)}.$$
		By Definition \ref{def:extension_restriction}, as $\mu^{\rm iv} = \mu_1 \cup \mu^{\rm iii}$, $\mu^{\rm i} = \mu_1 \cup \mu^{\rm ii}$ and $\mu^{\rm iii} \succeq \mu^{\rm ii}$, it holds $\mu^{\rm iv} \succeq \mu^{\rm i}$.
		Then, for $\mu^{\rm iv}$ as an extenstion of $\mu^{\rm i}$, from $\mu^{\rm i} \zadovoljava \kod{R}$, it holds:
		$$\mu^{\rm iv} \zadovoljava \kod{R}.$$
		Therefore, by Definition \ref{definition_sem}, it holds
		$$\mu^{\rm iv} \in \semdg{\kod{qpat}'}{\ag}{\mathit{df}(\ag)},$$
		i.e.~by Definition \ref{defn:query_evaluation},
		$$\mu^{\rm iv}_{\overline{\kod{dv}}} \in \semd{\kod{Q}'}{\dataq}.$$
		Let us prove that $\mu^{\rm iv}_{\overline{\kod{dv}}} = \mu$.
		By Definition \ref{def:extension_restriction}, from $\mu^{\rm iv} \succeq \mu^{\rm i}$, it holds $\mu^{\rm iv}_{\overline{\kod{dv}}} \succeq \mu^{\rm i}_{\overline{\kod{dv}}}$, i.e.~$\mu^{\rm iv}_{\overline{\kod{dv}}} \succeq \mu$.
		Also, it holds:
		\begin{align*}
			&\hspace*{18mm} dom(\mu) \hspace*{-20mm} & = \;\\
			&\text{(by Def \ref{def:extension_restriction})} &=\;& dom(\mu^{\rm i}) \cap \overline{\kod{dv}} \\
			& \text{(as } \mu^{\rm i} = \mu_1 \cup \mu^{\rm ii} \text{)}&=\;& (dom(\mu_1) \cup dom(\mu^{\rm ii})) \cap \overline{\kod{dv}}\\
			&\text{(by dist.of } \cap &=\;&  (dom(\mu_1) \cap \overline{\kod{dv}} ) \cup\\
			&\text{ over} \cup \text{)}&& (dom(\mu^{\rm ii}) \cap \overline{\kod{dv}})\\
			& \text{(by Def \ref{def:extension_restriction})} &=\;& (dom(\mu_1) \cap \overline{\kod{dv}} ) \cup\\
			&&& (dom(\mu^{\rm iii}) \cap \overline{\kod{dv}_\kod{sq}} \cap \overline{\kod{dv}})\\
			& \text{(by Lemma \ref{thm:pattern_domain})} &=\;& (dom(\mu_1) \cap \overline{\kod{dv}} ) \cup\\
			&&& (\kod{var}(\kod{qpat}'_\kod{sq}) \cap \overline{\kod{dv}_\kod{sq}} \cap \overline{\kod{dv}})\\
			& \text{(from (\ref{eq:subquery2}) )} &=\;& (dom(\mu_1) \cap \overline{\kod{dv}} ) \cup\\
			&&& (\kod{var}(\kod{qpat}'_\kod{sq}) \cap \overline{\kod{dv}})\\
			& \text{(by Lemma \ref{thm:pattern_domain})} &=\;& (dom(\mu_1) \cap \overline{\kod{dv}} ) \cup\\
			&&& (dom(\mu^{\rm iii}) \cap \overline{\kod{dv}})\\
			&\text{(by dist.of} \cap \text{over} \cup \text{)} \hspace*{-4mm} &=\;& (dom(\mu_1) \cup dom(\mu^{\rm iii})) \cap \overline{\kod{dv}}\\
			& \text{(as } \mu^{\rm iv} = \mu_1 \cup \mu^{\rm iii} \text{)}&=\;& dom(\mu^{\rm iv}) \cap \overline{\kod{dv}}\\
			& \text{(by Def \ref{def:extension_restriction})}&=\;& dom(\mu^{\rm iv}_{\overline{\kod{dv}}})
		\end{align*}
		Therefore, it holds $\mu^{\rm iv}_{\overline{\kod{dv}}} = \mu$, and 
		$$\mu \in \semd{\kod{Q}'}{\dataq}.$$
		
		($\supseteq$)
		Let $\mu$ be a mapping such that $\mu \in \semd{\kod{Q}'}{\dataq}$.
		By Definition \ref{defn:query_evaluation}, there exists a mapping $\mu^{\rm i}$, such that
		$$\mu = \mu^{\rm i}_{\overline{\kod{dv}}} \text{~~~and~~~} \mu^{\rm i} \in \semdg{\kod{qpat}'}{\ag}{\mathit{df}(\ag)}.$$
		Therefore, by Definition \ref{definition_sem}, it holds
		$$\mu^{\rm i} \in \semdg{\kod{\pat}_1}{\ag}{\mathit{df}(\ag)} \bowtie \semdg{\kod{qpat}'_\kod{sq}}{\ag}{\mathit{df}(\ag)} \text{~~~and~~~} \mu^{\rm i} \zadovoljava \kod{R}.$$
		By Definition \ref{def:union_join}, there exist compatible mappings $\mu_1$ and $\mu^{\rm ii}$ such that
		$$\mu^{\rm i} = \mu_1 \cup \mu^{\rm ii} \text{~~and~~} \mu_1 \in \semdg{\kod{\pat}_1}{\ag}{\mathit{df}(\ag)} \text{~~and~~} \mu^{\rm ii} \in \semdg{\kod{qpat}'_\kod{sq}}{\ag}{\mathit{df}(\ag)}.$$
		By Definition \ref{defn:query_evaluation}, it holds
		$$\mu^{\rm ii}_{\overline{\kod{dv}_\kod{sq}}} \in \semd{\kod{Q}'_\kod{sq}}{\ag}.$$
		By Definitions \ref{def:compatible_mappings} and \ref{def:extension_restriction}, as $\mu_1 \comp \mu^{\rm ii}$, it holds $\mu_1 \comp \mu^{\rm ii}_{\overline{\kod{dv}_\kod{sq}}}$.
		Let $\mu^{\rm iii}$ denote mapping $\mu_1 \cup \mu^{\rm ii}_{\overline{\kod{dv}_\kod{sq}}}$.
		By Definition \ref{def:union_join}, it holds
		$$\mu^{\rm iii} \in \semdg{\kod{\pat}_1}{\ag}{\mathit{df}(\ag)} \bowtie \semd{\kod{Q}'_\kod{sq}}{\ag}.$$
		By Definition \ref{def:extension_restriction}, it holds $\mu_1 \cup \mu^{\rm ii} \succeq \mu_1 \cup \mu^{\rm ii}_{\overline{\kod{dv}_\kod{sq}}}$, i.e.~$\mu^{\rm i} \succeq \mu^{\rm iii}$.
		Note that $dom(\mu^{\rm i}) = dom(\mu_1) \cup dom(\mu^{\rm ii})$ and $dom(\mu^{\rm iii}) = dom(\mu_1) \cup (dom(\mu^{\rm ii}) \cap \overline{\kod{dv}_\kod{sq}})$, i.e.~by Lemma \ref{thm:pattern_domain}, 
		$$dom(\mu^{\rm i}) = \kod{var}(\kod{\pat}_1) \cup \kod{var}(\kod{qpat}'_\kod{sq}),$$
		$$dom(\mu^{\rm iii}) = \kod{var}(\kod{\pat}_1) \cup (\kod{var}(\kod{qpat}'_\kod{sq}) \cap \overline{\kod{dv}_\kod{sq}}).$$
		Then,
		$$dom(\mu^{\rm i}) \setminus dom(\mu^{\rm iii}) = \kod{var}(\kod{qpat}'_\kod{sq}) \setminus \overline{\kod{dv}_\kod{sq}}.$$
		All of the variables from $dom(\mu^{\rm i}) \setminus dom(\mu^{\rm iii})$ are renamed, and they cannot belong to $\kod{var}(\kod{R})$, because of (\ref{eq:subquery0}).
		Therefore, from $\mu^{\rm i} \zadovoljava \kod{R}$ it holds
		$$\mu^{\rm iii} \zadovoljava \kod{R}.$$
		Then, by Definition \ref{defn:query_evaluation}, it holds
		$$\mu^{\rm iii}_{\overline{\kod{dv}}} \in \semd{\kod{Q}}{\dataq}.$$
		Let us prove that $\mu^{\rm iii}_{\overline{\kod{dv}}} = \mu$.
		By Definition \ref{def:extension_restriction}, from $\mu^{\rm iii} \preceq \mu^{\rm i}$, it holds $\mu^{\rm iii}_{\overline{\kod{dv}}} \preceq \mu^{\rm i}_{\overline{\kod{dv}}}$, i.e.~$\mu^{\rm iii}_{\overline{\kod{dv}}} \preceq \mu$.
		Also, it holds:
		\begin{align*}
			&\hspace*{20mm} dom(\mu) \hspace*{-20mm} & = \;\\
			&\text{(by Def \ref{def:extension_restriction})} &=\;& dom(\mu^{\rm i}) \cap \overline{\kod{dv}} \\
			& \text{(as } \mu^{\rm i} = \mu_1 \cup \mu^{\rm ii} \text{)}&=\;& (dom(\mu_1) \cup dom(\mu^{\rm ii})) \cap \overline{\kod{dv}}\\
			&\text{(by dist.of } \cap &=\;&  (dom(\mu_1) \cap \overline{\kod{dv}} ) \cup\\
			&\text{ over } \cup \text{)}&& (dom(\mu^{\rm ii}) \cap \overline{\kod{dv}})\\
			&\text{(by Lemma \ref{thm:pattern_domain})} &=\;& (dom(\mu_1) \cap \overline{\kod{dv}} ) \cup\\
			&&& (\kod{var}(\kod{qpat}'_\kod{sq}) \cap \overline{\kod{dv}})\\
			&\text{(from (\ref{eq:subquery2}) )} &=\;& (dom(\mu_1) \cap \overline{\kod{dv}} ) \cup\\
			&&& (\kod{var}(\kod{qpat}'_\kod{sq}) \cap \overline{\kod{dv}_\kod{sq}} \cap \overline{\kod{dv}})\\
			&\text{(by Lemma \ref{thm:pattern_domain})} &=\;& (dom(\mu_1) \cap \overline{\kod{dv}} ) \cup\\
			&&& (dom(\mu^{\rm ii}) \cap \overline{\kod{dv}_\kod{sq}} \cap \overline{\kod{dv}})\\
			&\text{(by Def \ref{defn:query_evaluation})} &=\;& (dom(\mu_1) \cap \overline{\kod{dv}} ) \cup\\
			&&& (dom(\mu^{\rm ii}_{\overline{\kod{dv}_\kod{sq}}}) \cap \overline{\kod{dv}})\\
			&\text{(by dist.of} \cap \text{over} \cup \text{)} \hspace*{-4mm} &=\;& (dom(\mu_1) \cup dom(\mu^{\rm ii}_{\overline{\kod{dv}_\kod{sq}}})) \cap \overline{\kod{dv}}\\
			& \text{(as } \mu^{\rm iii} = \mu_1 \cup \mu^{\rm ii}_{\overline{\kod{dv}_\kod{sq}}} \text{)}&=\;& dom(\mu^{\rm iii}) \cap \overline{\kod{dv}}\\
			& \text{(by Def \ref{def:extension_restriction})}&=\;& dom(\mu^{\rm iii}_{\overline{\kod{dv}}})
		\end{align*}
		Therefore, it holds $\mu^{\rm iii}_{\overline{\kod{dv}}} = \mu$, and 
		$$\mu \in \semd{\kod{Q}}{\dataq}.$$
	\end{proof}
\end{lemma}

If a query \kod{Q} contains more than one subquery, then the elimination of subqueries should be applied starting from the inner most subquery. 
This way, the containment problem concerning queries with subqueries is reduced to the containment problem of queries that do not contain subqueries. 

For the containment problem, we assume that there is no subqueries in the query $\kod{Q}_2$ or in cases where they are present in $\kod{Q}_2$, all their variables are projected up to $\kod{Q}_2$.
Otherwise, the construction of an equivalent query $\kod{Q}_2'$ from the proof of Lemma \ref{thm:subquery} would make projections present within $\kod{Q}_2'$, leading to an undecidable problem. 
When considering the subsumption relation, this restriction is not necessary, as the query subsumption problem is decidable in cases when a super-query contains projections \cite{Pichler2014}.

%--------------------------------------------------------------------
\section{Conclusions and Further Work}
\label{sec:conclusions}

In this paper we proved correctness of the approach described in \cite{solving_with_specs} and implemented by a tool \specs{} \cite{specssolver}. Correctness, i.e.~soundness and completeness, are proved for modeling conjunctive queries for both  containment and subsumption relation. Soundness and completeness proofs are extended to cover non-conjunctive queries containing operator \kod{union}, operator \kod{optional} and subqueries in cases when well designed patterns and \kod{filter}-normal form are considered. 

There are several possible directions for further work: extending the language coverage while keeping the soundness and completeness of the approach if possible, considering containment and subsumption in the context of \rdfs{} \textsc{schema}, considering containment and subsumption when a wider set of axioms is included (e.g.~$\mathcal{SHI}$ axioms \cite{10.1007/3-540-48242-3_11,Chekol:2012:SQC:2900728.2900730} or code refactoring context \cite{spasic2020verification}), applying a similar approach for different graph query languages (e.g.~XPath \cite{Schwentick2004,Miklau2004}, GQL \cite{hartig2018}), and making the presented proofs formal within a proof assistant \cite{isabelle,coq}.

	%%%%%%%%%%% The bibliography starts:
	
	%%  ios1.bst will be used to                               %%
	
	%\nocite{*} 
	
\bibliographystyle{cas-model2-names}

\bibliography{sparql}
	
	% or include bibliography directly:
	%\begin{thebibliography}{0}
	%\bibitem{r1} F. Author, Information about cited object.
	%
	%\bibitem{r2} S. Author and T. Author, Information about cited object.
	%\end{thebibliography}
	
	%\appendix
	%\setlength{\abovedisplayskip}{4pt}
	%\setlength{\belowdisplayskip}{4pt}
	%\setlength{\abovedisplayshortskip}{4pt}
	%\setlength{\belowdisplayshortskip}{4pt}
	%\section{Proofs}
	%\label{sec:appendix}
	
	%\input{appendix.tex}

\end{document}